\documentclass[numbook,natbib,final,runningheads]{svjour3}
\usepackage{natbib}
\usepackage{epsfig}
\usepackage{graphicx}
\usepackage{amssymb}
\usepackage{grffile}
\usepackage{epstopdf}

\usepackage{makeidx}
\usepackage{subfigure}
\makeindex

\newcommand{\degr}{\ifmmode^\circ\else$^\circ$\fi}
\newcommand{\lapprox} {\, \lower3pt\hbox{$\sim$}\llap{\raise2pt\hbox{$<$}}\,}
\newcommand{\gapprox} {\, \lower3pt\hbox{$\sim$}\llap{\raise2pt\hbox{$>$}}\,}

\begin{document}



\title{Deducing Electron Properties From Hard X-Ray
Observations}

\author{E.~P.~Kontar$^{1}$,
        J.~C.~Brown$^{1}$,
        A.~G.~Emslie$^{2}$,
        W.~Hajdas$^{3}$,
        G.~D.~Holman$^{4}$,
        G.~J.~Hurford$^{5}$,
        J.~Ka\v{s}parov{\'a}$^{6}$,
        P.~C.~V.~Mallik$^{1}$,
        A.~M.~Massone$^{7}$,
        M.~L.~McConnell$^{8}$,
        M.~Piana$^{9}$,
        M.~Prato$^{7}$,
        E.~J.~Schmahl$^{11}$, and
        E.~Suarez-Garcia$^{3,12}$
        }

\institute{$^{1}$ Department of Physics and Astronomy, University
                  of Glasgow, Kelvin Building, Glasgow, G12 8QQ,
                  U.K.\\
           $^{2}$ Department of Physics, Oklahoma State University, Stillwater, OK 74078, U.S.A., and
           Western Kentucky University, 1906 College Heights Blvd., Bowling Green, KY 42101\\
           $^{3}$ Paul Scherrer Institut, Villigen PSI, Switzerland\\
           $^{4}$ NASA Goddard Space Flight Center, Code 671,
                  Greenbelt, MD 20771, U.S.A.\\
           $^{5}$ Space Sciences Laboratory, University of California, Berkeley, CA 94720, U.S.A.\\
           $^{6}$ Astronomick\'{y} \'{u}stav AV \v{C}R, v.v.i., Fri\v{c}ova 298,
                  Ond\v{r}ejov, 251 65, Czech Republic \\
           $^{7}$ CNR-SPIN, Via Dodecaneso 33, I-16146 Genova, Italy  \\
           $^{8}$ Space Science Center, University of New Hampshire, Durham, NH 03824, U.S.A. \\
           $^{9}$ Dipartimento di Matematica, Universita di Genova, Via Dodecaneso 35, I-16146 Genova, Italy \\
           $^{10}$ Dipartimento di Matematica Pura ed Applicata, Universita di Modena e Reggio Emilia, Via Campi 213/b, I-41125 Modena, Italy\\
           $^{11}$ NWRA/CoRA, Boulder, Colorado, 80301, U.S.A. \\
           $^{12}$ Department of Nuclear and Particle Physics,
           University of Geneva, Quai Ernest Ansermet 24, 1212 Geneva, Switzerland
             }


\date{Version: \today  $\;\;\;$ Received XXX; accepted XXX}

\authorrunning{E.P.  Kontar et al.}
\titlerunning{Deducing electron properties}
\maketitle

\begin{abstract}
X-radiation from energetic electrons is the prime diagnostic
of flare-acceler\-ated electrons. The observed X-ray flux (and polarization state)
is fundamentally a convolution of the
cross-section for the hard X-ray emission process(es) in question
with the electron distribution function, which is in turn a
function of energy, direction, spatial location and time. To
address the problems of particle propagation and acceleration one
needs to infer as much information as possible on this electron
distribution function, through a deconvolution of this fundamental
relationship. This review presents recent progress toward this goal using
spectroscopic, imaging and polarization measurements, primarily
from the \textit{Reuven Ramaty High Energy Solar Spectroscopic Imager} ({\em
RHESSI}). Previous conclusions regarding the energy, angular
(pitch angle) and spatial distributions of energetic electrons in
solar flares are critically reviewed. We discuss the role and
the observational evidence of several radiation processes: free-free electron-ion,
free-free electron-electron, free-bound electron-ion bremsstrahlung,
photoelectric absorption and Compton back-scatter (albedo),
using both spectroscopic and imaging techniques. This unprecedented quality
of data allows for the first time inference of the angular distributions
of the X-ray-emitting electrons using albedo, improved model-independent
inference of electron energy spectra and emission measures
of thermal plasma. Moreover, imaging spectroscopy has revealed hitherto unknown details
of solar flare morphology and detailed spectroscopy of coronal, footpoint
and extended sources in flaring regions. Additional attempts to measure
hard X-ray polarization were not sufficient to put constraints on the degree of anisotropy
of electrons, but point to the importance of obtaining good quality
polarization data.
\end{abstract}

\setcounter{tocdepth}{3}
\tableofcontents

\section{INTRODUCTION\label{sec:7intro}} 

X-ray emission, because it is produced promptly in an optically-thin environment,
is one of the most direct methods with which to study energetic
electrons in solar flares.
\index{optically-thin source}
Such remote radiation measurements are
generally functions of photon energy $\epsilon$, direction ${\bf\Omega}$,
and time $t$. A common description of incoherent and partially
polarized X-ray radiation typical of solar flares is in terms of its
{\it photon intensity}\index{intensity!photon}
$I(\epsilon,{\bf \Omega},t)$
(photons~s$^{-1}$~cm$^{-2}$~sr$^{-1}$~keV$^{-1}$;
CGS units are used throughout the paper)
and the fractional degree and orientation of linear
polarization\index{polarization}\index{hard X-rays!polarization} $(P,\Psi)$.
The emergent photon intensity is the number of photons $dN$ that escape from the
source in the time interval $t$ to $t+dt$ in the energy range
$\epsilon$  to $\epsilon +d\epsilon$, from a solar source
with a direction and angular size defined by the elementary cone
$d{\bf \Omega}$, oriented with respect to the normal of the
detector of area $dS$:

\begin{equation}\label{eqn:into1}
dN= I(\epsilon,{\bf \Omega},t) \, d\epsilon \, \mbox{d} {S} \,
\mbox{d}{\Omega}\, \mbox{d}t.
\end{equation}
In other words, the intensity is the number of photons emitted by
a unit solid angle source (sr) on the Sun and received in a unit of time (1~s)
in a unit of energy band (1~keV) by a unit detector area (1~cm$^2$)
at the Earth.

For optically-thin radiation, the emergent intensity of the
radiation is simply a linear convolution of the cross-section for
the pertinent emission process and the electron phase-space
distribution function\index{electrons!distribution function} $f_{e}({\bf v},{\bf r},t)$
(electrons~cm$^{-3}$~[cm~s$^{-1}]^{-3}$), or equivalently the
particle {\it flux} $F(E,{\bf \Omega}^\prime,{\bf r}, t)$
(electrons~cm$^{-2}$~s$^{-1}$~sr~$^{-2}$~keV$^{-1}$),\index{electrons!energy distribution} differential in energy
$E$, and velocity solid angle ${\bf \Omega}^\prime$. Since $dE =
m_e v \, dv$ in the non-relativistic regime, it follows that $F(E,{\bf \Omega}^\prime,{\bf r}, t)
= v^2 f_e({\bf v},{\bf r},t)/m_e $, where $m_e$ (g) is the
electron mass.

For an elementary bremsstrahlung\index{bremsstrahlung} source of ambient plasma density
$n({\bf r})$, located at position ${\bf r}$ on the Sun, along the
line of sight ${\bf \Omega}$, which is subjected to an electron
flux spectrum $F(E,{\bf \Omega}^\prime, {\bf r}, t)$, the emergent
photon flux spectrum\index{hard X-rays!flux spectrum} at distance $R$
is the convolution
\begin{eqnarray}
I(\epsilon,{\bf \Omega}, t) = \int_{\ell}\int_{\Omega'}
\int_\epsilon^\infty n({\bf r}) F(E,{\bf r}, {\bf \Omega}', t)
Q({\bf \Omega},{\bf \Omega'},\epsilon, E) \, \mbox{d}E \,
\mbox{d}{\bf\Omega '} \, \mbox{d}\ell, \label{eqn:intro2}
\end{eqnarray}
where $\ell$ is the distance along the line of sight, $Q({\bf
\Omega},{\bf \Omega'},\epsilon,E)$
(cm$^2$~keV$^{-1}$~sr$^{-1}$) is the cross-section for the
pertinent hard X-ray emission process(es), differential in
$\epsilon$ and ${\bf \Omega}$. The dominant hard
X-ray emission process\index{bremsstrahlung} in solar flares is {\it bremsstrahlung}
radiation associated with electron deceleration in the Coulomb
field of an ion or other electron. For this process, the angular
dependence of $Q$ depends only on the angle $\theta^\prime =
\widehat{{\bf\Omega'}{\bf \Omega}}$ between the incoming electron
${\bf\Omega'}$ and the emitted photon ${\bf\Omega}$ directions, so
that $Q = Q(\epsilon, E, \theta')$.

To deduce the physical properties of energetic particles in solar
flares from observed hard X-ray quantities, the electron flux
spectrum\index{electrons!flux spectrum} $F(E,{\bf r}, {\bf \Omega}', t)$ must be deconvolved from
the emission cross-section $Q(\epsilon,E,\theta')$ in this
integral. Although the relation between the observable quantity
$I(\epsilon,{\bf \Omega}, t)$ and the physical electron flux $F(E,{\bf r}, {\bf
\Omega}', t)$ is linear, it is still nontrivial and the deconvolution
requires some rather insightful techniques. This fundamental problem
and the progress toward the solution of the problem using spectroscopic,
imaging and polarization measurements from the \textit{Reuven Ramaty High Energy Solar Spectroscopic Imager}\index{RHESSI@\textit{RHESSI}}
\citep[\textit{RHESSI},][]{2002SoPh..210....3L} are addressed in this review.
\index{satellites!RHESSI@\textit{RHESSI}}

In Section \ref{sec:7emission}, we review the physical processes
leading to X-ray emission in solar flares, including free-free
electron-ion bremsstrahlung (Section \ref{sec:7ei}), free-bound
\index{free-bound emission}
electron-ion emission (Section \ref{sec:7fb}), and free-free
electron-electron bremsstrahlung
\index{free-free emission}
\index{bremsstrahlung!electron-electron}
(Section \ref{sec:7ee}).
This review, however, does not discuss bound-bound transitions,
\index{bound-bound emission}
nor the emissions due to energetic ions such as nuclear gamma-ray lines,
\index{gamma-rays!nuclear line radiation}
ion gamma-ray continuum, and pseudo-continuum\index{gamma-rays!continuum}.

Since the emission mechanisms are well established, hard X-rays (HXR) are often
viewed as one of the most direct (i.e., least affected by propagation effects)
diagnostics of solar flare electrons. However, it must not be forgotten that
downward-propagating X-ray photons are effectively scattered
toward the observer by electrons in the dense layers of the solar
atmosphere, thereby complicating the diagnostic potential of hard
X-ray radiation. Section \ref{sec:7albedo} reviews spectroscopic
and imaging techniques to infer this Compton backscattered
(albedo)\index{albedo}\index{photospheric albedo}  component.
Progress in the deduction of primary, i.e., directly
flare-emitted, and photospherically-reflected X-rays from the
observed spectrum, is discussed.

Section \ref{sec:7fbar} presents an overview of the results
obtained using purely spectroscopic data (i.e., data integrated
over the source volume). Both forward fitting\index{inverse problem!forward fit} and regularized
inversion\index{inverse problem!regularized inversion} techniques to deduce the energy dependence of the {\it mean source electron flux spectrum} ${\bar F}(E)$ (averaged over
volume and solid angle) are reviewed\index{electrons!mean flux spectrum}.
Properties of the electron
flux distribution, deviations from power-law forms, low-energy
cutoffs\index{electrons!flux spectrum!low-energy cutoff}\index{low-energy cutoff} and interpretation
\index{particles!energy spectra!low-energy cutoff}
in terms of a thermal source are discussed.

Section \ref{sec:7angular} provides methods and results for the
case of an anisotropic angular distribution ${\bar F}(E, {\bf
\Omega})$ of electrons. The role of X-ray Compton scattering in
the solar atmosphere
(solar albedo)\index{hard X-rays!albedo}\index{albedo}\index{photospheric albedo} in deducing this angular
distribution is discussed. Recent {\it RHESSI} polarization measurements
and their implications for electron anisotropy\index{electrons!anisotropy}\index{anisotropy!electron} are also reviewed.

Section \ref{sec:7spatial} focuses on the spatial structure
$F(E,{\bf r})$ of the electron flux, using imaging spectroscopy
observations from {\em RHESSI}. Recently-developed,
visibility-based techniques to optimize the inference of electron
maps are discussed. Section \ref{sec:7discussion} highlights the major
finding from {\em RHESSI} and discusses the open questions.

The role of these results in the context of multi-wavelength
observations of solar flares is discussed by \citet{chapter2}.
The implications of these findings for electron transport,
and acceleration models are discussed by \citet{chapter3},
\citet{chapter6}, and \citet{chapter8}.



\section{X-Ray emission processes and energetic electrons}\label{sec:7emission}


When energetic electrons are deflected in close encounters with
ambient particles (both electrons and ions), a {\it
bremsstrahlung} (literally, ``braking radiation'') photon is
produced. \index{bremsstrahlung!electron-ion}  As stated in
Section \ref{sec:7intro}, the cross-section $Q(\epsilon, E,
\theta')$ for electron-ion free-free bremsstrahlung\index{free-free emission}
is a function of the
emitted photon energy $\epsilon$, the pre-collision electron
energy $E$, and the angle $\theta'$ between the direction of the
pre-collision electron and the outgoing photon \citep[see,
e.g.,][]{1959RvMP...31..920K}. For simplicity we often consider
only the solid-angle-integrated form of the cross-section
$Q(\epsilon, E)$ (and the corresponding scalar electron flux $F[E,
{\bf r}, t]$); however, it must be remarked that the effects of
the angular dependence of the cross-section can, for highly-beamed
electron distributions, be quite significant, leading to
substantial differences in the number of electrons required to
produce a given hard X-ray flux -- see Section~\ref{sec:7angular}.

The photons of energy from a few keV to a few hundred~keV
under consideration are mostly produced by collisional
electron-ion bremsstrahlung in the solar atmosphere.
Bremsstrahlung emission from energetic electrons
is more efficient than inverse Compton scattering
\index{hard X-rays!inverse Compton scattering}\index{Compton scattering!inverse}\index{inverse Compton radiation}
or synchrotron emission\index{synchrotron emission}
from the same electron population \citep{1967SvA....11..258K}\index{collisions!and column density}.
The responsible electrons have kinetic energies $E$~up to a few hundreds of~keV,
and so will collisionally stop within a column density
$N \approx E^2/6 \pi e^4 \Lambda
\sim 10^{17} \, E^2 \lapprox 10^{21}$~cm$^{-2}$, where $e$ (esu)
is the electronic charge and $\Lambda$ the Coulomb logarithm
\citep[e.g.,][]{1978ApJ...224..241E}.
Thus, the column density in a solar coronal loop of density
$10^{10}$ ~cm$^{-2}$ and length $10^{9}$~cm
is $10^{19}$~cm$^{-2}$, which stops electrons up to $\sim$10~keV,
while the upper chromosphere with typical column
densities $10^{20}$-$10^{22}$~cm$^{-2}$ can stop electrons
with the energies of $10-300$~keV.
The cross-section for
scattering the emitted photons is of the order of the Compton\index{cross-sections!Compton scattering}
cross-section\index{Compton scattering!cross-section}\index{Compton scattering} $\sigma _C \approx \pi r_o^2 \approx 2 \times
10^{-25}$~cm$^2$, where $r_o = e^2/m_e c^2$ is
the classical electron radius,
so that the optical depth $\tau = N \, \sigma_C
\approx 10^{-4}$ and there is negligible self-absorption in the
source, i.e., the source is optically thin\index{hard X-rays!optical depth}.
In Section \ref{sec:7albedo}, we shall address the issue
of so-called ``albedo''
\index{hard X-rays!albedo}
\index{albedo}\index{photospheric albedo}  photons -- photons
that are emitted downward toward the solar photosphere, which is optically thick,
and subsequently backscattered toward the observer. For now, we
consider only ``primary'' photons, i.e., those initially directed
toward the observer.  Integrating the photon intensity $I(\epsilon, \Omega)$
over the solid angle subtended by the source of area $A$, $d{\Omega}=dA/R^2$ and
making the volume element substitution
$d{\bf \Omega}\, d\ell = d^3{r}/R^2$, the fundamental
Equation~(\ref{eqn:intro2}) shows that the observed bremsstrahlung
flux (at time $t$; hereafter understood) at the Earth
(photons~s$^{-1}$~keV$^{-1}$~cm$^{-2}$ of detector area) is

\begin{equation}
I(\epsilon) = {1 \over 4 \pi R^2} \int_\epsilon^\infty \, \int_V
\, n({\bf r}) \, F(E, {\bf r}) \, Q(\epsilon, E) \, dE \, d^3{\bf
r}, \label{eqn:emslie_fund}
\end{equation}
where the second integral is taken over the source volume $V$.

\subsection{Electron-ion bremsstrahlung}\label{sec:7ei} 
\index{bremsstrahlung!electron-ion}

For electron-ion
bremsstrahlung\index{bremsstrahlung!electron-ion}, the full form
of the cross-section $Q(\epsilon, E)$ is given by formula 3BN of
\citet{1959RvMP...31..920K}. Numerical computations may be
facilitated by use of the simplified form published by
\citet{1997A&A...326..417H}. A frequently-used analytic
approximation to the cross-section $Q(\epsilon, E)$ is the Kramers
\index{bremsstrahlung!Kramers approximation} form

\begin{equation}
Q(\epsilon,E) = Z^2{\sigma_o \over \epsilon E},
\label{eqn:emslie_kramers}
\end{equation}
where $\sigma_o$ = $(8 \alpha/3) \, (m_e c^2) \, r_o^2= 7.9 \times
10^{-25}$~cm$^2$~keV.  Here $\alpha \simeq 1/137$ is the fine
structure constant, $m_e$ is the electron mass, $Z$ is the mean
ion charge, $c$ is the speed of light. A more accurate analytic form,
valid in the non-relativistic limit, is the Bethe-Heitler form
\index{bremsstrahlung!Bethe-Heitler cross-section}
\index{cross-sections!Bethe-Heitler}

\begin{equation}
Q(\epsilon,E) = Z^2 {\sigma_o \over \epsilon E} \ln {1 + \sqrt{1 -
\epsilon/E} \over 1 - \sqrt{1 - \epsilon/E} }.
\label{eqn:emslie_bethe-heitler}
\end{equation}

For purely spectral observations, a spatially-integrated form of
the basic Equation~(\ref{eqn:emslie_fund}) is appropriate.  In
this case, we can write \citep[e.g.,][]{1971SoPh...18..489B}

\begin{equation}
I(\epsilon) = {1 \over 4 \pi R^2} \int_\epsilon^\infty \,
[{\overline n} \, V \, {\overline F}(E)] \, Q(\epsilon, E) \, dE,
\label{eqn:emslie_fbar}
\end{equation}
where ${\overline n} = (1/V) \int_V n({\bf r}) \, d^3{\bf r}$ and
${\overline F}(E) = (1/{\overline n \, V}) \, \int_V n({\bf r}) \,
F(E, {\bf r}) \, d^3{\bf r}$.  The quantity ${\overline F}(E)$
(electrons~cm$^{-2}$~s$^{-1}$~keV$^{-1}$) is termed the {\it mean
electron flux spectrum}\index{mean electron flux} \citep{2003ApJ...595L.115B};
\index{electrons!flux spectrum} it has also been termed the ``X-ray
emitting electron spectrum'' by \citet{1992SoPh..142..219J,
1992SoPh..137..121J}. Since the quantity ${\overline n} \, V$ (the
number of target particles in the emitting volume) is
dimensionless, the units of the quantity $[{\overline n} \, V \,
{\overline F}(E)]$ are the same as those for electron flux, viz.,
electrons~cm$^{-2}$~s$^{-1}$~keV$^{-1}$. For large events, typical
values of ${\overline F}(E)$ and $[{\overline n} \, V \,
{\overline F}(E)]$ at a representative energy $E \simeq 20$~keV
are of order $10^{18}$ and $10^{55}$, respectively.

The inference of $[{\overline n} \, V \, {\overline F}(E)]$
corresponding to an observed $I(\epsilon)$ may be accomplished in
several ways, which are described in detail later in this
chapter\index{electrons!flux spectrum}.
A review of the different degrees of effectiveness
of these techniques in discerning the overall magnitude of,
overall spectral shape of, and form of ``local'' features in
${\overline F}(E)$ has been presented by
\citet{2006ApJ...643..523B}.

It is of crucial importance to note that the quantity $[{\overline
n} \, V \, {\overline F}(E)]$ is the {\it only} quantity that can
be inferred unambiguously (i.e., without additional model assumptions)
from the source-integrated
bremsstrahlung emission $I(\epsilon)$.  Use of
Equation~(\ref{eqn:emslie_fbar}) to obtain $[{\overline n} \, V \,
{\overline F}(E)]$ for a given $I(\epsilon)$ is therefore a
fundamental issue in the interpretation of solar hard X-ray
spectra.  Once $[{\overline n} \, V \, {\overline F}(E)]$ has been
determined, the actual magnitude of ${\overline F}(E)$ depends on
the values\footnote{The astute reader will note that as the source
volume $V \rightarrow \infty$, the value of ${\overline F}(E)
\rightarrow 0$.  This formal difficulty may be removed in practice
by the truncation of the emission volume $V$ at some reasonable
upper limit.} of ${\overline n}$ and $V$.

As an example of the use of the mean source electron spectrum to
determine physical properties of solar flares, let us consider the
inference of the {\it accelerated} electron flux
spectrum
\index{electrons!flux spectrum}
${\cal F}_0(E_0)$.
\index{particles!inference of flux}
The bremsstrahlung yield, the number of bremsstrahlung
photons emitted between $\epsilon$ and $\epsilon + d\epsilon$
from an electron of initial energy $E_0$ in a plasma
of density $n({\bf r})$, may be written

\begin{equation}
\nu(\epsilon,E_0) = \int_\epsilon^{E_0} {n({\bf r}) \, Q(\epsilon,
E) \, v(E) \, dE \over \vert dE/dt \vert}, \label{eqn:emslie_nu}
\end{equation}
where $dE/dt$ is the energy loss rate (here assumed a function of
$E$ only).
\index{collisions!energy losses}
For energy losses in a cold target due to binary
collisions with the background electrons (radiation energy losses are much smaller and can be ignored), $dE/dt
= - (K/E) \, n({\bf r}) \, v(E)$, where $K = 2 \pi e^4 \, \Lambda
= 2.6 \times 10^{-18}$~cm$^2$~keV$^2$ \citep{1978ApJ...224..241E},
and so the total observed flux from an {\it injected} distribution\index{electrons!injected distribution}
with energy spectrum ${\cal F}_0(E_0)$
(electrons~cm$^{-2}$~s$^{-1}$~keV$^{-1}$) is

\begin{eqnarray}
I(\epsilon) = {A \over 4 \pi R^2} \, \int_{E_0 = \epsilon}^\infty
\, {\cal F}_0(E_0) \, \nu(\epsilon, E_0) \, dE_0
=\;\;\;\;\;\;\;\;\;\;\;\;\;\;\;\;\;\;\;\;\;\;\;\;\;\; \cr
\;\;\;\;\;\;\;\;\;\;\;\;\;\;\;\;\;={A \over 4 \pi R^2} \, {1 \over
K} \int_{E_0 = \epsilon}^\infty \, {\cal F}_0(E_0) \, dE_0\,
\int_{E = \epsilon}^{E_0} E \, Q(\epsilon, E) \, dE,
\label{eqn:emslie_ifo}
\end{eqnarray}
where $A$~(cm$^2$) is the area of the flare.  Reversing the order of
integration in~(\ref{eqn:emslie_ifo}) gives

\begin{equation}
I(\epsilon) = {A \over 4 \pi R^2} \, {1 \over K} \int_{E =
\epsilon}^\infty \, E \, Q(\epsilon, E) \,dE\, \int_{E_0 =
E}^\infty {\cal F}_0(E_0) \, dE_0, \label{eqn:emslie_int-rev}
\end{equation}
and comparing this with the fundamental
Equation~(\ref{eqn:emslie_fbar}) yields the result

\begin{equation}
{\overline n} \, V \, {\overline F}(E) = {A \over K} \, E \,
\int_{E_0 = E}^\infty {\cal F}_0(E_0) \, dE_0.
\label{eqn:emslie_fbar-fo}
\end{equation}
From this it follows straightforwardly that

\begin{equation}
A \, {\cal F}_0(E_0) = - K \, {d \over dE} \left ( {{\overline n}
\, V \, {\overline F}(E) \over E} \right )_{E = E_0},
\label{eqn:emslie_fo}
\end{equation}
permitting the determination of the quantity $A \, {\cal
F}_0(E_0)$ (electrons~s$^{-1}$~keV$^{-1}$); this quantity
represents the {\it rate} of injection of electrons per unit
energy \citep[e.g.,][]{2003ApJ...595L..97H}. Clearly ${\cal
F}_0(E_0)$ is a nonnegative function, and so
Equation~(\ref{eqn:emslie_fo}) constrains ${\overline F}(E)$ to
either be a decreasing function of $E$ or, at worst, a function
that increases more slowly than $E$.  As we shall see below,
certain recovered forms of $[{\overline n} \, V \, {\overline
F}(E)]$ \citep[e.g.,][]{2003ApJ...595L.127P} can have difficulty
satisfying this constraint; however it is also shown that
such difficulties may be removed if the effects of
photospherically-backscattered (albedo)\index{albedo}\index{photospheric albedo}
photons \citep[e.g.,][]{2006A&A...446.1157K} are taken into account in Equation~(\ref{eqn:emslie_fbar}).

\subsection{Free-bound emission}\label{sec:7fb}


In fitting or inferring mean source electron flux spectra ${\overline F}(E)$,
free-bound \index{free-bound emission}
recombination emission by nonthermal electrons had always been neglected
compared with free-free electron-ion bremsstrahlung, as argued by
\cite{1967SvA....11..258K} and \cite{1973SoPh...29...93L}.
\index{bremsstrahlung!electron-ion}\index{recombination radiation!non-thermal}
For {\it hot} plasma hard X-ray sources (coronal or in soft X-ray footpoints)
this is inconsistent with inclusion of recombination
as significant for thermal electrons of similar energies in thermal spectrum
modeling \citep[e.g.,][]{1969MNRAS.144..375C, 1970MNRAS.151..141C}.
Of importance also is the fact that the estimated coronal abundance $A_Z$ for Fe is now much
higher than in \cite[e.g.][]{1973SoPh...29...93L}. The recombination emission
rate $\propto Z^4A_Z$ for hydrogenic ions of charge $Ze$ and abundance
$A_Z$ with $A_ZZ^4 \approx 1$ for H and $\approx 40$ for Fe$^{25+}$.
\citet{2010A&A...515C...1B} have therefore re-examined
the importance of nonthermal electron recombination.

In the hydrogenic Kramers approximation\index{cross-sections!Kramers approximation}, the free-bound emission rate\index{free-bound emission!Kramers approximation}
from a plasma of proton density $n_p$ and volume $V$ from nonthermal electrons
with mean source electron spectrum ${\bar F}(E)$ is given
by \citep{2010A&A...515C...1B}
\begin{equation}\label{eqn:bromal_3}
J_{R}(\epsilon) \approx \frac{32\pi}{3{\sqrt
3}\alpha}\frac{r_e^2\chi^2}{\epsilon} \, n_p \, V \, \sum_{Z_{\rm
eff}} \sum_{n \geq n_{\rm min}} \, {p_n \over n^3} \, Z_{\rm
eff}^4 \, A_{Z_{\rm eff}}\frac{{\bar F}(\epsilon-Z_{\rm eff}^2 \,
\chi/n^2)}{\epsilon-Z_{\rm eff}^2 \, \chi/n^2},
\end{equation}
where $\chi = 13.6$~eV is the H~ionization potential,
$Z_{\rm eff}$ the effective charge on the ion, $r_e$ the
classical electron radius and $\alpha$ the fine-structure
constant.
$n$ is the principal quantum number of the empty shell
into which the electron recombines and $p_n$ is a `vacancy factor'\index{vacancy factor}
which takes into account the ratio of available to total states
in that shell. [Note that Equation~(\ref{eqn:bromal_3}) applies
for $\epsilon \ge E_c + Z_{\rm eff}^2 \, \chi/n^2$; for $\epsilon < E_c + Z_{\rm eff}^2 \,
\chi/n^2$, $J_R(\epsilon) = 0$.] Since the recombination
cross-section falls off as $1/n^3$, it
is adequate to take $n=n_{\rm min}$, the value for the first
empty $n$-shell. A key feature of free-bound emission, unlike
free-free, is that for any specific shell, each electron energy
value $E$ maps to a unique photon energy value $\epsilon$ so that $J_R(\epsilon)$
is a much more direct reflection of $\overline F(E)$ than is the
bremsstrahlung $J_B(\epsilon)$ convolution of $\overline F(E)$, the former preserving
features in ${\overline F}(E)$.

\begin{figure}
\begin{center}
\includegraphics[width=0.48\textwidth]{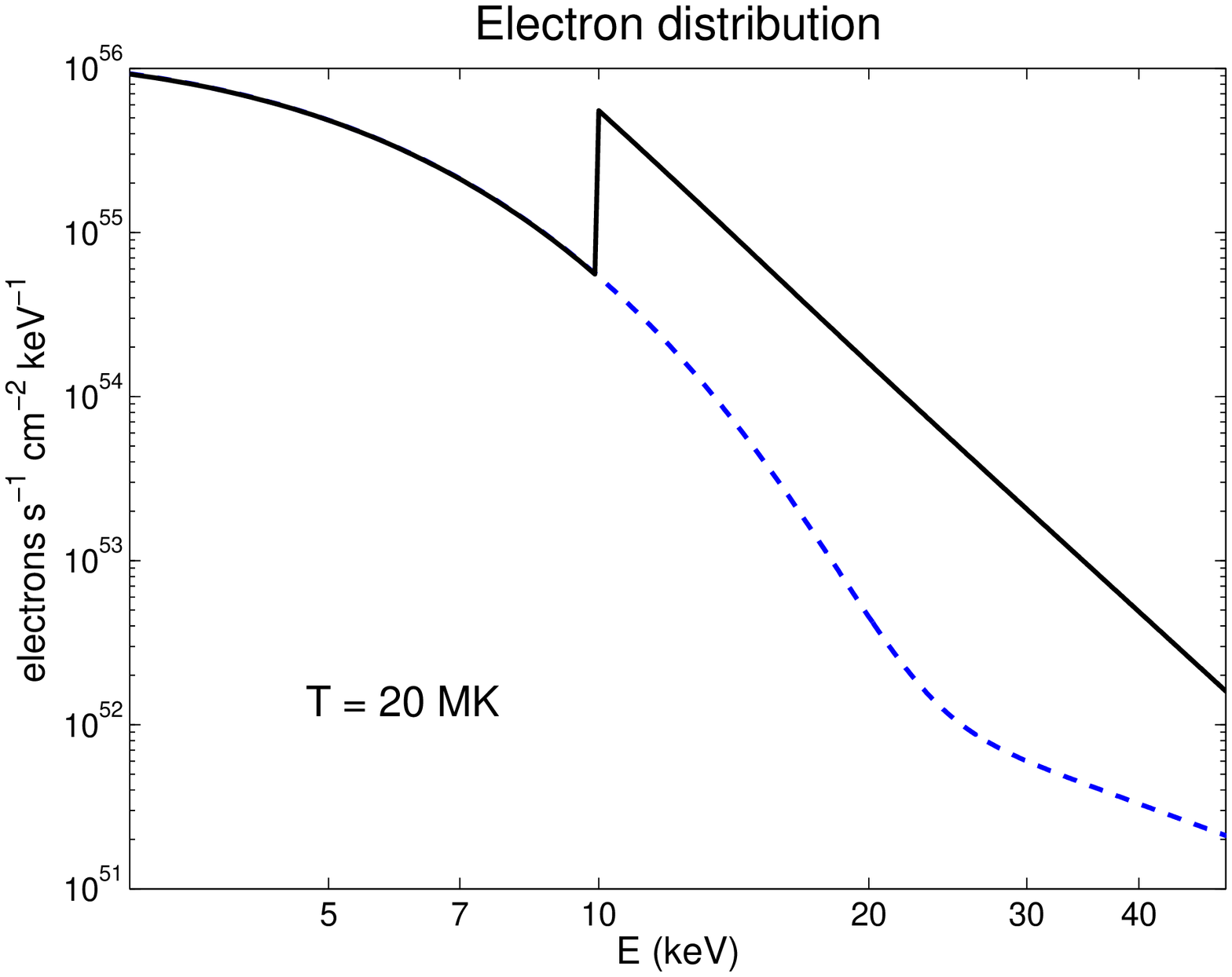}
\includegraphics[width=0.48\textwidth]{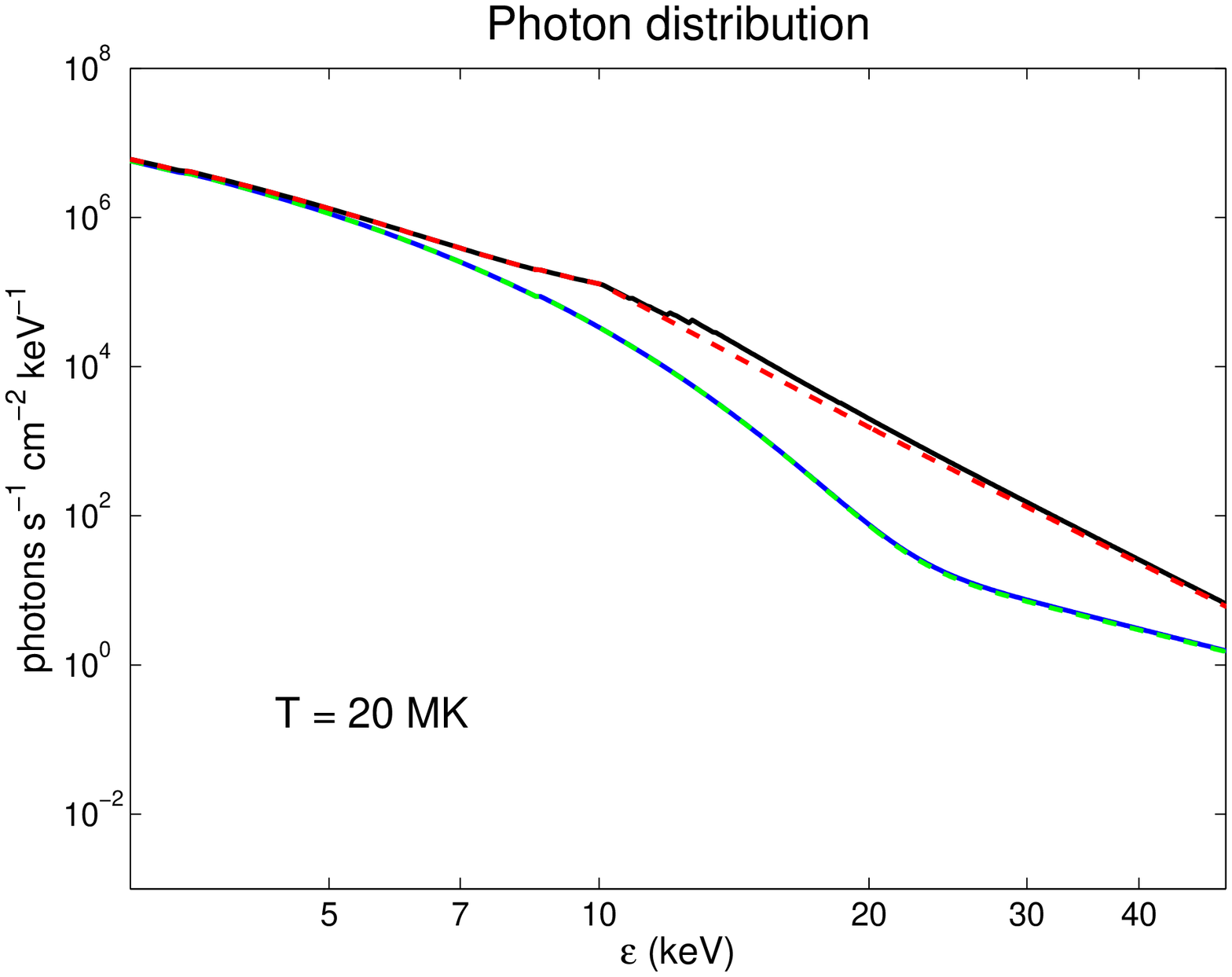}\\
\includegraphics[width=0.48\textwidth]{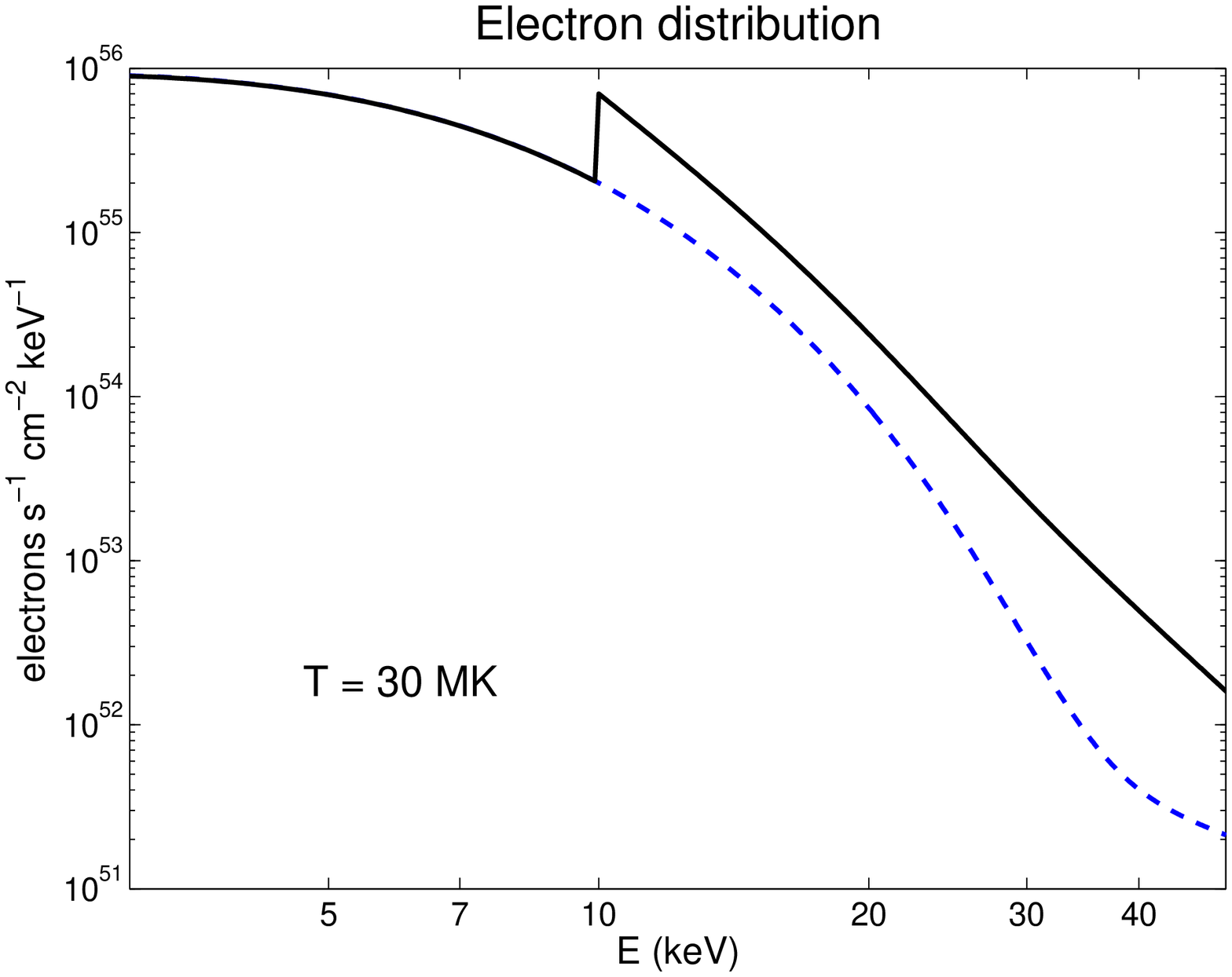}
\includegraphics[width=0.48\textwidth]{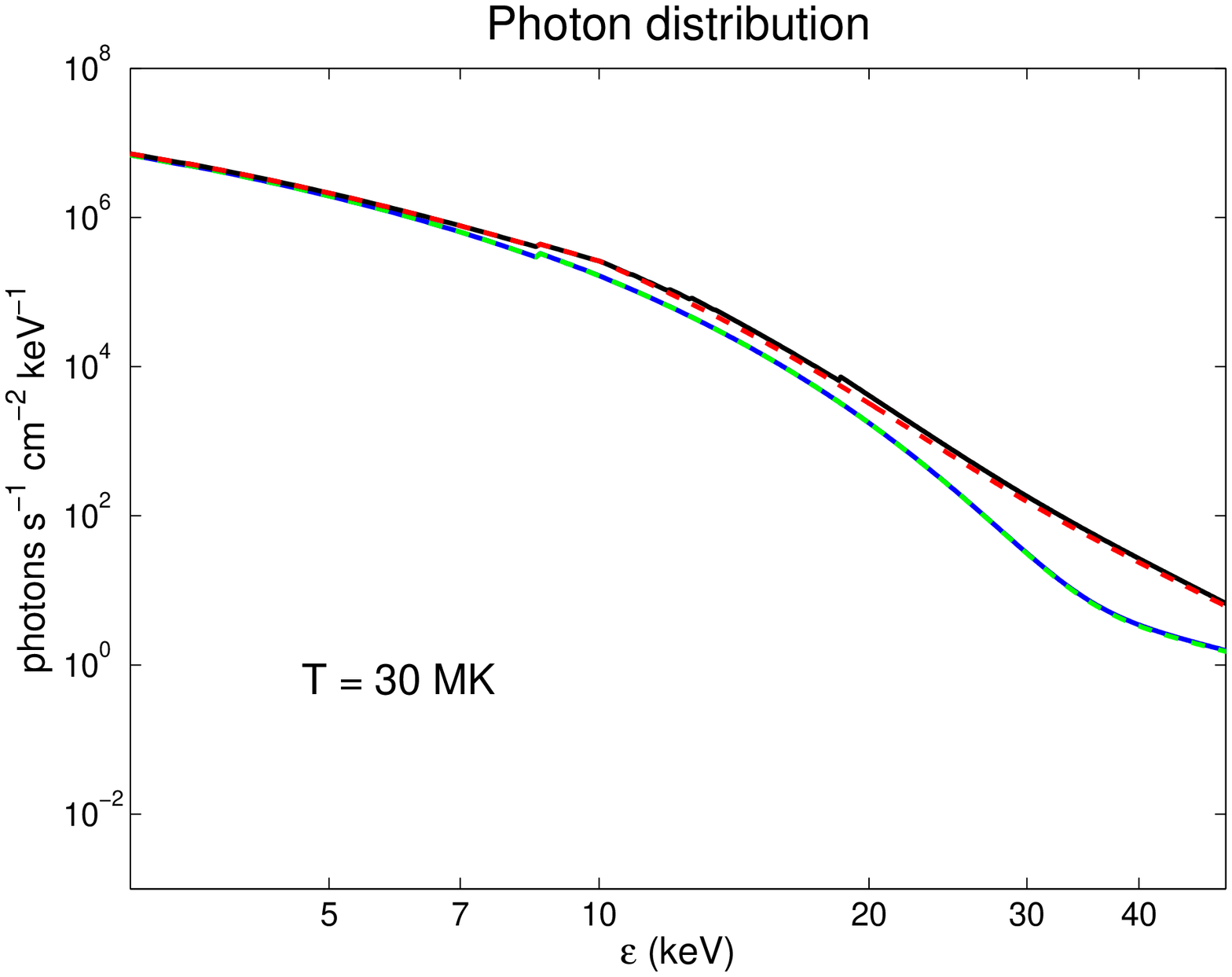}
\end{center}
\caption{\textit{Left:} the thin-target (including both thermal and nonthermal)
electron and photon (\textit{right}) spectra for two different plasma
temperatures, $20$~MK (\textit{top}) and $30$~MK (\textit{bottom}).
Electron nonthermal
spectra for various values of low energy cutoff $E_c$ and electron spectral index
$\delta$:  $E_c = 1$~keV and $\delta = 2$ (blue and dashed); $E_c = 10$~keV and $\delta = 5$ (black and solid);
The corresponding photon spectra: blue solid line for $\delta = 2$, $E_c = 1$~keV with free-bound
emission; green dashed line for $\delta = 2$, $E_c = 1$~keV without free-bound
emission; black solid line for $\delta = 5$, $E_c = 10$~keV with free-bound;
red dashed line for $\delta = 5$, $E_c = 10$~keV without free-bound.}
\label{fig:Brown_Figure2}
\index{free-bound emission!illustration}
\index{thin target!and free-bound emission}
\end{figure}

Comparing Equation~(\ref{eqn:bromal_3}) with the corresponding Kramers
free-free expression, \citet{2008A&A...481..507B,2009ApJ...697L...6B} concluded
that nonthermal free-bound emission is negligible in cold sources.  However, they concluded that in hot
plasmas ($T \approx 10-30$ MK), such as coronal sources and the soft X-ray
component of footpoint sources, free-bound emission can dominate for steep spectra
and low cut-off energies (Figure \ref{fig:Brown_Figure2}).
Subsequently, a serious error in \citet{2008A&A...481..507B,2009ApJ...697L...6B}, involving the use
of incorrect values of $n_{\rm min}$ in their Equation (13) (i.e., Equation~[\ref{eqn:bromal_3}] above)
was recognized \citep{2010A&A...515C...1B}.
This substantially reduced both the magnitude of the cross-section and the energy
shift $\epsilon - E$ involved in recombination.  Because of the typically steep
electron spectra involved, the latter substantially affects the number of electrons responsible for
emitting the photon in question.
 In the amended results  $J_{R}(\epsilon)$ is never dominant, even for for hot sources,
but can account for up to $\sim 30\%$ of the flux $J_B(\epsilon)+J_R(\epsilon)$
in the range $\sim$10-30~keV for the non-thermal component
dominating thermal as shown in Figure~\ref{fig:Brown_Figure2}.
{\it For such hot sources}, the free-bound emission could be comparable
with such effects as albedo and differences in bremsstrahlung cross-section used,
and could be used to diagnose the sharp features in $\overline F(E)$ from $J(\epsilon)$.
Even more important is the fact that $J_{R}(\epsilon)$ adds
edges to the total $J(\epsilon)$ with the result that,
in data with good signal to noise ratio, inversion (essentially differentiation)
of data $J(\epsilon)$ to yield $\overline F(E)$ based on bremsstrahlung
alone (essentially differentiation of $J$) could result in spurious
features in $\overline F(E)$ just as happens when albedo is ignored.

\subsection{Electron-electron bremsstrahlung}\label{sec:7ee} 
\index{bremsstrahlung!electron-electron}


Energetic electrons propagating in the solar atmosphere encounter
ions and electrons (both free and bound) and hence can produce
X-ray emission via both electron-ion and electron-electron
bremsstrahlung\index{bremsstrahlung!electron-ion}.

When the maximum electron energy is much larger than the photon
energies under consideration, the photon spectrum resulting from a
power-law spectrum\index{spectrum!power-law} of electrons ${\overline F}(E) \propto
E^{-\delta}$ is also close to the power-law form $I(\epsilon) \propto
\epsilon^{-\gamma}$ \citep{1989A&A...218..330H}.
However, while
for pure electron-ion bremsstrahlung $\gamma \simeq \delta+1$, for
pure electron-electron bremsstrahlung\index{bremsstrahlung!electron-electron}
 a significantly shallower
photon spectrum, with $\gamma \simeq \delta$, results. Thus, the
importance of the electron-electron bremsstrahlung
\index{bremsstrahlung!electron-electron}
contribution
increases with photon energy and the enhanced emission per
electron leads to a flattening of the photon spectrum
$I(\epsilon)$ above $\sim$300~keV produced by a given ${\overline F}(E)$
\citep{1975SoPh...45..453H} or, equivalently, a steepening of the
${\overline F}(E)$ form required to produce a given $I(\epsilon)$.
\citet{2007ApJ...670..857K} provide a discussion of the essential
differences between electron-electron
and electron-ion bremsstrahlung processes.
We here provide a succinct summary of that discussion and note that
the properties of electrons with the energies above $\sim$400~keV\index{electrons!relativistic}
are also crucial for ion diagnostics and radio emission.

\begin{figure}[pht]
\begin{center}
\includegraphics[width=0.6\textwidth]{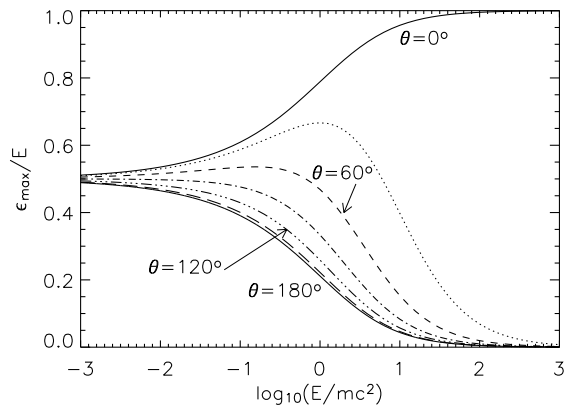}
\end{center}
\caption{Maximum photon energy $\epsilon_{\rm max}$ produced by
electron-electron bremsstrahlung, expressed as a fraction of the
incident electron energy $E$ (in units of the electron rest mass
$mc^2$), for various values of $\theta$, the angle between the
incoming electron and the outgoing photon trajectories. For
clarity, only curves for $\theta = 0^\circ, 60^\circ, 120^\circ,$
and $180^\circ$ are labeled; the curves for $\theta = 30^\circ,
90^\circ$ and $150^\circ$ lie between these
\citep[after][]{2007ApJ...670..857K}.} \label{ee:epsmax}
\end{figure}

As is well known
\citep{1959RvMP...31..920K},\index{bremsstrahlung!electron-electron}
the cross-section for electron-ion brems\-strahlung scales as $Z^2$.
\index{bremsstrahlung!electron-ion}
Further, when considering electron-electron bremsstrahlung, the possible binding of target
electrons to their host ions in a neutral or partially-ionized
medium is not significant. Hence, in a quasi-neutral target
of particles with atomic number
$Z$, the bremsstrahlung cross-section per atom for emission of a
photon of energy $\epsilon$ by an electron of energy $E$ is in
general equal to
\begin{equation}
Q (\epsilon, E) = Z^2 Q_{e-p} (\epsilon, E) + Z Q_{e-e} (\epsilon,
E), \label{eq:ee_qtot}
\end{equation}
where $Q_{e-p} (\epsilon, E)$ and $Q_{e-e} (\epsilon, E)$ are the
cross-sections, in the laboratory frame, for electron-proton\index{bremsstrahlung!electron-ion}, and electron-electron bremsstrahlung\index{bremsstrahlung!electron-electron},
as given by
\citet{1959RvMP...31..920K} and \citet{1989A&A...218..330H},
respectively\footnote{Note that there is a typographical error in
the form of $Q_{e-e}$ in \citet{1989A&A...218..330H}; see
\citet{2007ApJ...670..857K} for details.}. It is also important to
note that while the electron-ion cross-section is finite for all
$\epsilon < E$, the cross-section for electron-electron
bremsstrahlung vanishes above a maximum photon energy, due to the
necessarily finite energy carried by the recoiling target electron
\citep{1975SoPh...45..453H, 2007ApJ...670..857K}. For highly {\it
non}-relativistic electrons ($E \ll m_ec^2$), $\epsilon_{\rm max}
\rightarrow E/2$; only for highly {\it relativistic} electrons ($E
\gg m_ec^2$) can $\epsilon_{\rm max} \rightarrow E$ (see
Figure~\ref{ee:epsmax}).  As discussed by
\citet{2007ApJ...670..857K}, this result has important
implications for the form of the photon spectrum produced by
electron-electron bremsstrahlung.

In Section~\ref{sec:7fbar}, we discuss the application to the
recovery of the electron spectrum at pertinent
(mildly-relativistic) energies\index{electrons!mildly relativistic}.

\section{Primary and Compton backscattered X-rays} \label{sec:7albedo}
\index{albedo}

As discussed in Section~\ref{sec:7emission}, the atmosphere {\it
above} the X-ray bremsstrahlung-producing region is optically
thin. However, the very dense lower photospheric layers {\it
below} the primary source are not.  Consequently, photons emitted
downwards are efficiently Compton backscattered\index{Compton
scattering!photospheric} by atomic electrons in the photosphere.
The observed hard X-ray (HXR) flux spectrum from solar flares is
therefore a combination of primary bremsstrahlung photons with a
spectrally modified component from photospheric Compton
backscatter\index{Compton scattering!photospheric} of the
downward-directed primary emission. This backscattered component
can be significant, creating new features and/or distorting or
masking features in the primary spectrum, and so substantially
modifying key diagnostics such as the electron energy budget.

Photons of energy above $\sim$100~keV penetrate so deeply that
they are lost to the observer, while below $\sim$10~keV they are
mostly photoelectrically absorbed
\citep[e.g.,][]{1972ApJ...171..377T}. Therefore, the reflectivity
of the photosphere has a broad hump in the range 10-100~keV, with
a maximum around 30-40 keV. At some energies and view angles the
reflectivity approaches 90\%, so the observed spectrum may be very
substantially affected by backscatter. This effect is well-known
in solar physics (and more generally in X-ray astronomy
\citep[see][]{1995MNRAS.273..837M}, and there have been several
discussions of its influence on observed X-ray spectra
\citep[e.g.,][]{1972ApJ...171..377T,1973SoPh...29..143S,1978ApJ...219..705B}, and on the electron spectra inferred from them
\cite[e.g.,][]{1992SoPh..137..121J,2002SoPh..210..407A,2005SoPh..232...63K,2006A&A...446.1157K}.

Understanding and modeling backscatter (albedo)
\index{hard X-rays!albedo}\index{albedo}\index{photospheric albedo} has become even
more important with the advent of high quality X-ray spectra from
{\it RHESSI}
\citep{2002SoPh..210....3L} with spectral resolution as high as
$\simeq$1~keV, in combination with uncertainties as low as a few
percent (for strong flares).
\index{RHESSI@\textit{RHESSI}!spectroscopy}
Generally, contamination of the observed X-ray spectrum by
reflected photons leads to a flattening of the spectrum and hence
to an underestimation of the electron spectral index
\index{spectrum!electrons!spectral index} if the
contribution of backscattered photons is not taken into account
\citep{1978ApJ...219..705B}. Extrapolation to low electron
energies using such an underestimated spectral index leads to a
substantial underestimation of the total electron energy in a
flare.  Indeed, a low-energy cutoff
\index{particles!energy spectra!low-energy cutoff}
\index{low-energy cutoff} in an uncorrected
electron spectrum is not even required if a ``true'' primary
(albedo-corrected) electron spectrum can be used
\citep{2006A&A...446.1157K}\index{albedo!spectroscopy}.

\subsection{Spectroscopy of the photospheric albedo}\label{sec:7albedo_spectr} 
\index{albedo!spectroscopy}


Downward-emitted photons are either absorbed or Compton-scattered,
with some of the latter returned toward the observer, adding to
the total flux detected (X-ray albedo).
\index{hard X-rays!albedo}
Scattering takes place on electrons, whether free or atomic.
To account for elements heavier than
hydrogen\index{abundances!and Compton scattering}, the Compton cross-section is multiplied by an effective mean atomic number $Z=1.2$ \citep[e.g.,][for element abundances]{2005psci.book.....A}.
The detailed density structure of the medium is irrelevant \citep{1972ApJ...171..377T}.

Absorption,\index{absorption!photoelectric} on the other hand,
does depend strongly on chemical composition, and the best
estimate of photospheric abundances should be included. The heavy
elements Fe/Ni play the most important role from $6-8$~keV up to
$\sim$30~keV, while lighter elements contribute below $6$~keV
\citep{1983ApJ...270..119M}.

\subsubsection{Green's function approach} \label{Ko1_Green}
\index{albedo!Green's function approach}\index{Green's functions}

Propagation, absorption, and Compton scattering of primary hard X-rays
can be straightforwardly
studied using Monte-Carlo simulations
\citep[e.g.,][]{1978ApJ...219..705B}. This technique is ideal for
obtaining the reflected, and hence the total observed, photon
spectrum for a given form of the primary X-ray spectrum. However,
the primary spectrum is generally unknown and unlikely to be an exact
power-law, as is sometimes assumed. Therefore, an approach
independent of the primary spectrum is required
\citep{2006A&A...446.1157K}.

\begin{figure}
\begin{center}
\includegraphics[width=0.6\textwidth]{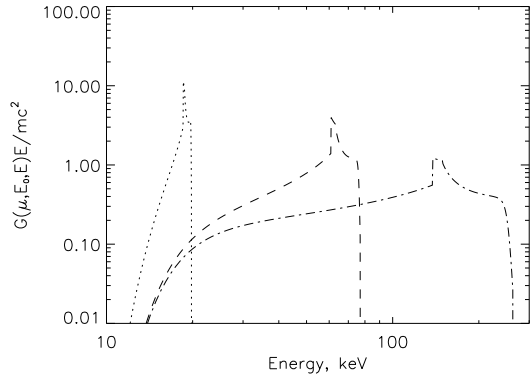}
\end{center}
\caption{Green's functions $G(\mu, \epsilon,\epsilon_0)$ [ keV$^{-1}$],
including Compton scattering and photoelectric absorption, for
three primary photon energies $\epsilon _0 =20,80,300$~keV and
$\mu =0.7$ ($\theta \approx 45^o$) calculated using approximations
\citep{1995MNRAS.273..837M} for solar conditions
\citep[after][]{2006A&A...446.1157K}.} \label{Ko1_greenf}
\end{figure}

For any isotropic primary spectrum $I_{P}(\epsilon_0)$,
(photons~cm$^{-2}$~s$^{-1}$~keV$^{-1}$), we can write the
secondary, backscattered spectrum $I_S (\epsilon)$ as

\begin{equation}\label{Ko1_eq1}
I_{S}(\epsilon,\mu)=\int\limits_{\epsilon}^{\infty}I_{P}(\epsilon_0)
\, G(\mu,\epsilon,\epsilon_0) \, d\epsilon_0,
\end{equation}
where $G(\mu,\epsilon,\epsilon_0)$ is an angle-dependent Green's
function and $\mu =\cos \theta$, where $\theta$ is the
heliocentric position angle of the primary hard X-ray source,
or the angle between the Sun center - observer and Sun
center - X-ray source lines. The
observed spectrum, at photon energy $\epsilon$ and direction
$\mu$, is $I_S(\epsilon) + I_P(\epsilon)$. The importance of not
averaging over viewing angle may be seen from
\cite{1978ApJ...219..705B}. Using Green's function
analytical fits to Monte Carlo simulations
derived by \citet{1995MNRAS.273..837M},
\citet{2006A&A...446.1157K} calculated
functions $G(\mu,\epsilon,\epsilon_0)$ shown in Figure~\ref{Ko1_greenf}
for solar flare parameters. Green's functions, as
calculated, account for Compton scattering
and bound-free absorption.

The shape of the Green's function depends on the energy of the
primary photon. For primary photons with low energies $\epsilon
_{0}<30$ keV, the Green's function has a rather simple structure
close to a Dirac delta-function (Figure ~\ref{Ko1_greenf}),
showing that backscattering is dominated by the first scattering
(especially at low energies), with the contributions
from higher orders of scattering being generally small.

\index{albedo!Green's function approach!illustration}

\subsubsection{Reflected X-ray photon spectrum}

\begin{figure}
\begin{center}
\includegraphics[width=0.6\textwidth]{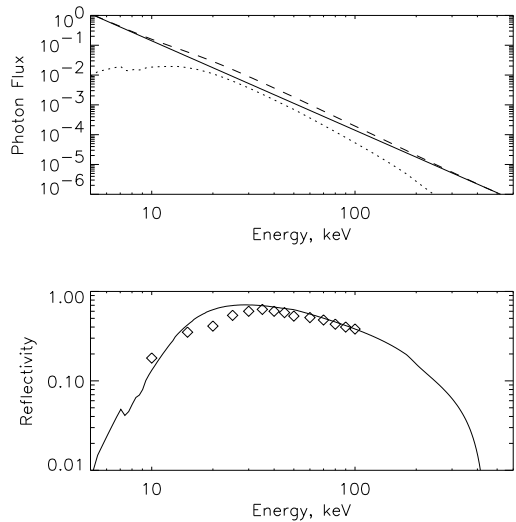}
\end{center}
\caption{{\it Upper panel:} Primary (solid line), reflected (dotted
line) and total (dashed line) photon spectra calculated assuming a
primary spectrum $I_P(\epsilon)\propto \epsilon ^{-3}$, and using the
Green's function for an X-ray source at heliocentric angle $\theta
= 45^o$. {\it Lower panel:} Reflectivity, defined as the
ratio of reflected to primary fluxes $R(\epsilon, \theta
=45^o)=I_R(\epsilon)/I_P(\epsilon)$.
The reflectivity taken from
\citet{1978ApJ...219..705B} is shown with diamonds.
\index{bound-free absorption!illustration}
Two absorption edges
of Fe~at $7.1$ keV and Ni~at $8.3$ keV are seen in the reflected
component \citep[after][]{2006A&A...446.1157K}.}
\label{Ko1_a_test}
\end{figure}

Because the reflectivity is spectrum-dependent (Figure
\ref{Ko1_a_test}), the albedo spectrum also depends on the shape
of the primary spectrum \citep{2006A&A...446.1157K,2007PASJ...59.1161K}.  While
previous studies considered only the results for prescribed
power-law or thermal primary spectra
\citep[cf.][]{1978ApJ...219..705B,1992SoPh..137..121J,2002SoPh..210..407A},
the Green's function method allows more general forms of the
primary spectrum.

The total observed spectrum $I(\epsilon)$ is given by

\begin{equation}\label{Ko1_itot}
I(\epsilon,\mu)=I_P(\epsilon)+\int\limits_{\epsilon}^{\infty}
I_{P}(\epsilon_0) \, G(\mu,\epsilon,\epsilon_0) \, d\epsilon_0.
\end{equation}
For a measured $I(\epsilon,\mu)$\index{hard X-rays!flux spectrum!observed},
we may obtain the primary spectrum $I_P(\epsilon)$
\index{hard X-rays!flux spectrum!primary} by solving the integral equation
(\ref{Ko1_itot}). In practice measurements yield discrete
quantities and the integral equation (\ref{Ko1_itot}) is used in
the matrix form
\begin{equation}\label{Ko1_eq2}
  I(\epsilon_i,\mu)=I_{P}(\epsilon_i)+G_{ij}(\mu)I_{P}(\epsilon_j),
\end{equation}
where we have used the summation convention for repeated indices,
and introduced the {\it Green's matrix}

\begin{equation}\label{Ko1_eq3}
  G_{ij}(\mu)=\int\limits_{\epsilon_{j}}^{\epsilon_{j+1}}G(\mu,\epsilon _i,\epsilon _0)
  \, d\epsilon _0.
\end{equation}
Due to sharp features in the Green's function
(Figure~\ref{Ko1_greenf}), the integration in
Equation~(\ref{Ko1_eq3}) is best performed via a change of
variable to the wavelength (reciprocal energy) domain.

     \subsection{Imaging of photospheric albedo}\label{sec:7albedo_image} 
\index{albedo!imaging}

\subsubsection{Expected spatial signatures of albedo}

Before considering observational approaches to the spatial
isolation of albedo,
it is instructive to consider an elementary
model of the solar backscattering process.
\index{hard X-rays!albedo}
\index{albedo}\index{photospheric albedo}
To do this, we make
four simplifying assumptions \citep[e.g.,][]{1975A&A....41..395B}: first, that the primary
bremsstrahlung is generated as a point source at a height $h$
above a planar scattering surface; second, that this surface is
perpendicular to the line of sight (e.g., disk center flare); third, that the primary X-ray
emission and backscattering processes are isotropic; and fourth,
that absorption of the scattered photons can be neglected.
With such a scenario, the albedo source
would extend out to the horizon as seen from the primary source.
\index{albedo}\index{albedo!imaging}
The albedo surface brightness
would be determined by the distance from the primary source and
the scattering location and by a cosine illumination factor and so the
surface brightness would fall off as $(1+[r/h]^2)^{-3/2}$,
where $r$ is the radial distance between the scattering location
and the sub-source point on the scattering surface.  Note that the
scale of the radial profile is determined by the height of the
primary source.

\begin{figure}
\begin{center}$
\begin{array}{cc}
\includegraphics[width=0.49\textwidth]{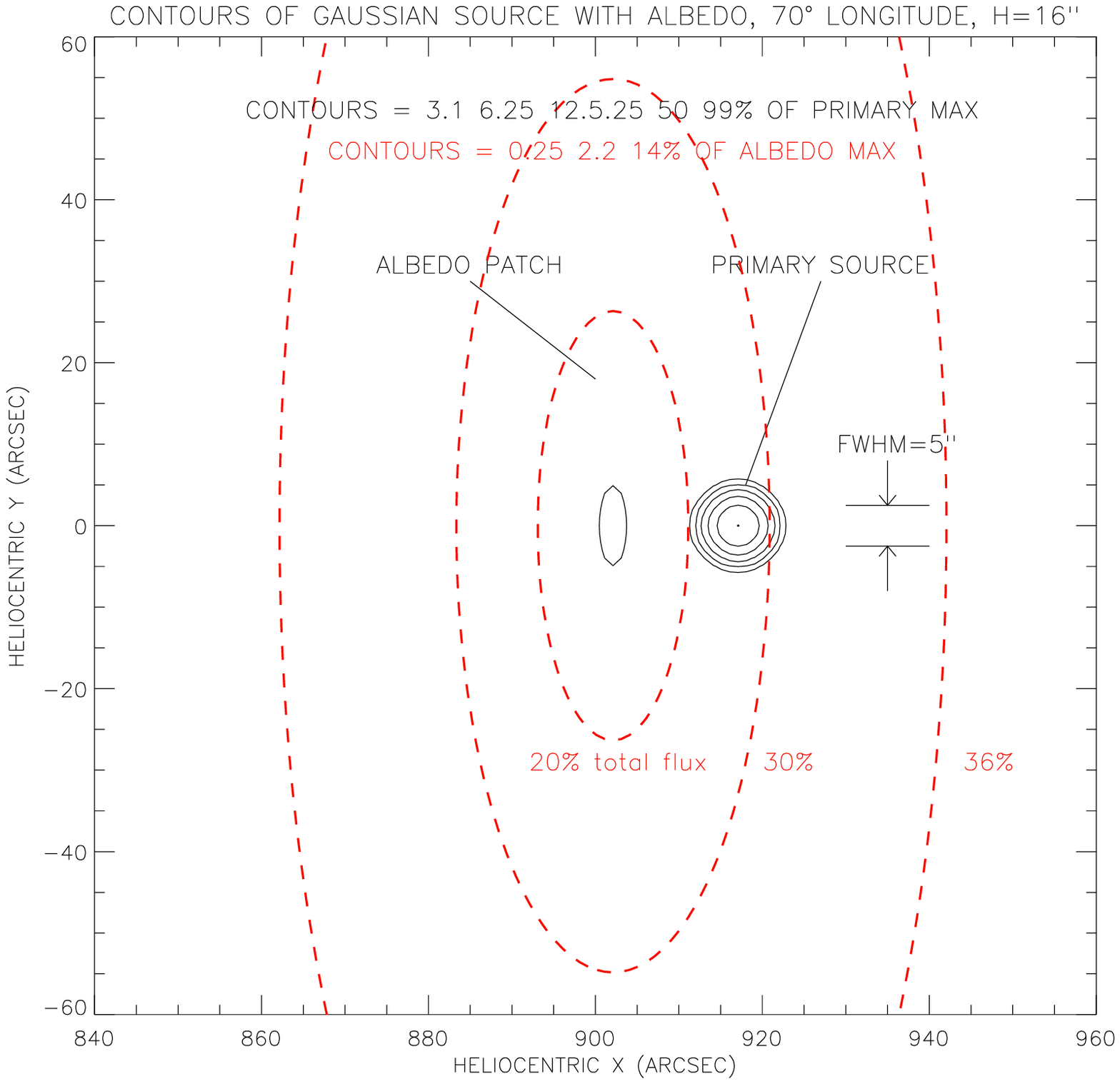} &
\includegraphics[width=0.49\textwidth]{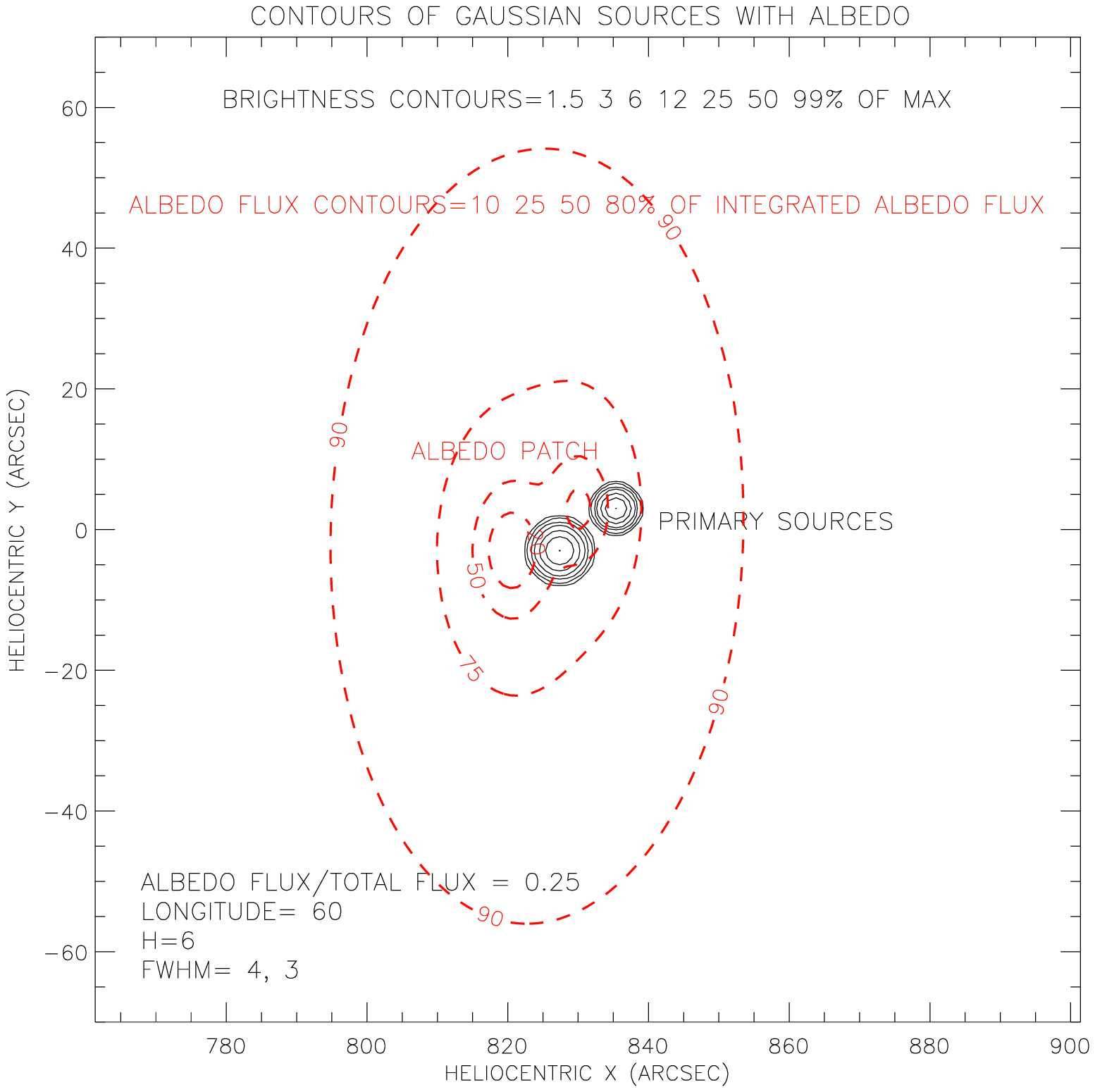}
\end{array}$
\end{center}
\caption{{\it Left panel:} Model of a single primary
source with albedo contours. The solid contours at
logarithmically-spaced intervals show the primary source down to
$6.25$\% of the maximum, and the brightest portion of the albedo
patch at $3.1$\% of the peak of the primary. The dashed red
contours show the integrated flux of the albedo patch. Note that
$50$\% of the albedo flux arises from an area about 10 times
larger than the primary source, given its height of $12$~Mm. {\it
Right panel:} A model of a double source $10$~Mm above the
photosphere with the resultant albedo patch shown by flux contours
(red dashes). There is considerable overlap of the primary
source and albedo patch in this case
\citep[after][]{2002SoPh..210..273S}.} \label{fig:schmahl_fig2}
\index{hard X-rays!albedo}\index{albedo}\index{photospheric albedo!illustration}
\end{figure}

The implications of easing some of the restrictions in this simple
model are illustrated in Figure~\ref{fig:schmahl_fig2}.  For a
primary source with finite size or structure, the resulting albedo
patch would be a convolution of the primary source with the
aforementioned profile. For a primary source located away from disk
center, the center of the albedo patch
\index{hard X-rays!albedo}\index{albedo}
\index{photospheric albedo}
is displaced toward disk center and presents an elliptical shape
oriented parallel to the limb.

More rigorous calculation of the spatial properties of an albedo
patch are provided by \citet{1978ApJ...219..705B} who show, for
example, that limb darkening would make the albedo more easily
detected near disk center. In addition, the albedo patch will
be energy- and primary-spectrum-dependent \citep{2010A&A...513L...2K}.
Nevertheless, the simplified model does suggest that the albedo
has three potentially observable spatial signatures:
first, for sufficiently high primary source
altitudes, the albedo
\index{hard X-rays!albedo}
\index{albedo}\index{photospheric albedo}
would be significantly larger in extent than the primary
source, with a size scale that increases with source height;
second the albedo source would be displaced toward disk center by
a distance $h \sin \theta_\odot$, where $\theta_\odot$ is the
heliocentric angle; third, the albedo source would be elongated
parallel to the limb with a minor to major axis ratio of $\cos
\theta_\odot$.  Observationally, however, because of the
relative size scales of the primary source and its
albedo patch,
the albedo surface brightness would be only a small fraction
(typically only a few percent) of that of the primary source.
\index{hard X-rays!albedo}\index{albedo}\index{photospheric albedo}
This would pose a potential challenge for conventional imaging systems
because of scattered light; for reconstructed {\em RHESSI} hard
X-ray images, typically limited in dynamic range to about 10:1, this
would seem to make the spatial detection of albedo even more
problematic.

\subsubsection{The spatial-frequency signature of albedo}
\index{albedo!spatial-frequency signature}\index{frequency!spatial}

The potential observational difficulties posed by the low surface
brightness of the albedo source can be eased if we Fourier
transform the source of size $\sim$$d$ with the distribution
$\exp(-x^2/d^2)$ and consider the amplitude
of the Fourier components as a function of spatial frequency $\propto \exp(-k^2d^2)$,
where $k$ is the spatial frequency.
Compact primary sources have Fourier amplitudes that fall off at
high spatial frequencies $kd \gg 1$.  For spatial periods large compared to the
source dimension $kd \ll 1$, however, the amplitudes are effectively constant.
Further, the Fourier transform of the $(1+[r/h]^2)^{-3/2}$ profile of an
albedo source implies an amplitude that varies as $e^{-kh}$, where $k$ is the
spatial frequency.  As illustrated in the lower half of
Figure~2 in \citet{2002SoPh..210..273S}, this could be readily
distinguished from that of the primary source.  In effect. the
Fourier transform integrates over the faint, distributed albedo source so that
it becomes  potentially detectable. The
most obvious albedo signature would then be an ``excess'' in
Fourier amplitude at low spatial frequencies over that expected
for the compact primary source. The excess would be comparable to
the reflected fraction, viz., up to several tens~\%
\citep{1978ApJ...219..705B,2006A&A...446.1157K}.
This is relevant to {\em RHESSI}, because with nine logarithmically-distributed spatial
frequencies, its rotating modulation collimators
(RMCs) directly measure the Fourier components of the source distribution.
\index{RHESSI@\textit{RHESSI}!rotating modulation collimators}

\subsubsection{Detection of a ``halo'' component}
\index{albedo!halo component}

\begin{figure}
\begin{center}
\includegraphics[width=0.9\textwidth]{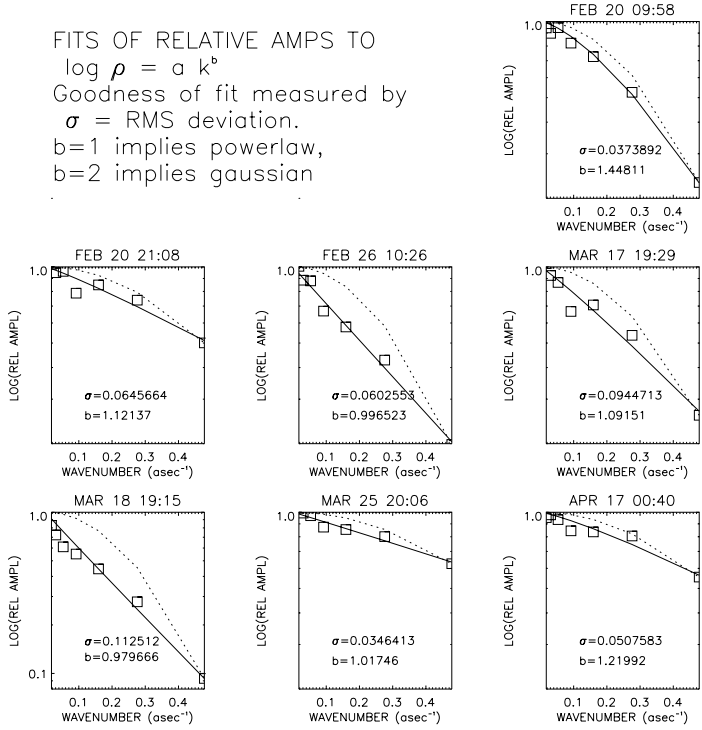}
\end{center}
\caption{
Relative amplitudes of 7 flares vs. spatial frequency in the energy
range 12-25 keV. The departures of the profiles from a Gaussian shape are
interpreted as a ``halo'' component. Since (for
computational simplicity) the fits were assumed to be azimuthally
symmetric, the profiles must be considered to be polar
averages over 360 degrees of rotation
\citep[after][]{2002SoPh..210..273S}.
\index{flare
(individual)!SOL2002-02-20T09:59 (M4.3)!albedo patch}
\index{flare
(individual)!SOL2002-02-20T09:59 (M4.3)!illustration}
\index{flare
(individual)!SOL2002-02-20T21:07 (M2.4)!illustration}
\index{flare
(individual)!SOL2002-02-26T10:27 (C9.6)!illustration}
\index{flare
(individual)!SOL2002-03-17T19:31 (M4.0)!illustration}
\index{flare
(individual)!SOL2002-03-18T19:18 (C8.9)!illustration}
\index{flare
(individual)!SOL2002-03-18T19:18 (C8.9)!albedo patch}
\index{flare
(individual)!SOL2002-03-25T20:08 (C9.8)!albedo patch}
\index{flare
(individual)!SOL2002-04-17T00:40 (C9.9)!illustration}
}\index{frequency!spatial!illustration}
\label{fig:schmahl_fig3}
\end{figure}

Applying such considerations to flare observations, early analysis of {\em
RHESSI} data \citep{2002SoPh..210..273S} directly fit the observed modulated light curves
to derive an average modulation amplitude for each RMC\index{RHESSI@\textit{RHESSI}!rotating modulation collimators (RMCs)}.
This technique, although simple, assumed circularly symmetric
sources and was subject to potential statistical issues since it
directly fit the counts which were only sparsely populating the short time bins.
Nevertheless, the analysis showed (Figure~\ref{fig:schmahl_fig3})
that the modulation amplitude continued to increase toward the
coarsest subcollimator even though the latter had a 183$''$
spatial resolution \citep{2002SoPh..210...61H}.  This was taken as evidence of a hard X-ray
``halo'' component, consistent with expectations from albedo.

\begin{figure}
\begin{center}
\includegraphics[width=0.9\textwidth]{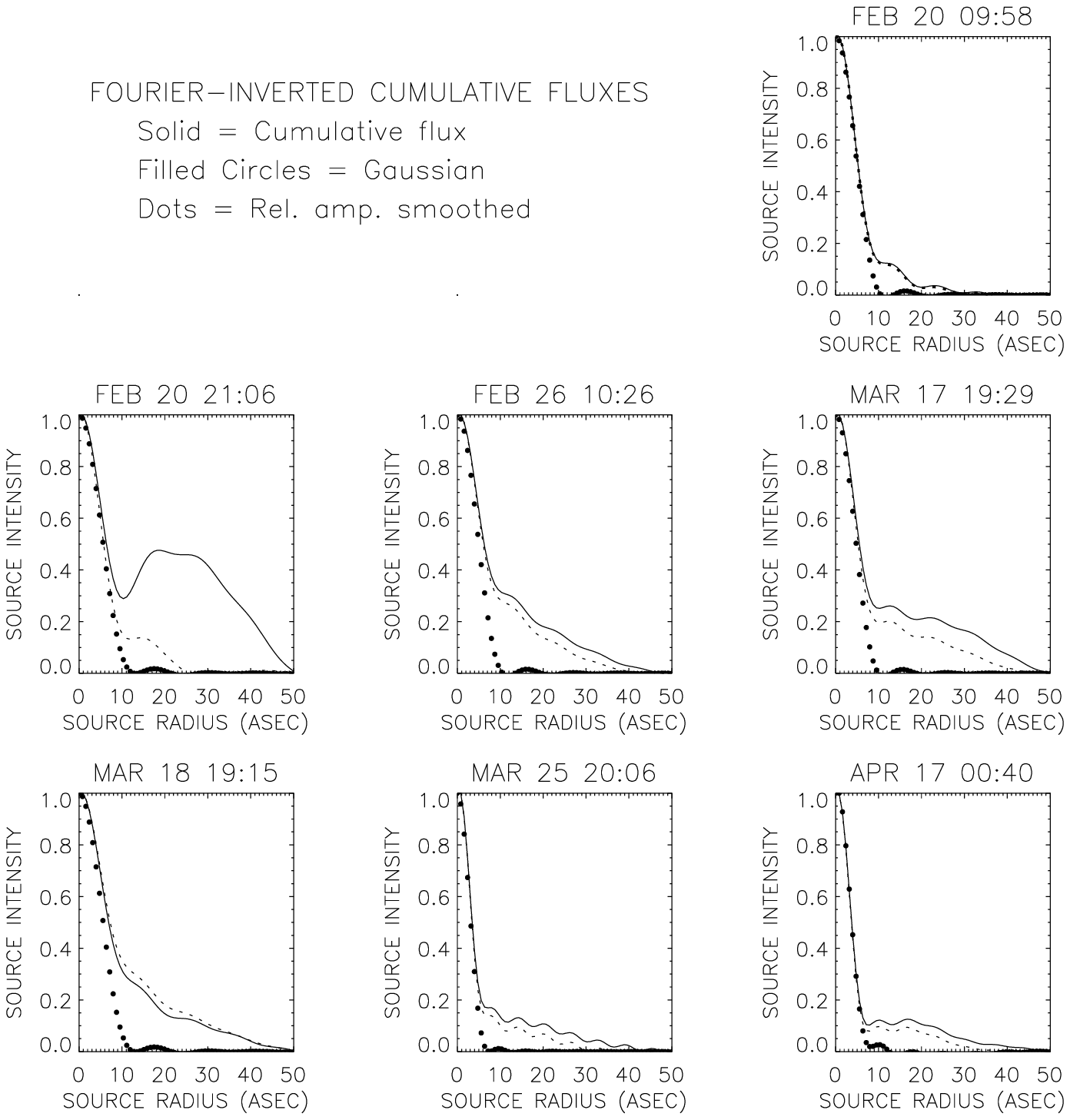}
\end{center}
\caption{A Fourier inversion produces the brightness profile in the energy
range $12-25$~keV. Since the albedo surface brightness is very low, the
cumulative integrated flux from $r$ to $\infty$ is plotted as a function of $r$. If the primary
source profile is Gaussian, the integrated flux profile is
Gaussian. The clear departures from a Gaussian shape in these
profiles indicates the presence of extended sources inferred to be
albedo patches \citep[after][]{2003AdSpR..32.2477S}.
\index{flare
(individual)!SOL2002-02-20T09:59 (M4.3)!illustration}
\index{flare (individual)!SOL2002-02-20T21:07 (M2.4)!illustration}
\index{flare (individual)!SOL2002-02-26T10:27 (C9.6)!illustration}
\index{flare (individual)!SOL2002-03-17T19:31 (M4.0)!illustration}
\index{flare (individual)!SOL2002-03-18T19:18 (C8.9)!illustration}
\index{flare (individual)!SOL2002-03-25T20:08 (C9.8)!illustration}
\index{flare (individual)!SOL2002-04-17T00:40 (C9.9)!illustration}
}
\label{fig:schmahl_fig4}
\end{figure}

The analysis was refined by \citet{2003AdSpR..32.2477S}, who formed
individual back-projectionimages with the nine RMCs.
\index{imaging algorithms!back-projection}
In this case, the peak calibrated intensity of each back-projection image
corresponded to the azimuthally-averaged modulation amplitude for
the corresponding spatial frequency.
The results, illustrated in
Figure~\ref{fig:schmahl_fig4}, confirmed the presence of a
non-Gaussian large-scale component consistent with a halo source.
\index{hard X-rays!halo component}\index{frequency!spatial!halo source}
This technique alleviates potential
statistical concerns associated with directly fitting the
sparsely-populated time bins, but still requires circular sources
for unambiguous interpretation.

\subsubsection{Direct use of visibility measurements}
\index{albedo!use of visibilities}

Since the initial reports of the halo component, analysis
techniques have been developed that enable the {\em RHESSI} data
to be directly transformed into calibrated measurements of the
visibilities
(specific Fourier components) as a function of
spatial period and orientation \citep{2002SoPh..210..273S,2007SoPh..240..241S}.
\index{visibilities!electron}
\index{hard X-rays!visibilities}
\index{hard X-rays!halo component}
There are several reasons why
visibilities are more appropriate for albedo determination: the
visibilities are fully calibrated so that instrumental issues can
be cleanly separated from solar issues; the expected visibilities
can be directly calculated from source models; each visibility
measurement is independent and has well-determined statistical
errors whose propagation can guide subsequent conclusions;
visibilities can be readily calculated using code that is now an
integral part of the {\em RHESSI} object-oriented software
package; visibilities are determined from linear transforms of the
observed count rates, so that they can be combined in time or
energy as desired; visibilities
are well-suited to more complex
sources since an observed visibility is the sum of the
corresponding visibilities of its components.
\index{visibilities!electron}
\index{hard X-rays!visibilities}
\index{visibilities!electron}
\index{hard X-rays!visibilities}

\begin{figure}
\begin{center}
\includegraphics[width=0.6\textwidth]{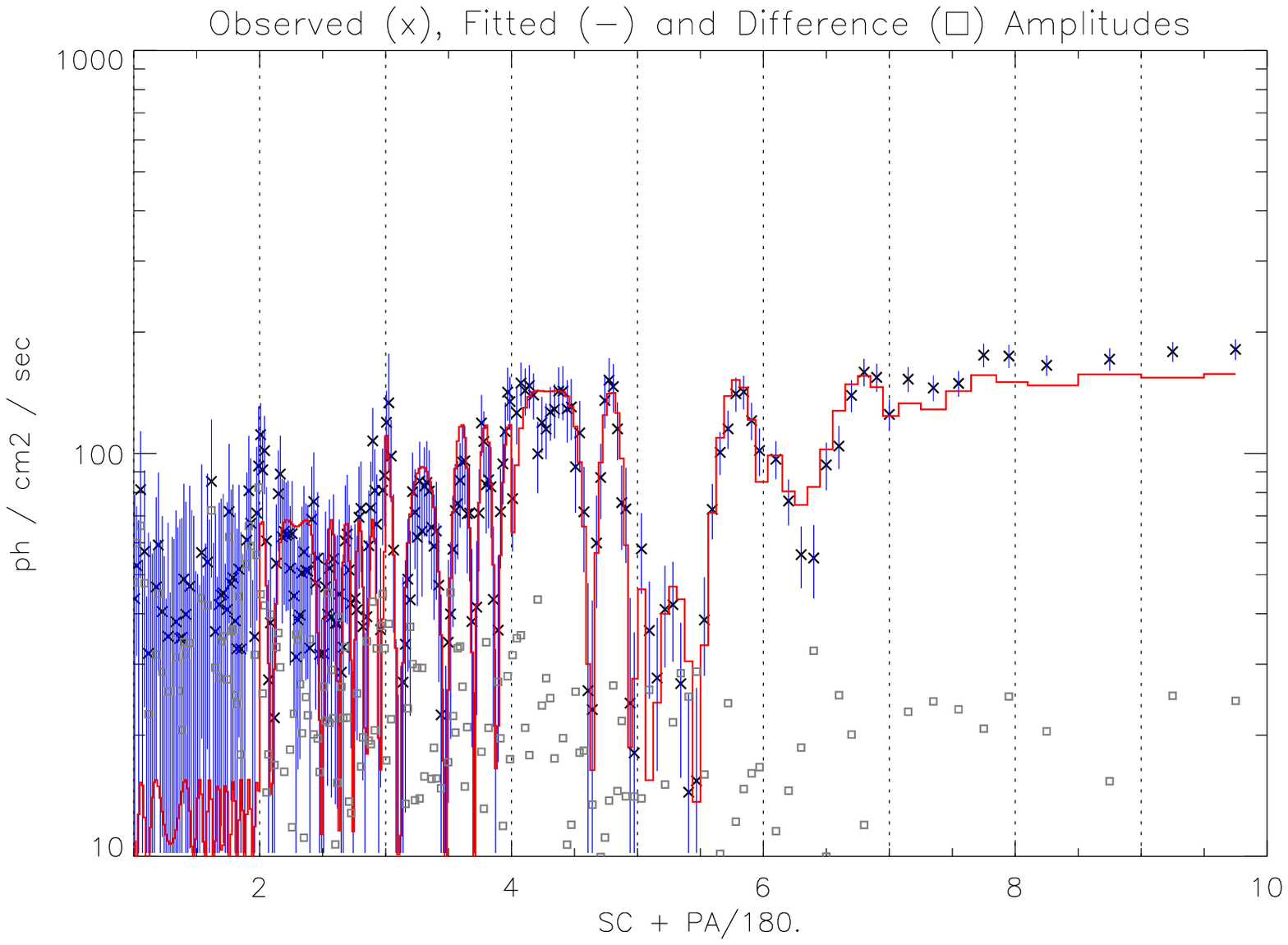}
\end{center}
\caption{Observed visibility amplitudes (crosses with blue error bars) for a flare
interval as a function of subcollimator
(SC=1-9) and position angle (PA = $0-180^{\circ}$)
of the grids in the energy range $12-25$~keV. Each of the 9 vertical
panels shows the amplitude as a function of PA for one
subcollimator (labeled by digits below the X-axis). The red
curve represents a model using two Gaussian sources, and the
squares show the residuals relative to the model. For a given
subcollimator ($5$ and $6$ are good examples), the amplitude rises
and falls while the grids rotate from PA=0 to PA$ = 180^{\circ}$.
Such patterns can correspond to an
extended or double source.} \label{fig:schmahl_fig5}
\end{figure}
\index{RHESSI@\textit{RHESSI}!subcollimators (SC)!visibilities}
\index{visibilities!X-ray!illustration}

Figure~\ref{fig:schmahl_fig5} illustrates observations expressed
in terms of visibilities\index{visibilities!electron}\index{hard X-rays!visibilities}.
This event occurred at about
$60^{\circ}$ longitude, so the albedo emission is expected to be
relatively weak, although there may be a signature of albedo for
subcollimators\index{RHESSI@\textit{RHESSI}!subcollimators (SC)} $8$ and $9$, as indicated by the excess observed
flux over the model flux.

\subsubsection{Future prospects for visibility-based albedo measurements with RHESSI}

In the context of albedo, visibilities\index{visibilities!electron}
\index{hard X-rays!visibilities} are used in two ways: first,
to generate maps of the source using a direct visibility
imaging method [such as back-projection
(MEM-NJIT) \citep{2007SoPh..240..241S} to obtain
primary source positions and fluxes;  and second, to use a visibility-based
forward-fit algorithm\index{imaging algorithms!visibility forward fit} to parameterize simple models of the primary source and
the corresponding parameters of the albedo component (size,
intensity, ellipticity, location).
\index{imaging algorithms!back-projection} or maximum entropy
\index{imaging algorithms!maximum entropy}
\index{RHESSI@\textit{RHESSI}!maximum entropy}

The MEM-NJIT\index{imaging algorithms!MEM-NJIT} mapping algorithm is useful for obtaining the
qualitative source configuration, although it is less reliable for
quantitative measurement of source sizes.  On the other hand the
forward-fit algorithm requires a good starting point in parameter
space, but can yield reliable source parameters.

There are practical difficulties that remain to be overcome.  As
suggested by Figure~\ref{fig:schmahl_fig5}, the relative
calibration (to within $\sim$5~\%) of the different subcollimators
(especially the coarse ones) is critical to the isolation of
albedo. Regrettably, the detector-to-detector calibration cannot
currently support such an objective\index{caveats!RHESSI@\textit{RHESSI}!detector-to-detector calibration}.  However, using relative
visibilities\index{visibilities!electron}
(viz., normalizing each RMC's visibilities to that
detector's spatially integrated response) eliminates the effects
of the detectors' relative efficiencies
and so provides a potential approach to
bypassing the calibration issue.
\index{hard X-rays!visibilities}
Confirmation of this approach
could be achieved with near-limb flares (presumed to be without
significant albedo) or at energies where the albedo component is
minimal. Other potential improvements are corrections for
azimuthal averaging in each visibility, improvements in the
present forward fit algorithm, perhaps with the use of an
alternative search algorithm and with the use of third harmonic
to add additional spatial frequencies.
visibilities\index{visibilities!electron}
\index{hard X-rays!visibilities}
\index{RHESSI@\textit{RHESSI}!rotating modulation collimators!modulation harmonics}

The first flares for which albedo\index{albedo!imaging}\index{Compton scattering!photospheric}
can be parameterized will
necessarily be spatially simple (single, compact, and strong) and
located within a few arc min of disk center. Eventually, it is
expected that spectroscopic tools for albedo
will be combined with imaging tools to provide comprehensive albedo information for a
large subset of {\em RHESSI} flares.  Since the albedo intensity
and location depends on electron directivity,
the potential reward of spatially-based albedo diagnostics is well worth the effort
required to refine the analysis tools.
\index{electrons!directivity} \citep{2006ApJ...653L.149K}

\section{The electron energy spectrum}\label{sec:7fbar}

In this section, the angular dependence of the bremsstrahlung
cross-section and the angular/spatial/temporal characteristics of
the electron distribution are neglected. The primary source photon
spectrum $I_P(\epsilon)$ may be therefore treated simply as the
convolution in electron energy of the solid-angle-averaged
bremsstrahlung cross-section and the mean source electron flux
(equation~\ref{eqn:emslie_fbar}).

In general, we first correct for instrumental effects
\citep[see][]{2002SoPh..210...33S,2002SoPh..210..165S}
such as pulse pileup\footnote{Pulse pileup is an issue for large solar flare spectra
with high count rate in {\it RHESSI} detectors.
Un-physical counts are recorded when pairs (or more) of low-energy photons,
arriving nearly simultaneously, are detected as a single energy count
at higher energies \citep[for details, see][]{1976SSI.....2..239D,1976SSI.....2..523D}.
Current pileup corrections for spatially integrated spectra have
limited precision that might be inadequate for the events with extremely
high count rates. As of April 2011, no standard image pileup corrections
yet exist.},
then correct the observed hard X-ray spectrum for albedo effects
(Section \ref{sec:7albedo_spectr}) to obtain the
primary source spectrum $I_P(\epsilon)$.
\index{pulse pileup}
As explained in
Section~\ref{sec:7emission}, the functional form of the primary
photon energy spectrum $I(\epsilon)$ (subscript `$P$' hereafter
understood) contains crucial information on the form of the {\it
mean source electron flux spectrum} ${\overline F}(E)$,
information that may in turn be used to reveal properties of the
electron acceleration and propagation processes.
\index{acceleration!and mean flux spectrum}

We first discuss forward fitting (Section \ref{sec:7ff}) and
regularized inversion (Section \ref{sec:7inversion}) methods of
extracting the electron energy spectrum ${\overline F}(E)$ from
noisy hard X-ray data $I(\epsilon)$.
    \subsection{Forward fitting}\label{sec:7ff} 
    \index{inverse problem!forward fit}\index{forward fit}


Forward fitting is the process of quantitatively comparing a
parameterized model with observational data.  Criteria are
established to determine acceptable fits and, if the model is
capable of providing acceptable fits, a best fit, that gives the
most probable values of the model parameters, is determined by
minimizing chi-squared ($\chi^2$), the sum of the squares of the
normalized residuals\footnote{A normalized residual is defined as
the difference between the measured and model-predicted value,
divided by the uncertainty in the measured value. The division by
the uncertainty gives greater weight to measured values with
smaller relative uncertainties \citep[e.g., Chapter 15
of][]{1992nrfa.book.....P}}.
\index{residuals!normalized}

Models may be based on the apparent structure of the data, a
physical model, or a combination of the two.  At the lowest X-ray
energies, \textit{RHESSI} flare spectra can contain {\it thermal
bremsstrahlung}\index{bremsstrahlung!thermal} and {\it
free-bound}
\index{free-bound emission} (recombination) continua
from the hottest plasma in flares, in addition to spectral lines
\citep[e.g.,][]{2006ApJ...647.1480P}. The X-rays at higher
energies (normally $\gapprox$10-20~keV) are dominated by
electron-ion bremsstrahlung\index{bremsstrahlung!electron-ion}
from energetic, nonthermal\index{bremsstrahlung!nonthermal}
electrons (Section \ref{sec:7ei}).
Free-bound radiation
\index{free-bound emission}\index{recombination radiation!non-thermal} from nonthermal
electrons\index{electrons!nonthermal}
may also contribute (Section~\ref{sec:7fb}).  At higher energies,
electron-electron
bremsstrahlung\index{bremsstrahlung!electron-electron}
can become
significant (Section~\ref{sec:7ee}). At $\gamma$-ray energies,
spectral lines excited by energetic, nonthermal ions and
positronium continuum emission can be present.
\index{positronium}\index{continuum!positronium}
The quality of a fit is also dependent on a careful subtraction
of background counts before obtaining the
spectral fit.
Spectral fits are only reliable\index{caveats!spectral fits and background subtraction} over the range of photon energies for which the flare emission is well above the
background.

\begin{figure}
\centering \scalebox{0.5}{\includegraphics{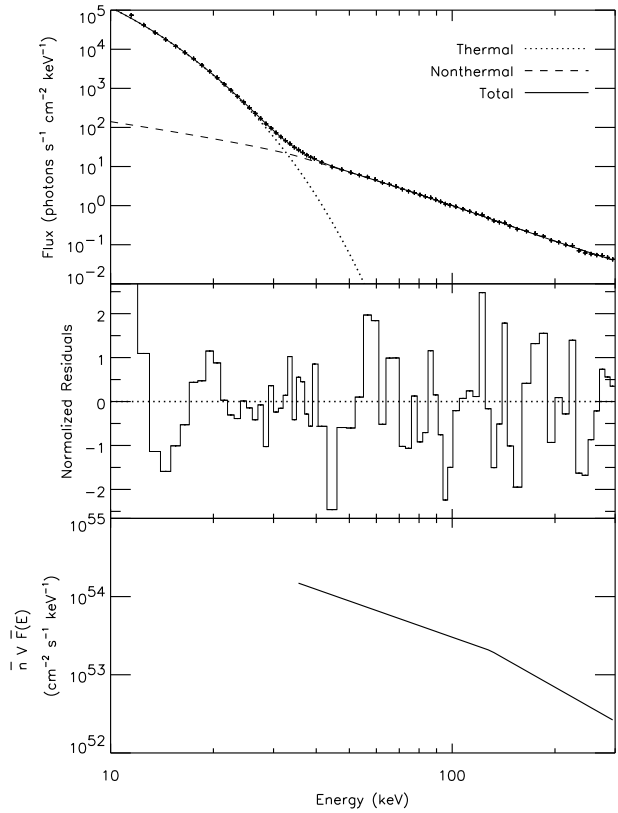}}
\vspace{30pt} \caption{Fit to a spatially-integrated spectrum from
SOL2002-07-23T00:35 (X4.8).  {\em Top Panel:} Photon flux
spectrum ({\em plus signs}) integrated over the 20~s time period
00:30:00--00:30:20~UT.  A fit to the spectrum ({\em solid curve})
consisting of the sum of the bremsstrahlung from an isothermal
plasma ({\em dotted curve}) and the bremsstrahlung from a double
power-law mean electron flux\index{mean electron flux} distribution with a low-energy
cutoff.  {\em Middle Panel:} Residuals from the fit in the top
panel (observed flux minus model flux divided by the 1$\sigma$
uncertainty in the observed flux).  {\em Bottom Panel:} Best fit
mean electron flux distribution times the mean plasma density and
source volume, plotted as a function of electron energy in keV
\citep[after][]{2003ApJ...595L..97H}. Note that pulse pileup
\citep{2002SoPh..210...33S,2003ApJ...595L.123K} might be an issue for this flare.}
\label{fig:Holman_July23_spectrum}
\end{figure}

\textit{RHESSI}'s high spectral resolution often makes it easy to
distinguish the thermal component of an X-ray spectrum from the
nonthermal component\index{electrons!nonthermal}, especially in large flares.
An example of this is shown in the top panel of
Figure~\ref{fig:Holman_July23_spectrum}, a spatially-integrated
spectrum from SOL2002-07-23T00:35 (X4.8)\index{flare (individual)!SOL2002-07-23T00:35 (X4.8)!electron spectrum}; below a photon
energy $\sim$40~keV, the thermal component\index{spectrum!electrons!thermal}
clearly dominates over
the flatter nonthermal\index{spectrum!electrons!nonthermal} component\index{nonthermal electrons}.
For many spectra, however, the
thermal and nonthermal components\index{electrons!nonthermal} are not so clearly
distinguishable. This was the case during the early rise of SOL2002-07-23T00:35 (X4.8), for example \citep{2003ApJ...595L..97H}.
For these spectra, the most likely model can sometimes be deduced from
the time evolution of the flare spectra and/or flare images
\citep[cf.][]{2005ApJ...626.1102S}.

\index{flare (individual)!SOL2002-07-23T00:35 (X4.8)!forward fit}
\index{flare (individual)!SOL2002-07-23T00:35 (X4.8)!photon spectrum}
\index{flare (individual)!SOL2002-07-23T00:35 (X4.8)!illustration}
\index{spectrum!photons!illustration}

The thermal component of the spectrum\index{soft X-rays!isothermal continuum approximation}
can typically be well fitted
with the form of $I(\epsilon)$ from an isothermal plasma at
temperature $T: I(\epsilon) \propto (EM/\epsilon T^{1/2}) \,
\exp(-\epsilon/kT)$, where $k$ is Boltzmann's constant and $EM$ is
the emission measure ($\int{n_e n_i dV}$).  The fit in
Figure~\ref{fig:Holman_July23_spectrum}, for example, gave a
temperature of 37~MK and an emission measure of $4.1 \times
10^{49}$~cm$^{-3}$.  A fit to thermal bremsstrahlung alone is
often adequate, but this does not account for the spectral line
complexes at $\sim$6.7~keV\index{Fe lines} and $\sim$8~keV or for recombination
radiation\index{RHESSI@\textit{RHESSI}!non-diagonal spectral response}\footnote{RHESSI spectroscopy in the range of energies
below $\sim$10~keV is often complicated by
non-diagonal
instrument response \citep{2002SoPh..210...33S,2002SoPh..210..165S}
and some instrumental features \citep[][]{2006ApJ...647.1480P}}.
\index{Chianti}
The thermal component of {\it RHESSI} spectra is now
routinely fitted (included into standard {\it RHESSI} software)
using the latest version of Chianti
\citep{2006ApJS..162..261L}, which incorporates all the emission
mechanisms important at low energies. The multithermality
of plasma and the corresponding emission measure differential in temperature
are addressed in Section \ref{sec:7thermal}.

The nonthermal component\index{electrons!nonthermal} of the spectra can usually be fitted
adequately with either a single or a double power-law photon
flux\index{spectrum!power-law!single component} spectral model\index{hard X-rays!simple power-law continuum approximation}.
Sometimes a third, flatter power-law component\index{spectrum!power-law!single
component} is included at low energies to simulate a
low-energy cutoff in the electron distribution
\index{spectrum!electrons!low-energy cutoff}\index{low-energy cutoff}. Such fits are useful for
examining the evolution of flare spectra with time. They do {\bf
not}, however, contain any direct physical information about the
electrons responsible for the observed emission. It is therefore
more interesting to fit the photon spectra with the radiation from
a model {\it electron} distribution\index{electrons!energy
distribution}, typically assumed to have the form of a double
power law with a possible low- and/or high-energy cutoff. This
form allows sharp breaks\index{spectrum!electrons!spectral break}
\index{spectral break} in the electron distribution (either a
mean electron flux\index{mean electron flux} ${\overline F}(E)$;
Equation~[\ref{eqn:emslie_fbar}] or, for a thick-target
model\index{thick-target model}, an injected electron distribution
${\cal F}_0(E_0)$; Equation~[\ref{eqn:emslie_int-rev}]).
However, due to the filtering of the bremsstrahlung cross-section
$Q(\epsilon,E)$, such breaks\index{spectrum!electrons!spectral break}
\index{spectral break} are generally smoothed out in the corresponding photon spectrum $I(\epsilon)$.

The nonthermal\index{electrons!nonthermal} part of the flare spectrum in
Figure~\ref{fig:Holman_July23_spectrum} is fit with the
bremsstrahlung from a double power-law mean electron flux
\index{mean electron flux}
distribution with a low-energy cutoff\index{low-energy cutoff}:

\begin{equation}
{\overline F}(E) = \left\{ \begin{array}{ll} 0 ;& E < E_c \\ A \,
E^{-\delta_1} ;& E_c < E < E_b \\ A \, E_b^{\delta_2 - \delta_1}
\, E^{-\delta_2} ;& E_b < E.
\end{array} \right.
\end{equation}
(bottom panel). The highest value of the low-energy
cutoff\index{particles!energy spectra!low-energy cutoff}
\index{spectrum!low-energy cutoff} $E_c$ consistent with a good fit
to the data was used; the value of $E_c$ is not constrained below
this value because of the dominance of thermal radiation.  This
fit therefore provides a lower limit to the energy in nonthermal
electrons.  The spectrum could not be acceptably fit with a single
power law; note that the location of the break\index{electrons!flux spectrum!spectral break}\index{spectral break} energy
$E_b$ is at a higher energy than the apparent location of the break in the
photon spectrum; this is because all electrons with energies above
a given photon energy contribute to the radiation at that photon
energy.

The photon spectrum residuals (using a sum of the isothermal and
nonthermal models) are shown in the middle panel of
Figure~\ref{fig:Holman_July23_spectrum}. Besides providing a reduced
$\chi^2$ ($\chi^2$ divided by the number of degrees of freedom in
the fit) close to $1$, the residuals from a good fit should be
random and uncorrelated and have a near-normal distribution
$N(0,1)$.
For the event in question, the residuals do {\it not}
exhibit this desired behaviour, with significant deviation at
photon energies between $10$ and $15$~keV.
This is not a well-understood issue and the explanation could be
either due to unaccounted non-diagonal response
\citep[see][]{2002SoPh..210...33S} or the presence of a "superhot" component
in the thermal continuum spectrum \citep[][]{2010ApJ...725L.161C}.
\index{superhot component}
Practically, broad spectral ``line'' is often included in the model to account for this feature;
with the inclusion of this \textit{ad hoc} feature, it is generally possible above $\sim$10~keV to obtain
good fits to the \textit{RHESSI} spectra without assuming the presence of any systematic
uncertainty in the data above the level of Poisson noise.

\begin{figure}
\centering
\includegraphics[width=.49\hsize]{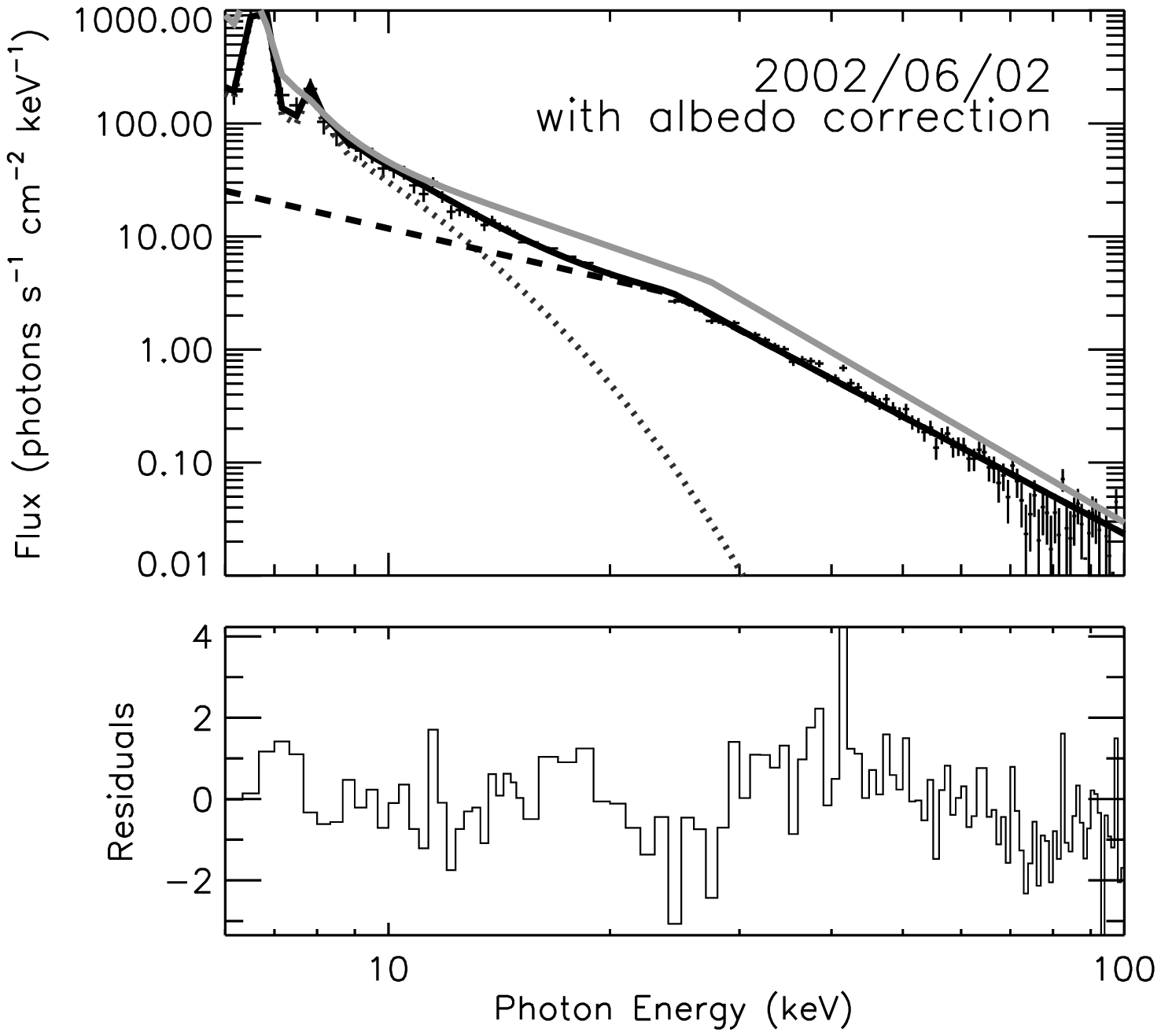}
\includegraphics[width=.49\hsize]{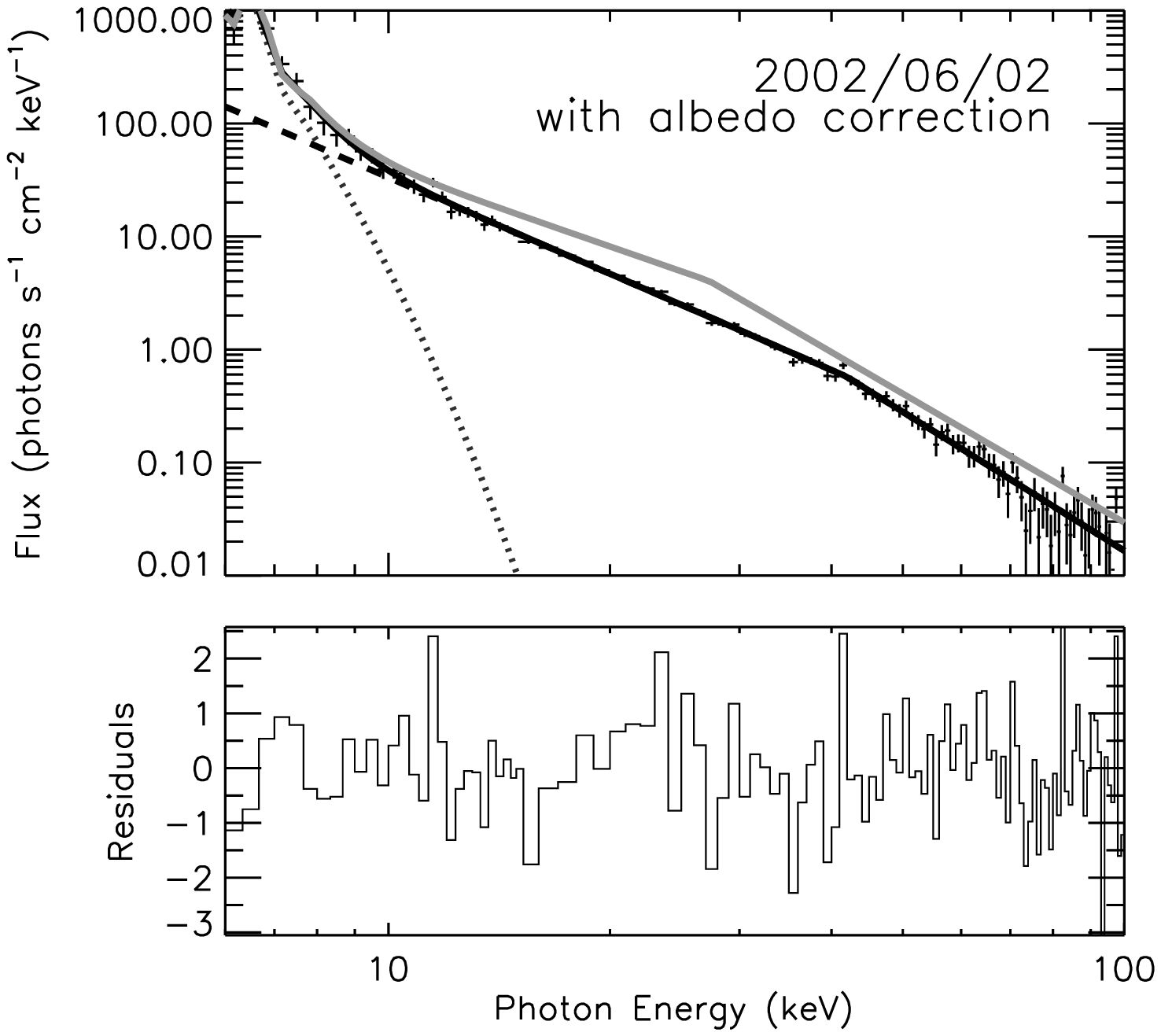}
\vspace{0.15in} \caption{Albedo-corrected \textit{RHESSI} spectrum (crosses
with error bars) at the hard X-ray peak (11:44:28--11:44:32~UT) of
SOL2002-06-02T20:44 (M1.0). The solid line shows the combined
isothermal (dotted line) plus double power-law (dashed line)
spectral fit. The spectral fit before albedo correction is
overlaid (gray, solid line). {\em Left panel:} The nonthermal part
of the spectrum is fitted to a double power-law model with a break
energy at $\sim$21~keV. {\em Right panel:} The same spectrum is
fitted to a double power-law model with a break energy at
$\sim$40~keV. The reduced $\chi^{2}$ values of the fits in the
left and right panels are 1.4 and 1.0, respectively. The
normalized residuals are plotted below each spectrum
\citep[after][]{2007ApJ...670..862S}.} \label{fig:Holman_Jun2_fits}
\index{flare (individual)!SOL2002-06-02T20:44 (M1.0)!electron spectrum}
\index{flare (individual)!SOL2002-06-02T20:44 (M1.0)!illustration}
\index{flare (individual)!SOL2002-06-02T20:44 (M1.0)!albedo}
\index{flare (individual)!SOL2002-06-02T20:44 (M1.0)!low-energy cutoff}
\index{flare (individual)!SOL2002-06-02T20:44 (M1.0)!forward fit}
\index{flare (individual)!SOL2002-06-02T20:44 (M1.0)!spectral break}
\index{spectrum!electrons!spectral break}
\index{spectrum!electrons!low-energy cutoff}
\index{spectral break}
\index{hard X-rays!albedo}\index{albedo}\index{photospheric albedo}
\end{figure}

A comparison of a poor forward fit (left panels) to a good forward
fit (right panels) is shown in Figure~\ref{fig:Holman_Jun2_fits}.
Both fits are to the same spectrum and in both the model is
bremsstrahlung from an isothermal plasma plus a double power-law
photon spectrum.  In the fit on the left, the break energy $E_b$ is
fixed at a value $\sim$21~keV, so the model is a single power
law above this energy.
\index{spectrum!electrons!spectral break}\index{spectral break}
In the fit on the right, the break energy
is allowed to adjust to a value that gives the best fit to the
data.
The reduced $\chi^2$ for the fit on the right is 1.0; in the
fit on the left it is 1.4, still consistent with an ``acceptable''
fit. However, the ``long wavelength'' oscillation in the residuals
for the fit on the left provides the most obvious clue that the
model is {\it not} adequate; the overall shape of the spectrum is
not consistent with the assumed model form.  Adding more
parameters (in this case a variable break energy $E_b$)
results in a more acceptable distribution of residuals, as in the right
panel (Figure~\ref{fig:Holman_Jun2_fits}).
\index{spectrum!electrons!spectral break}\index{spectral break}

The gray curves in Figure~\ref{fig:Holman_Jun2_fits} are fits to
the data before albedo, assumed to be from isotropically-emitted
photons, is taken into account (see Section~\ref{sec:7albedo_spectr}).
They demonstrate the significant impact that albedo can have
on the inferred spectrum of the emitted radiation.
\index{hard X-rays!albedo}\index{albedo}\index{photospheric albedo}
\index{hard X-rays!albedo}\index{albedo}\index{photospheric albedo}

An inadequately resolved issue with forward fitting is determining
the uncertainty in the fit parameters and the resulting model
function.  Since, in general, the fit is not linear, the fit
parameters are not independent and the uncertainties are not
necessarily even symmetric around the best-fit values.  In
Figure~\ref{fig:Holman_Jun2_fits}, for example, notice that the
temperature of the isothermal component is adjusted
to a higher value to compensate for the low break
energy in the double power-law component\index{spectrum!power-law!two-component}
\citep[][]{2007ApJ...670..862S}.
\index{spectrum!electrons!spectral break}\index{spectral break}
The uncertainty in the value of the
low-energy cutoff to the mean electron flux\index{mean electron flux} fit function in
Figure~\ref{fig:Holman_July23_spectrum} is small in the positive
(higher-energy) direction, but indefinitely large in the negative
direction!  As long as a good initial choice is made for the fit
parameters, the process of obtaining the best fits is relatively
quick.  An efficient method for determining the uncertainties in
the fit parameters and function is not in place, however. Bayesian
Monte Carlo approaches to determining these uncertainties are
robust but slow.  A practical solution to this important issue is
badly needed.
    \subsection{Regularized inversion}\label{sec:7inversion} 
    \index{inverse problem!regularized inversion}



The unprecedented energy resolution of {\em RHESSI} hard X-ray
spectra has introduced the need, perhaps for the first time in
solar hard X-ray spectroscopy, to apply sophisticated mathematical
tools for information retrieval in order to fully exploit the
physical significance of the recorded spectra. It is well
established that the most effective mathematical framework for
this problem lies in the theory of linear inverse problems
\citep[e.g.,][]{1986ipag.book.....C}. In this setting, linear integral
equations of the first kind relate the photon spectrum (``data'') to
the electron spectrum (``source function').  Such equations are
usually ill-posed\index{inverse problem!ill-posedness}
in the sense of Hadamard -- the effects of the
experimental noise can be strongly amplified by the intrinsic
numerical instability of the model \citep[for details, see][]{1985InvPr...1..301B}.
\index{Hadamard, J.}
\index{inverse problem}

Standard approaches to obtain a solutions of inverse problem
in solar X-ray spectroscopy are based on
forward-fitting \citep[e.g.,][]{2003ApJ...595L..97H} the photon
flux spectrum\index{spectrum!photons} with parametric forms of the electron flux spectrum
(see Section \ref{sec:7ff}). However, in forward-fitting\index{inverse problem!forward fit}, the
number of parameters utilized in the input form is generally
small.  This imposes severe, possibly artificial,
constraints\index{inverse problem!constraints} on the allowable
form for the source function and is the main reason
why inversion techniques
\citep[e.g.,][]{1985InvPr...1..301B,1994A&A...288..949P}, which find the best
model-free non-parametric fit to the data subject to physically
sound constraints, are currently a very
promising approach to data analysis in solar hard X-ray
spectroscopy.

A particularly promising technique is regularized
inversion\citep[e.g.,][]{1985InvPr...1..301B}\index{inverse problem!regularized inversion}.
The essence of the regularization technique
is to seek a least-squares solution of the pertinent integral equation (e.g.,
Equation [\ref{eqn:emslie_fbar}]) within a subset of the solution
space which accounts for some measure of {\it a priori}
information on the source function. Consider the linear system
\begin{equation}\label{eqn:Piana_my_eq8}
{\bf{g}}={\bf{A}} {\bf{f}},
\end{equation}
which represents a discretized version of the (Volterra) integral
equation (\ref{eqn:emslie_fbar}), where
\begin{eqnarray}\label{eqn:} A_{ij} = \frac{1}{4\pi
R^2} \, Q\left(\frac{\epsilon_{i}+\epsilon
_{i+1}}{2},\frac{E_{j}+E_{j+1}}{2}\right) \, \delta E_j
~,i=1,...,N;~j=1,...,M;
\end{eqnarray}
${\bf f}$ is the ``source vector''
${\overline{n}}V({\overline{F}}(E_1),\ldots,{\overline{F}}(E_M))$,
${\bf{g}}$ is the ``data vector''
$(g(\epsilon_1),\ldots,g(\epsilon_N))$ (with $M \ge N$), and the
$\delta \epsilon_i$ and $\delta E_j$ are appropriate weights. The
values $g(\epsilon_i)$ correspond to a set of discrete photon
counts in energy bands $\epsilon_i \rightarrow \epsilon_i + \delta
\epsilon_i$, while the ${\overline F}(E_j)$ are the corresponding
values of the mean electron flux\index{mean electron flux} in energy bands $E_j \rightarrow
E_j + \delta E_j$. Owing to the strong smoothing properties of the
integral operators, the matrix ${\bf{A}}$ is quasi-singular and
standard inversion routines cannot be effectively applied.
However, Tikhonov\index{inversion algorithms!Tikhonov
regularization} regularization theory \citep{ti63} obtains sufficiently smooth
source functions as the (unique) solution of the minimization
problem
\begin{equation}\label{eqn:Piana_my_eq9}
\|{\bf{g}}-{\bf{A}} {\bf{f}}\|^2 + \lambda \|{\bf{C}} {\bf{f}}\|^2
= {\rm minimum},
\end{equation}
where $\lambda$ is the {\it regularization parameter}, which tunes
the trade-off between the fitting term $\|{\bf{g}}-{\bf{A}}
{\bf{f}}\|^2$ and the penalty term $\|{\bf{C}} {\bf{f}}\|^2$. If
${\bf C}={\bf I}$, the method is termed {\it zero-order
regularization}\index{inversion algorithms!Tikhonov
regularization!zero-order} \citep{2003ApJ...595L.127P}, while if
${\bf{C}}$ is the matrix corresponding to numerical
differentiation, the method is termed {\it first-order
regularization}\index{inversion algorithms!Tikhonov
regularization!first-order} \citep{2004SoPh..225..293K}. The
optimal choice of the parameter $\lambda$ can be accomplished by
means of some optimization approach or by means of a
semi-heuristic technique based on a statistical analysis of the
cumulative residuals \citep{2003ApJ...595L.127P}.

The main disadvantage of using Tikhonov
regularization\index{inversion algorithms!Tikhonov regularization}
is that for noisy data, solutions with negative (unphysical) values
might result. A possible solution to this is provided by the
projected Landweber method\index{inverse problem!Landweber method} \citep{1997InvPr..13..441P}
\begin{equation}\label{eqn:Piana_my_eq10}
{\bf{f}}_{n+1} = P_{+} [{\bf{f}}_n + \tau {\bf{A}}^{T} ({\bf{g}} -
{\bf{A}} {\bf{f}}_n)]~~~,~~~{\bf{f}}_0 = 0,
\end{equation}
where $P_{+}$ sets to zero all negative components at each
iteration and $\tau$ is a relaxation parameter. In this framework,
the tuning between stability and fitting is realized by applying
some optimal stopping rule to the iterative procedure.

\subsubsection{Validation of regularization techniques}
\label{sec:Piana_Validation}
\index{inverse problem!regularized inversion!validation}

\begin{figure}    
\begin{center}
\includegraphics[width=0.8\textwidth]{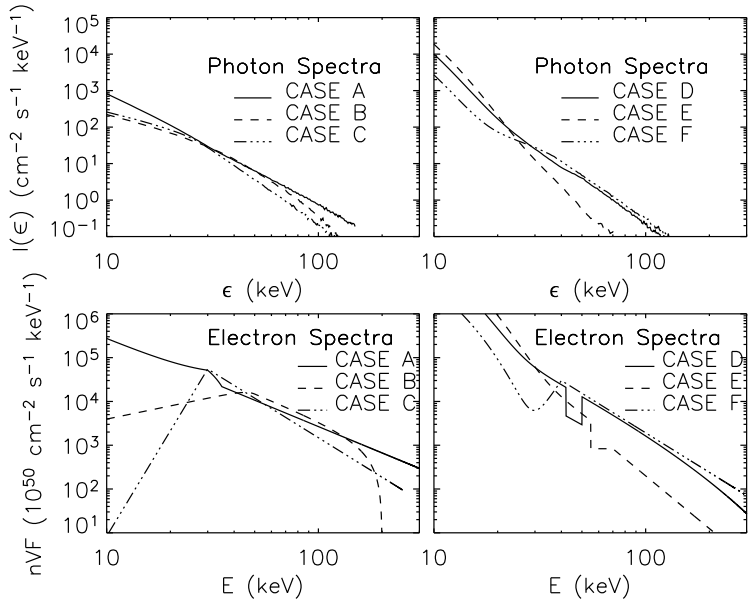}
\end{center}
\caption{Synthetic mean source electron spectra (bottom) and the
corresponding photon spectra (top) in a ``blind'' experiment for
assessing the effectiveness of different reconstruction methods in
\textit{RHESSI} X-ray spectroscopy \citep[after][]{2006ApJ...643..523B}.}
\label{fig:Piana_fig1}
\end{figure}

The effectiveness of different inversion algorithms, including a
standard forward-fitting technique, have been tested by
\citet{2006ApJ...643..523B} using synthetic data. In this test six
forms of the mean source electron spectrum ${\overline{F}}(E)$ in
Figure \ref{fig:Piana_fig1} (bottom), each one characterized by
specific features like bumps or cutoffs, were used to generate
the corresponding hard X-ray spectra $I(\epsilon)$ in
Figure~\ref{fig:Piana_fig1} (top), and a realistic amount of Poisson
noise added. In all cases, the photon spectra look smooth and quite
similar in their shape, while the corresponding mean electron
spectra exhibit very irregular behavior that is filtered out by
the smoothing effect of the bremsstrahlung cross-section.
These differences epitomize the mathematical concept of ill-posedness.
\index{inverse problem!ill-posedness}

The comparisons used four different techniques: zero-order
\citep{1994A&A...288..949P} and first-order
\citep{2004SoPh..225..293K} Tikhonov regularization, triangular matrix row elimination
with variable energy binning \citep{1992SoPh..137..121J} and forward-fitting with a parametric form consisting of
a double power law with low- and high-energy cutoffs plus an
isothermal component \citep[e.g.,][]{2003ApJ...595L..97H}.
These tests were done ``in the blind'' to recover ${\overline{F}}(E)$ for later
comparison with the true forms \citep{2006ApJ...643..523B}.
\index{inversion algorithms!Tikhonov regularization!zero-order}
\index{inversion algorithms!Tikhonov regularization!first-order}
\index{inversion algorithms!variable binning}
\index{inverse problem!regularized inversion!validation!list of methods}
All of these approaches
were able to reconstruct the general magnitude and form of ${\overline{F}}(E)$,
although forward-fitting inevitably fails to recover small
features which are not coded within the parameterized form of the
model function.

\subsubsection{Application to \textit{RHESSI} data}
\label{sec:Piana_Applications}

\begin{figure}    
\begin{center}
\psfig{file=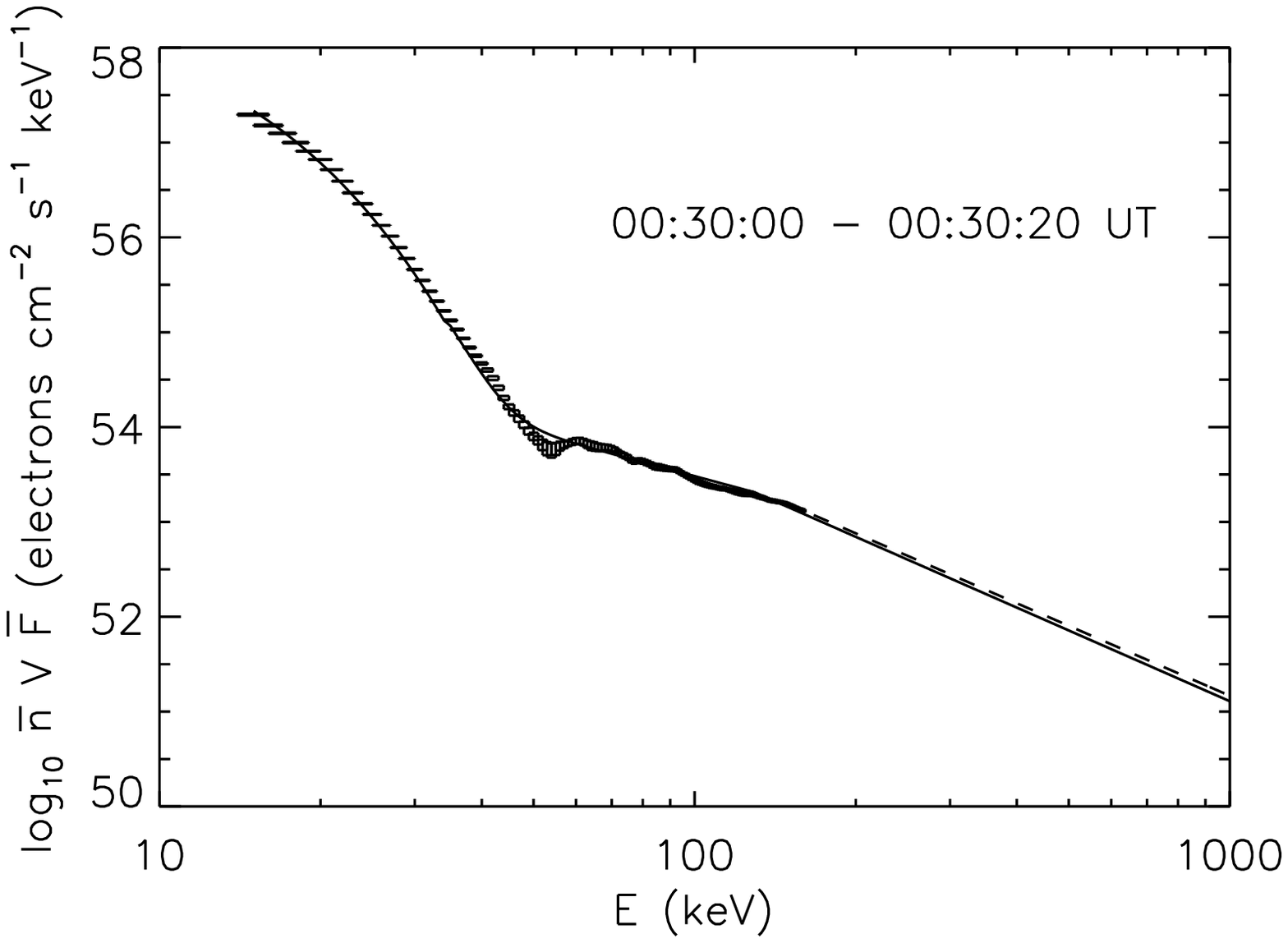,width=8.cm,clip=}
\end{center}
\caption{Regularized spectrum ${\overline{n}} V {\overline{F}}$
versus electron energy $E$ for the time interval shown in SOL2002-07-23T00:35 (X4.8).
\index{flare (individual)!SOL2002-07-23T00:35 (X4.8)!electron energy spectrum}
The spectrum has been extended at
high energies using a power law of index $\delta = 2.45$ (dashed
lined). The vertical size of the error boxes reflects the
$3\sigma$ limit caused by statistical noise in the observed
$I(\epsilon)$. The spectrum obtained by a forward-fitting
procedure using the same bremsstrahlung cross-section is shown as
a solid line \citep[after][]{2003ApJ...595L.127P}.}
\label{fig:Piana_fig2}

\end{figure}

The Tikhonov regularization\index{inversion algorithms!Tikhonov regularization} method has been applied to hard X-ray
measurements recorded by {\em RHESSI} for SOL2002-02-20T11:07 (C7.5)\index{flare (individual)!SOL2002-02-20T11:07 (C7.5)!electron energy spectrum},
SOL2002-03-17T19:31 (M4.0)\index{flare (individual)!SOL2002-03-17T19:31 (M4.0)!electron energy spectrum},
SOL2002-08-06T12:59 (C7.9)\index{flare (individual)!SOL2002-08-06T12:59 (C7.9)!electron energy spectrum}
\citep{2003A&A...405..325M},
SOL2002-02-26T10:27 (C9.6)\index{flare (individual)!SOL2002-02-26T10:27 (C9.6)!electron
energy spectrum} \citep{2005SoPh..226..317K} and
SOL2002-07-23T00:35 (X4.8)\index{flare (individual)!SOL2002-07-23T00:35 (X4.8)!electron
energy spectrum} \citep{2003ApJ...595L.127P}. In this last paper a
``dip-hump'' feature in the recovered mean source electron spectrum
${\overline{n}}V{\overline{F}}$ was noted near $E=55$ keV (Figure
\ref{fig:Piana_fig2}); such a feature is (by construction) absent
in the superimposed forward-fit spectrum
\citep{2003ApJ...595L..97H} using the same
\citep{1997A&A...326..417H} bremsstrahlung
cross-section\index{bremsstrahlung!Haug cross-section}\index{cross-sections!Haug}.

The $3\sigma$ error bars plotted in Figure~\ref{fig:Piana_fig2}
clearly show that this ``dip-hump'' feature is statistically
significant.  Its physical interpretation is still an open issue,
but it may reflect the depletion of low-energy nonthermal
electrons due to the effect of Coulomb collisions for an
injected distribution\index{electrons!injected distribution} with a low-energy cutoff \citep{1984ApJ...279..882E}.
\index{low-energy cutoff}
\index{pulse pileup}
\index{hard X-rays!spectra!and pulse pileup}
\index{flare (individual)!SOL2002-07-23T00:35 (X4.8)!pulse pileup}
\index{RHESSI@\textit{RHESSI}!pulse pileup}
On the other hand, it may simply
reflect an inadequate correction for the pulse pileup
effect in the {\em RHESSI} detectors, with a resultant aliasing of
the photon spectrum used in the construction of
Figure~\ref{fig:Piana_fig2} \citep{2002SoPh..210...33S,2003ApJ...595L.123K}.

    \subsection{High-energy cutoffs in the electron distribution}\label{sec:7hecutoff} 
\index{hard X-rays!high-energy cutoff}
\index{electrons!distribution function!high-energy cutoff}

It should be noted that the matrix ${\bf A}$ in Equation
(\ref{eqn:Piana_my_eq8}) need not be square, so that the energy
range corresponding to the electron flux (source) vector ${\bf f}$
may extend over a larger range than the photon flux (data) vector
${\bf g}$.  Physically, this corresponds to the production
(through free-free bremsstrahlung only)
of photons of energy $\epsilon$ by electrons of energy $E >
\epsilon_{\rm max}$, where $\epsilon_{\rm max}$ is the largest
photon energy observed. It should be noted that, because the
bremsstrahlung cross-section $Q(\epsilon, E)$ is a non-diagonal matrix of $E$,
the form of the photon spectrum at photon energy
$\epsilon$ does provide information on the electron spectrum for
all energies $E > \epsilon$, including those with $E >
\epsilon_{\rm max}$. Consequently, a distinct advantage of the
regularization methodologies \citep[over, say, the more
straightforward matrix inversion method of][]{1992SoPh..137..121J}
is that some information can be obtained on the form of the electron
spectrum ${\overline F}(E)$ at energies beyond the maximum
photon energy observed.

A particular example of this is the possible existence of a
high-energy cutoff $E_{\rm max}$ in the {\it electron} spectrum
${\bar F}(E)$.  Although for free-free emission this would
correspond to a maximum photon energy $\epsilon_{\rm max} = E_{\rm
max}$, $\epsilon_{\rm max}$ may lie beyond the range of
statistically useful photon (or count) data. However, as shown by
\citet{2004SoPh..225..293K}, the regularized inversion method can,
in principle, detect the presence of the high-energy
cutoff\index{electrons!energy distribution!high-energy cutoff} at
$E = E_{\rm max}$ through its effect on the photon spectrum at
observable energies $\epsilon$ that are all significantly less
than $E_{\rm max}$, and this technique was indeed used to discern
a high-energy cutoff (or at least a very sharp downward spectral
break)\index{electrons!flux spectrum!spectral break}\index{spectral break}
in the flare of SOL2002-02-26T10:27 (C9.6)
\index{flare (individual)!SOL2002-02-26T10:27 (C9.6)!electron energy spectrum}
\index{flare (individual)!SOL2002-02-26T10:27 (C9.6)!high energy cutoff} by
\cite{2005SoPh..226..317K}. The ability of the regularization
technique to detect such high-energy cutoffs in the electron
spectrum was dramatically highlighted in the validation study of
\citet{2006ApJ...643..523B}. In that study, a synthetic electron
spectrum ${\bar F}(E)$ with a high-energy cutoff at 200~keV
(case~B in Figure \ref{fig:Piana_fig1}) was used to generate noisy
photon ``data'' in the range wholly below 100~keV. Analysis of
these ``data'' using both zero-order and higher-order
regularization techniques rather faithfully reproduced the
high-energy cutoff in the electron spectrum, with an accuracy
better than 30\%. High-energy cutoffs have had very limited
use to date as a parameter in forward-fitting methods
\citep{2002SoPh..210..245S,2003ApJ...595L..97H}, although both forward-fit
and direct matrix inversion methods \citep{1992SoPh..137..121J}
should be able to say something about the electron spectrum
above maximum photon ``data'' energy, in that if there is
a high energy cutoff $E_{\rm max}$ close to the maximum observed
photon energy the spectral shape is much different.

    \subsection{Spectral breaks in the electron distribution}\label{sec:7knee} 


\begin{figure}[pht]
\begin{center}
\includegraphics[width=0.49\textwidth]{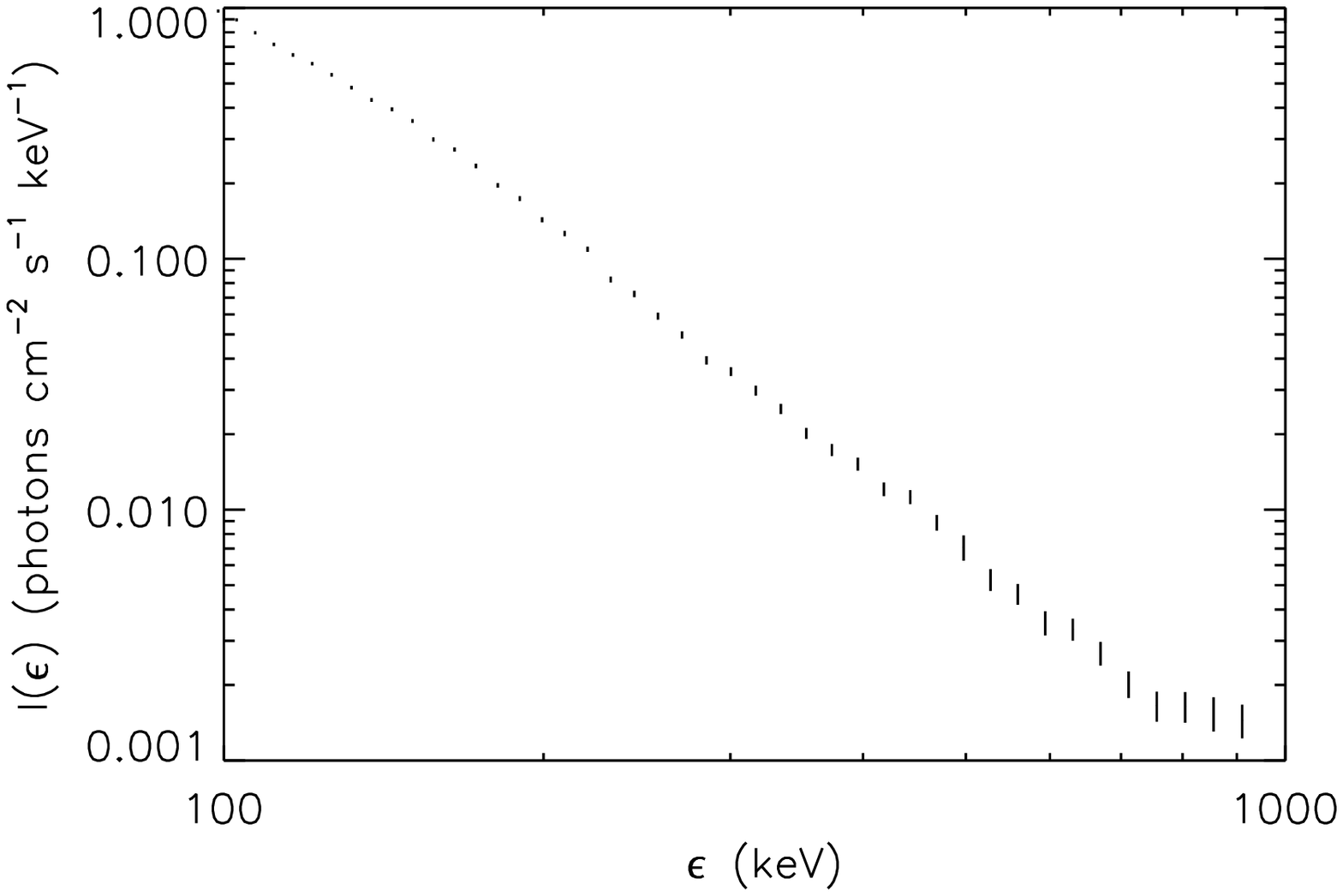}
\includegraphics[width=0.49\textwidth]{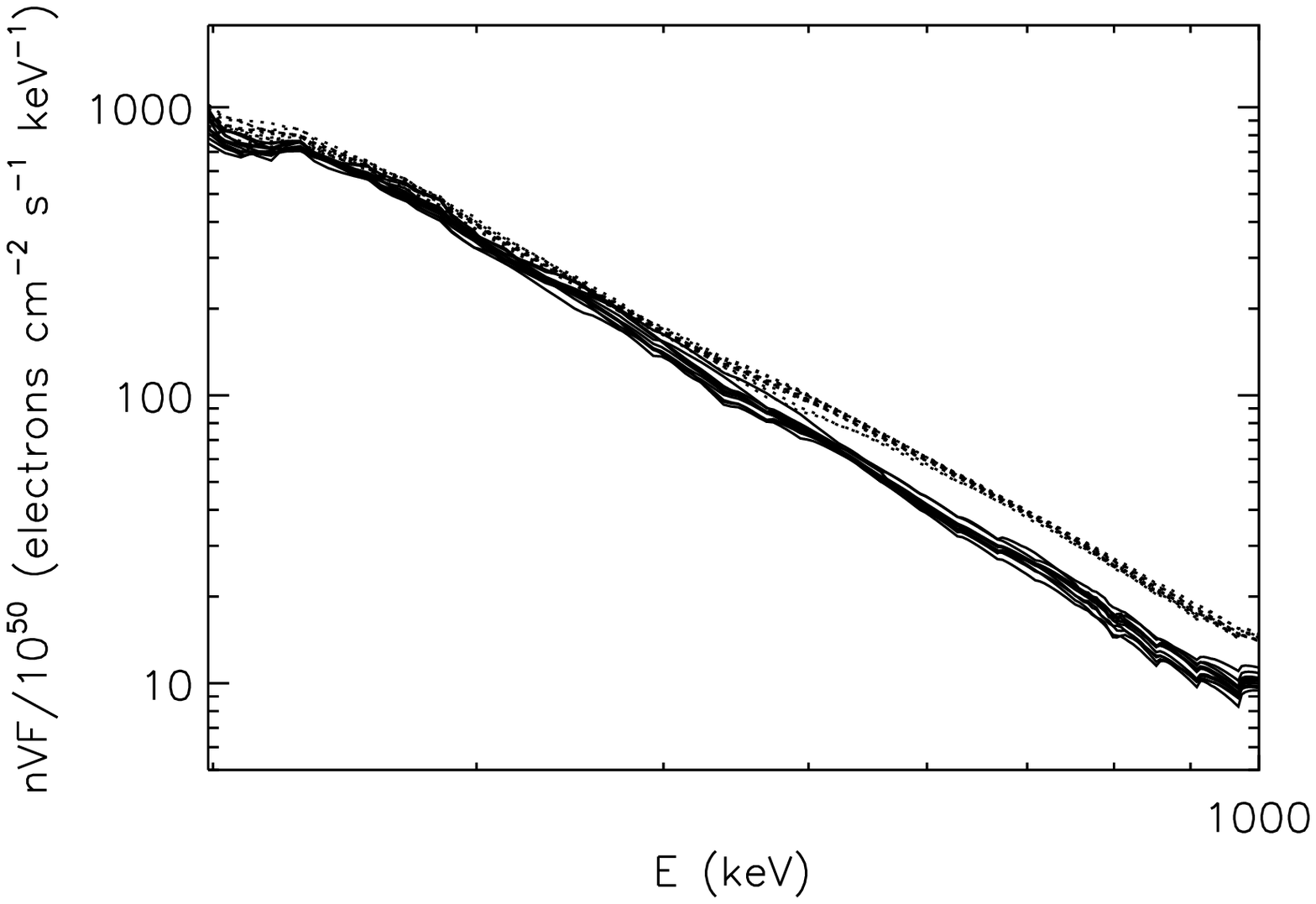}
\end{center}
\caption{{\it Left panel:} Photon spectrum for the time interval 09:43:16 --
09:44:24 UT in SOL2005-01-17T09:52 (X3.8), with gamma-ray lines
removed\index{gamma-rays!nuclear line radiation}. {\it Right panel:} Recovered forms of the quantity $\bar n \, V \,
{\overline F}(E)$ (in units of
$10^{50}$~electrons~cm$^{-2}$~s$^{-1}$~keV$^{-1}$; see
equation~[\ref{eqn:emslie_fbar}]) using a zero-order
regularization technique and presented as a ``confidence strip,''
i.e., a series of solutions, each based on a realization of the
data consistent with the size of the uncertainties. The dashed
lines assume electron-ion emission only; the solid lines include
the additional electron-electron emission term \citep[after][]{2007ApJ...670..857K}.}
\index{flare (individual)!SOL2005-01-17T09:52 (X3.8)!photon spectrum}
\index{flare (individual)!SOL2005-01-17T09:52 (X3.8)!electron energy spectrum}
\index{flare (individual)!SOL2005-01-17T09:52 (X3.8)!illustration}
\index{flare (individual)!SOL2005-01-17T09:52 (X3.8)!electron-electron bremsstrahlung}
\label{ee_knee}
\end{figure}

Figure~\ref{ee_knee} shows the photon spectrum for the time
interval 09:43:16 -- 09:44:24 UT (the time of approximate peak
flux) for SOL2005-01-17T09:52 (X3.8)
\citep[after][]{2007ApJ...670..857K}.
\index{gamma-rays!nuclear line radiation}
\index{flare (individual)!SOL2005-01-17T09:52 (X3.8)!electron energy spectrum}
\index{flare (individual)!SOL2005-01-17T09:52 (X3.8)!electron-electron bremsstrahlung}
This event, which produced several strong gamma-ray lines,
was previously studied by \citet{2006ApJ...653L.149K}, who
concluded that the pitch-angle distribution for electrons up to
$\sim$300~keV was close to isotropic.  There is evidence for an
upward break in the spectrum at energies $\gapprox$300~keV.
\index{spectrum!electrons!spectral break}\index{spectral break}
For completeness, we note that the energy range $\gapprox 300$ also contains
some pseudo-continuum $\gamma$-ray contributions from ions \citep[e.g.][]{2003ApJ...595L..81S}.

The mean source electron spectrum\index{electrons!mean source spectrum} ${\overline F}(E)$
recovered from the photon data, using two
different bremsstrahlung cross-sections -- electron-ion only and
(electron-ion + electron-electron) is shown in Figure~\ref{ee_knee}.  (The results are presented in
the form of two ``confidence strips,''\index{confidence strip} bundles of solutions using
different noisy realizations of the same data.) The ${\overline
F}(E)$ recovered using the full cross-section~(\ref{eq:ee_qtot}),
including electron-electron bremsstrahlung,
is, for $E \gtrsim$300~keV, steeper (spectral index greater by $\sim$0.4) than the
${\overline F}(E)$ recovered assuming purely electron-ion
emission.
\index{bremsstrahlung!electron-electron}
\index{spectrum!photons!spectral index}
\citet{2007ApJ...670..857K} point out that while the
upward break at $\epsilon \approx $300~keV in Figure~\ref{ee_knee}
is real, the break\index{spectrum!electrons!spectral break}
\index{spectral break} at $E \simeq 400$~keV in the ${\overline F}(E)$
recovered using the electron-ion bremsstrahlung cross-section
alone (Figure~\ref{ee_knee}) is an artifact of the neglect of
electron-electron emission at energies $\gtrsim$300~keV.  The
true form of ${\overline F}(E)$, obtained using the full
cross-section, has a rather straightforward power-law form over the
energy range from $200-1000$~keV.  As pointed out by
\citet{2007ApJ...670..857K}, the inclusion of electron-electron
bremsstrahlung may remove the need to explain ``break energies'' in several events
\citep[cf.][]{1985SoPh..100..465D,1986ApJ...311..474H,1995ApJ...443..855L,
1998A&A...334.1099T,1988SoPh..118...95V}.
\index{hard X-rays!break energies}
\index{bremsstrahlung!electron-electron}
\index{electrons!energy distribution!spectral break}\index{spectral break}
However electron-electron
bremsstrahlung cannot explain spectral breaks
\index{spectrum!electrons!spectral break}\index{spectral break} at low energies $\leq$300~keV
often observed in \textit{RHESSI} spectra \citep[e.g.,][]{2003A&A...407..725C}.

    \subsection{Low-energy cutoffs in the electron distribution}\label{sec:7cutoff} 


Given that nonthermal electron\index{electrons!nonthermal} spectra
\index{spectrum!electrons!nonthermal}
often have a form close to
a steep power law $E^{-\delta}$,
with $\delta > 2$ \citep{1985SoPh..100..465D}, an accurate value for the low-energy cutoff $E_c$
parameter is required to obtain values of the total energy in nonthermal electrons
\citep[see, however,][]{2003ApJ...595L.119E,2009ApJ...707L..45H}.\index{spectrum!electrons!low-energy cutoff}\index{low-energy cutoff} \index{spectrum!electrons!nonthermal}\index{spectrum!electrons!power-law}\index{electrons!nonthermal}
Thus, the determination of $E_c$ plays a key role in the interpretation of hard X-ray data
\citep[see also][]{chapter3}.

In most cases, the value of $E_c$ must be somehow disentangled
from the combined (thermal + nonthermal)
\index{electrons!nonthermal} form of ${\overline
F}(E)$, and the value of $E_c$ is, in general, only weakly
constrained by observations. Some flares require rather high
values of $E_c$ to explain observations.
The presence of a low-energy cutoff introduces a flattening of the photon spectrum
at energies below $E_c$ \citep[see, e.g.,][]{2003ApJ...586..606H,2009SoPh..257..323H,2009ChA&A..33..168H}.
\citet{1990ApJ...353..313N}
show that the spectrum of the impulsive component flattens toward
low energies, suggesting a value of $E_c$ as high as 50~keV.
\citet{1997A&A...320..620F}, using \textit{Yohkoh} data,
observed a few flares with flat spectra below 30~keV.
\index{Yohkoh@\textit{Yohkoh}}\index{Yohkoh@\textit{Yohkoh}!HXT}

\begin{figure}
\begin{center}
\includegraphics[width=0.6\textwidth]{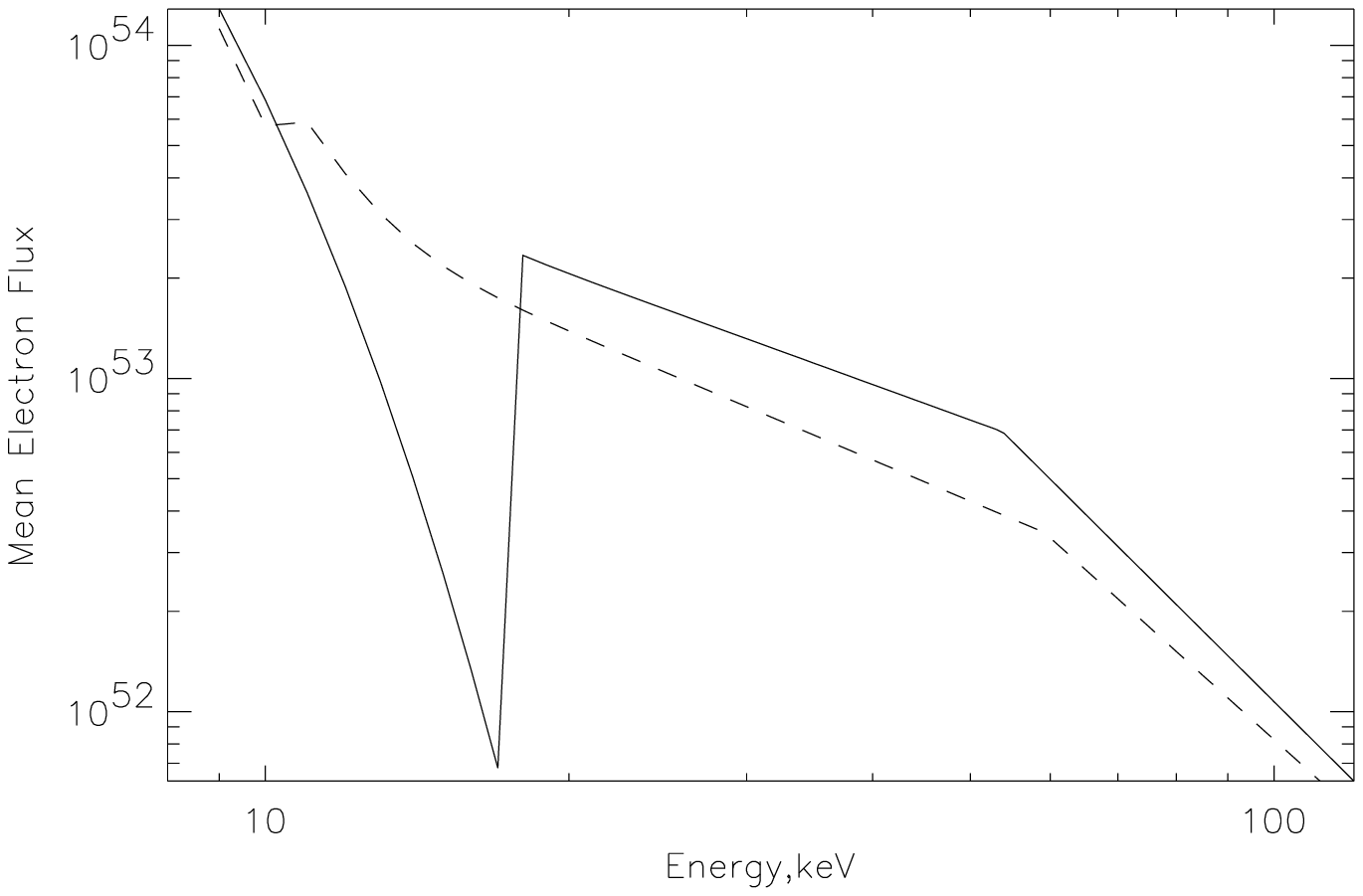}
\end{center}
\caption{Mean electron flux spectra ${\overline n}V{\overline F}(E)$ of SOL2002-09-17T05:54 (C2.0) recovered using forward fitting. The solid/dashed lines show the
spectrum without/with the albedo correction\index{albedo!correction}
\citep[after][]{2006A&A...446.1157K}.}
\index{flare (individual)!SOL2002-09-17T05:54 (C2.0)!albedo}
\index{flare (individual)!SOL2002-09-17T05:54 (C2.0)!low-energy cutoff}
\index{flare (individual)!SOL2002-09-17T05:54 (C2.0)!electron energy spectrum}
\index{flare (individual)!SOL2002-09-17T05:54 (C2.0)!illustration}
\label{Ko1_spectr2}
\end{figure}

Studies before {\em RHESSI} mostly assumed an arbitrary, fixed,
value of $E_c$. More recently, using {\em RHESSI}
high-spectral-resolution data (without albedo correction),
several studies have reported clear
evidence for a low-energy cutoff\index{spectrum!electrons!low-energy cutoff}\index{low-energy cutoff} or even a dip (or gap)
\citep[e.g.,][]{2006AdSpR..38..945K} in the
mean electron flux\index{mean electron flux} distribution ${\overline F}(E)$ between its
thermal and nonthermal
\index{electrons!nonthermal} components, leading to a clear
identification of a low-energy cutoff\index{spectrum!electrons!low-energy cutoff}
for the nonthermal\index{electrons!nonthermal} electron distribution (Figure \ref{Ko1_spectr2}).

Forward-fitting methods (Section \ref{sec:7ff}) usually assume a
strict low-energy cutoff, i.e., a sharp change in the ${\overline
F}(E)$.  However, as showed by, e.g., \citet{1996ApJ...466.1067K}
or \citet{2005SoPh..232...63K}, it is often not possible to
determine a unique best spectral fit; fit parameters can be varied
substantially without unacceptably large changes in the photon
spectrum, with other free parameters compensating. In particular,
because of the dominance of thermal emission\index{soft X-rays!spectral dominance}
at low energies,
forward-fit approaches can reliably infer only {\it upper limits}
to the value of $E_c$; values of $E_c$ well below such upper
limits provide equally good fits to an observed X-ray spectrum,
with a somewhat different value of the thermal source temperature
$T$.  Moreover, as mentioned in (Section \ref{sec:7ff})
determining confidence intervals for model parameters is an
extremely time-consuming task.

\begin{figure}
\begin{center}
\includegraphics[width=0.6\textwidth]{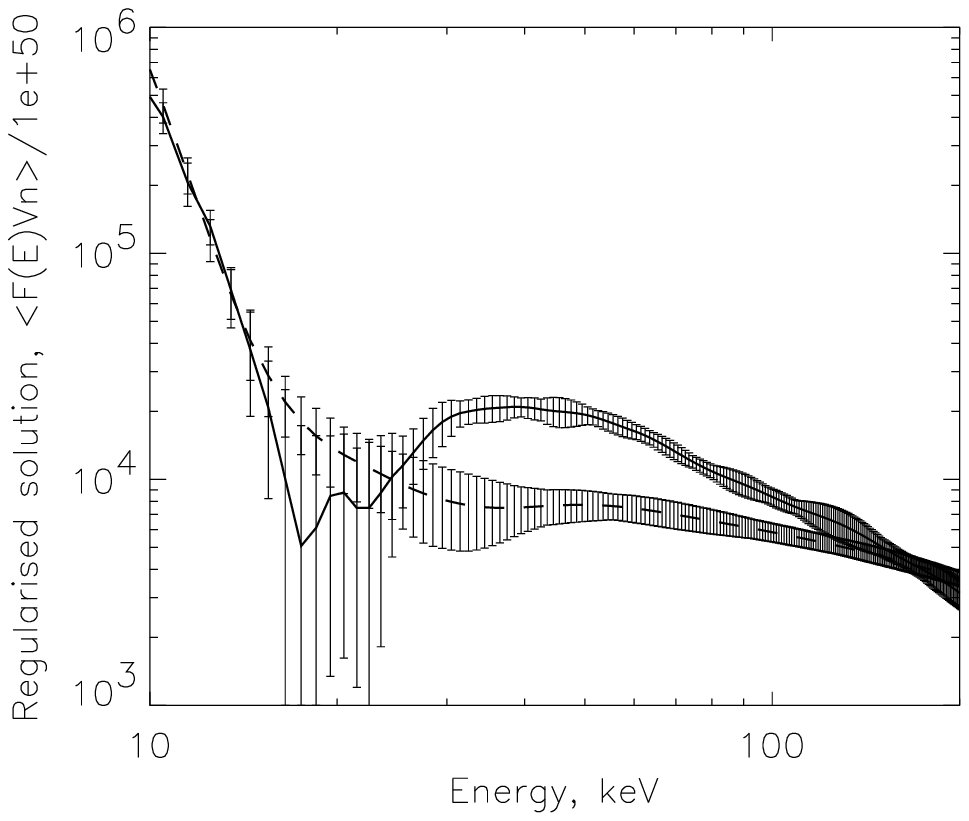}
\end{center}
\caption{Mean source electron flux spectrum ${\overline n}V{\overline F}(E)$ for SOL2002-08-20T08:25 (M3.4) for the time interval 08:25:20~-~08:25:40 UT recovered using regularized inversion.
\index{flare (individual)!SOL2002-08-20T08:25 (M3.4)!low-energy cutoff}
\index{flare (individual)!SOL2002-08-20T08:25 (M3.4)!illustration}
The
solid line shows the spectrum without albedo correction\index{albedo!correction}. The
confidence intervals represent the range of solutions found for
different statistical realizations of the photon spectrum
\citep[after][]{2006A&A...446.1157K}.}
\index{flare (individual)!SOL2002-08-20T08:25 (M3.4)!electron energy spectrum}
\index{flare (individual)!SOL2002-08-20T08:25 (M3.4)!low-energy cutoff}
\index{flare (individual)!SOL2002-08-20T08:25 (M3.4)!albedo}
\index{flare (individual)!SOL2002-08-20T08:25 (M3.4)!illustration}
 \label{Ko1_spectr3}
\end{figure}

On the other hand, regularized inversion methods\index{inversion algorithms} (Section
\ref{sec:7inversion}) have proven their ability to detect dips in
${\overline F}(E)$ \citep[see][]{2003ApJ...595L.127P}. These
methods also provide estimates of uncertainties in the solution,
through so-called {\it confidence strips} in which 99\% of
data-consistent solutions lie; see~\citet{2006ApJ...643..523B}.
Using this approach, clear low-energy cutoffs or dips in the mean
electron flux\index{mean electron flux} distribution
${\overline F}(E)$ at an energy around
20-40~keV were reported by \citet{2005SoPh..232...63K} for
SOL2002-08-20T08:25 (M3.4).
\index{flare (individual)!SOL2002-08-20T08:25 (M3.4)!low-energy cutoff}
\index{flare (individual)!SOL2002-09-17T05:54 (C2.0)!low-energy cutoff}
Forward-fit methods yielded $E_c = 44\pm 6$~keV, a somewhat higher value.
\citet{2006AdSpR..38..945K}
found a clear dip around 20~keV for the SOL2002-09-17T05:54 (C2.0) event.
For further examples of low-energy
cutoffs obtained through both forward-fitting and regularization,
see Figure~\ref{Ko1_spectr2} and Figure~\ref{Ko1_spectr3},
respectively.


Some flares, such as SOL2002-04-25T06:02 (C2.5), SOL2002-09-17T05:54 (C2.0),
or SOL2002-08-20T08:25 (M3.4), have quite hard (flat) spectra with a relatively
weak thermal components.
\index{flare (individual)!SOL2002-04-25T06:02 (C2.5)!albedo}
\index{flare (individual)!SOL2002-09-17T05:54 (C2.0)!albedo}
\index{flare (individual)!SOL2002-08-20T08:25 (M3.4)!albedo}
Such a flat form of the photon spectrum
can require \citep{2006AdSpR..38..945K} a low-energy cutoff
\index{spectrum!electrons!low-energy cutoff}\index{low-energy cutoff}, or
local minimum, in the corresponding mean electron distribution
${\overline F}(E)$. These local minima are particularly
interesting, since if ${\overline F}(E)$ is sufficiently steep
(steeper than $E^1$; see the remarks after
Equation~(\ref{eqn:emslie_fo})), it could have a form inconsistent
with the widely-used collision-dominated thick-target model
\index{thick-target model} for
X-ray production \citep{2006AdSpR..38..945K,2009A&A...508..993B}.

However, the effects of Compton backscattering
(Section \ref{sec:7albedo_spectr}) on the hard X-ray spectrum are most
pronounced for flares with such hard spectra.
\index{Compton scattering!photospheric}
\citet{2007A&A...466..705K} showed that the flares listed above
are all located close to solar disk center ($\mu> 0.5$,
$\theta < 60^o$; denoted as stars in
Figure~\ref{fig:kasparova_directivity_gamma0_mu}) and they
therefore attributed the flat spectra to the
heliospheric-angle-dependent albedo(Section \ref{sec:7albedo_spectr}).
\index{hard X-rays!albedo}\index{albedo}\index{photospheric albedo}
This result is consistent with earlier observations of
\citet{1990ApJ...353..313N} and \citet{1997A&A...320..620F}, who observed
several flares with flat spectra which were located not far from
disk center ($\mu\ge 0.6$) or from near center ($\mu \approx 1.0$),
respectively.  Interestingly, adding considerations of the albedo
\index{hard X-rays!albedo}\index{albedo}\index{photospheric albedo} in such events
\citep{2006A&A...446.1157K} removes the
spectral hardening, and hence the need for a low-energy cutoff in
this photon energy range -- see Figure \ref{Ko1_spectr2} and
\ref{Ko1_spectr3},
\cite{2005SoPh..232...63K},  and \cite{2006A&A...446.1157K}.
Recently, \citet{2008SoPh..252..139K} have
analyzed a large number of solar flares with weak thermal components and flat photon spectra.
It has been shown that if the isotropic albedo correction is applied, all low-energy cutoffs
in the mean electron spectrum are removed, and hence the low-energy cutoffs
in the mean electron spectrum of solar flares above $\sim$12~keV cannot be viewed as real features.
If low-energy cutoffs exist in the mean electron spectra, their energies should be less than $\sim$12~keV.
Thus, the apparent low-energy cutoff
\index{low-energy cutoff}
\index{spectrum!electrons!low-energy cutoff}
in the mean electron distribution is most likely to be
a feature connected with albedo;
it is {\it not} a true physical property \citep{2009ApJ...707L..45H}.
\index{hard X-rays!albedo}
\index{albedo}\index{photospheric albedo}
This result can substantially change the total electron energy requirements in a given flare.

In summary, our view on the existence and/or value of low-energy
cutoffs has been significantly broadened
since high quality \textit{RHESSI} observations have become available.
\index{spectrum!electrons!low-energy cutoff}\index{low-energy cutoff}
Yet, the determination and even the
existence of a low-energy cutoff remains a very complex issue.
First, the contribution of the photospheric albedo to the observed photon
spectrum must be taken into account and removed before the observed photon spectrum
is converted to the electron distribution.
Such correction removes apparent energy cutoffs in the mean electron distribution
above $\sim$12~keV \citep{2008SoPh..252..139K}.
Secondly, apparent evidence from hard X-ray emission for a low-energy cutoff
in the electron spectrum should be carefully combined with other
data sensitive to the low-energy cutoff in nonthermal electron
distribution, e.g., microwave spectra \citep{2003ApJ...586..606H}
or plasma radio emission \citep{1999SoPh..184..353M,2001SoPh..202..131K}.
Finally, effects which may lead to hard X-ray spectral flattening
\citep[e.g., albedo and non-uniform ionization; see][]{2002SoPh..210..419K,2003ApJ...595L.123K,2009ApJ...693..847L,2009ApJ...705.1584S}
should also be carefully assessed
before reaching conclusions on the value of the crucial parameter
$E_c$, with its attendant implications for the energy content in
nonthermal electrons \citep[e.g.,][]{2009A&A...500..901F}.
\index{electrons!nonthermal}\index{non-uniform ionization}

    \subsection{Temperature distribution of thermal plasma}\label{sec:7thermal} 
\index{soft X-rays!temperature distribution}

The thermal free-free (bremsstrahlung) continuum emission
(photons~s$^{-1}$~keV$^{-1}$) at photon energy $\epsilon$ from an
element of plasma of density $n$ (cm$^{-3}$), temperature $T$ (K)
and volume $dV$ is\index{bremsstrahlung!thermal}, neglecting
factors of order unity,
\begin{equation}
dI(\epsilon) = a \, {n^2 \, dV \over \epsilon \, T^{1/2}} \,
\exp(-\epsilon/kT), \label{eqn:emslie_isothermal_emission}
\end{equation}
where $k$ is Boltzmann's constant and $a$ is a constant. For an
extended source, with nonuniform density and temperature,
integration of (\ref{eqn:emslie_isothermal_emission}) over the
spatial extent of the source gives
\begin{equation}
I(\epsilon) = {a \over \epsilon }\int_0^\infty {1 \over T^{1/2}}
\, \xi(T) \, \exp(-\epsilon/kT) \, dT,
\label{eqn:emslie_dem_definition}
\end{equation}
where $\xi(T) = n^2 \, dV/dT$ (cm$^{-3}$~K$^{-1}$) is the {\it
differential emission measure}\index{emission measure!differential!continuum emission}\index{differential emission measure} at
temperature $T$ \citep[see equation (10) of][]{1976A&A....49..239C}.
As first pointed out by \citet{1974IAUS...57..395B},
equation~(\ref{eqn:emslie_dem_definition}) may be written as a
Laplace transform\index{Laplace transform!thermal
bremsstrahlung emission} with respect to the inverse temperature
variable $x = 1/kT$:
\begin{equation}
{k^{1/2} \over a} \, \epsilon \, I(\epsilon) = \int_0^\infty
e^{-\epsilon \, x}\, f(x) \, dx \equiv {\cal L}[f(x);
\epsilon],\label{eqn:emslie_dem_laplace}
\end{equation}
where $f(x) = \xi(1/kx) \, x^{-3/2}$ or, equivalently, $\xi(T) =
(kT)^{-3/2} \, f(1/kT)$.

A solution of equation~(\ref{eqn:emslie_dem_laplace}) for $f(x)$
(and hence $\xi(T)$) is formally possible for a large variety of
\citep[but, it should be noted, not {\it all\/}; see][]
{1988ApJ...331..554B} forms of $I(\epsilon)$.  Even when a
solution does exist, however, the solution of
equation~(\ref{eqn:emslie_dem_laplace}) is not a trivial task.
This integral equation is of Fredholm-type \citep{1985InvPr...1..301B}
and is highly
ill-posed\index{integral equations!ill-posed nature of}, with a
large class of solutions $\xi(T)$ corresponding to a given
$I(\epsilon)$ when (even very small) uncertainties in
$I(\epsilon)$ are taken into account.
\index{integral equations!Fredholm-type}
\index{inverse problem!ill-posedness}
Mathematically, this extreme ill-posedness arises from the very broad form of the Laplace
kernel $\exp(-\epsilon x)$. {\it Physically}, the problem exists
because of the broad range of temperatures $T$ that contribute to
the emission at a given photon energy $\epsilon$.
Unlike, for example, the bremsstrahlung photon-to-electron inversion
problem (see Section \ref{sec:7inversion}), in which only
electrons with energy $E > \epsilon$ contribute to the emission at
photon energy $\epsilon$ (and so the corresponding integral
equation is of Volterra type, and so less ill-posed), in the
electron-energy-to-temperature problem considered here
{\it all} temperatures $T$ contribute to the emission at energy
$\epsilon$ (and conversely a source at a single temperature $T$
produces emission at all photon energies).
\index{integral equations!Volterra-type}

An early study of the inversion of
equation~(\ref{eqn:emslie_dem_laplace}), using data from a
balloon-borne instrument \citep{1987ApJ...312..462L}, was carried out by
\citet{1995SoPh..156..315P}.  They deduced not only forms of
$\xi(T)$ but also (in a one-dimensional geometry) the
corresponding conductive flux $F_c(T)$ and its derivative
$dF_c/dx$.  However, later work by \citet{2006SoPh..237...61P}
examined the relation between $\xi(T)$ and the mean source
electron spectrum ${\overline F}(E)$, viz.,
\begin{equation}
{\overline F}(E) = b \, \int_0^\infty {1 \over T^{1/2}} \, \xi(T)
\, \exp(-E/kT) \, dT. \label{eqn:emslie_dem_definition2}
\end{equation}
This study utilized the Landweber method, which ensures positivity of $\xi(T)$
everywhere -- see equation~[\ref{eqn:Piana_my_eq10}] in
Section~\ref{sec:7inversion} -- and which is effective in
recovering narrow, $\delta$-function-like forms of the
differential emission measure $\xi(T)$.
It also showed that the form of $\xi(T)$ deduced from inversion of this (rigorous) equation is
much less well-determined than suggested by the earlier work of
\citet{1995SoPh..156..315P}, which involved inversion of the
(inexact) Equation~(\ref{eqn:emslie_isothermal_emission}).
\index{inversion algorithms!Landweber method}

The method was applied to three photon spectra emitted during the
flares SOL2002-08-21T01:41 (M1.4),
\index{flare (individual)!SOL2002-08-21T01:41 (M1.4)!temperature distribution}
SOL2003-11-03T09:55 (X3.9)\index{flare (individual)!SOL2003-11-03T09:55 (X3.9)!temperature distribution}, and
SOL2002-07-23T00:35 (X4.8).
\index{flare (individual)!SOL2002-07-23T00:35 (X4.8)!temperature distribution}
The photon spectra were first inverted by applying
zero-order Tikhonov regularizationto recover the form of
${\overline F}(E)$.
\index{Tikhonov regularization}
For the first two events, the recovered
$\xi(T)$ was consistent with a roughly isothermal low-temperature
plasma plus a very broad form of $\xi(T)$ at high temperatures.
However, for the SOL2002-07-23T00:35 (X4.8) event\index{flare (individual)!SOL2002-07-23T00:35 (X4.8)!temperature distribution}, the
reconstruction method produced unacceptably large residuals at low
temperatures, consistent with the fact that this same spectrum
fails to satisfy the derivative test \citep{1988ApJ...331..554B} necessary
for compatibility with a purely thermal interpretation of the event.
Observation of optically-thin lines is often used as an alternative
approach to infer the DEM of solar plasma \citep[as a review, see][]{2008uxss.book.....P}.

\section{The electron angular distribution}\label{sec:7angular}
\index{electrons!angular distribution}\index{electrons!pitch-angle distribution}


\label{polarization_intro}

Flare studies typically derive the properties of accelerated
particles in the target region from observations of the radiation
spectrum, but such radiation spectra are strongly dependent on the
angular distribution of the energetic particles.
\index{acceleration!distribution functions}
Consequently, knowledge of both the angular distribution and the energy
distribution of energetic particles, as they interact in the solar
material, is necessary for understanding the acceleration and
transport of particles in flaring regions.

Attempts to measure the angular distribution of the accelerated
electrons rely on the fact that an anisotropic ensemble of
bremsstrahlung-producing electrons will generate a radiation field
that is both polarized and anisotropic. Efforts to measure
electron beaming have therefore concentrated on studies of the
hard X-ray continuum emission, by looking at either the
directivity or the polarization of the emitted radiation.
\index{hard X-rays!directivity}
\index{polarization}\index{hard X-rays!polarization}
Theoretical studies have considered the evolution of the electron pitch-angle
distribution as the particles are transported along magnetic field lines \citep[e.g.,][]{1972SoPh...26..441B,1981ApJ...251..781L,1990ApJ...359..524M,1990ApJ...359..541M}.
It is generally expected that, even if the particles are strongly beamed
when injected, the net effect of the particle transport through the
atmosphere will be to broaden the angular
distribution \citep{chapter3}.
\index{pitch-angle scattering}
Ideally, the measurements of the electron
angular distribution should therefore be
performed as a function of both time and space.
No such measurements are available as yet.
However, those measurements that are available (as discussed below)
already indicate some evidence for electron beaming\index{electrons!beams!evidence for}.

The angular distribution of accelerated ions can be studied by
measuring the energies and widths of broad $\gamma$-ray lines
\citep[for details, see][]{2003ApJ...595L..81S}.
Studies of $\gamma$-ray line data from the \textit{Solar Maximum Mission (SMM)} Gamma Ray Spectrometer (GRS) suggest that protons and $\alpha$-particles are
likely being accelerated in a rather broad angular distribution
\citep{1997ApJ...485..409S,2002ApJ...573..464S}.
\index{satellites!SMM@\textit{SMM}!GRS}
It is currently still unclear whether electrons and ions are being
accelerated \citep{chapter8} in a similar fashion
or by the same process.

\subsection{Early results}

One technique for studying hard X-ray directivity\index{hard X-rays!directivity} is to look for
center-to-limb variations\index{center-to-limb variation} on a statistical basis.
 Correlations
between flare longitude and flare intensity or spectrum reflect
the anisotropy\index{hard X-rays!anisotropy}
of the X-ray emission and hence an associated
anisotropy of the energetic electrons.
\index{electrons!anisotropy}
\index{anisotropy!electron}
\index{anisotropy!X-ray}
For example, if the
radiation is preferably emitted in a direction parallel to the
surface of the Sun, then a flare located near the limb will look
brighter than the same flare near the disk center. Various
statistical studies of X-ray flares at energies below 300~keV
reported no significant center-to-limb variation of the observed
intensity
\citep[e.g.,][]{1974SoPh...39..155D,1974SoPh...35..431P}.  A
statistically significant center-to-limb variation\index{center-to-limb variation}
in the shape of the spectra of these events {\it was} found by
\citet{1975SoPh...40..165R}, suggesting that perhaps some
directivity may be present.

The large sample of flares detected at energies greater than
300~keV by \textit{SMM}/GRS allowed,
for the first time, a statistical search for
directivity at higher energies.
\index{hard X-rays!directivity}
Analysis of these data collected
during Solar Cycle~21 provided the first clear evidence for
directed emission, with a tendency for the high energy events to
be located near the limb
\citep{1987ApJ...322.1010V,1988ApJ...334.1049B}.
\index{satellites!Venera-13@\textit{Venera-13}}
\citet{1985SvAL...11..322B} and \citet{1991ApJ...379..381M}
also reported evidence for anisotropies at hard X-ray energies.
Observations from \textit{SMM}/GRS
during Solar Cycle~22 provided further support for directivity
\citep{1991ICRC....3...69V} at energies above 300~keV.
However, several high-energy events
were also observed near the disk center by a number of different
experiments during Cycle~22 (e.g., on \textit{GRANAT} and
the \textit{Compton Gamma-Ray Observatory, CGRO)}.
\citep[For a summary, see][]{1994ApJS...90..611V}; this perhaps suggests a
more complex pattern.  \citet{1995ApJ...443..855L} used data from
SMM to confirm the general results of \citet{1987ApJ...322.1010V}
and concluded that there was evidence for increasing directivity
with increasing energy.
Quantifying the magnitude of the
directivity\index{hard X-rays!directivity} from these statistical measurements is difficult.  For
example, one needs to know the size-frequency\index{frequency!flare occurrence} distribution for
flares as well as the form of the electron distribution to derive
the predicted limb fraction \citep[e.g.,][]{1985ApJ...299..987P}.
Furthermore, the results only represent an average for the flare
sample.  Different flaring regions are not likely to have
identical geometries; nor are individual flares likely to have
time-independent electron distributions.

Another, more direct, method for studying directivity\index{hard X-rays!directivity} in
individual flares is the stereoscopic technique
\citep{1973ApJ...185..335C}.\index{stereoscopic observations!HXRs}
This method compares simultaneous
observations made on two spacecraft that view the same flare from
different directions. \citet{1998ApJ...500.1003K} combined
observations from the \textit{Pioneer Venus Orbiter (PVO)} and \textit{ISEE-3} satellites
\index{satellites!ISEE-3@\textit{ISEE-3}}
\index{satellites!PVO@\textit{PVO}} to produce
stereoscopic flare observations of 39 flares in the energy range
from 100~keV to 1~MeV.  While the range of flux ratios measured by
\citet{1998ApJ...500.1003K} is consistent with the results of
statistical studies \citep{1987ApJ...322.1010V}, the deviations of
the ratio from unity show no clear correlation with increasing
difference in viewing angles.  Later studies concluded that there
was no clear evidence for directivity at hard X-ray energies
\citep{1998ApJ...500.1003K,1994ApJ...426..758L}. Stereoscopic
observations tend to suffer from cross-calibration issues between
different instruments.

The high quality hard X-ray data from {\em RHESSI} have opened new
new opportunities and diagnostic techniques for the study of electron
anisotropy\index{electrons!anisotropy}\index{anisotropy!electron} in solar flares.

    \subsection{Anisotropy of X-ray bremsstrahlung emission} 



\begin{figure}    
\begin{center}
\includegraphics[width=0.6\textwidth]{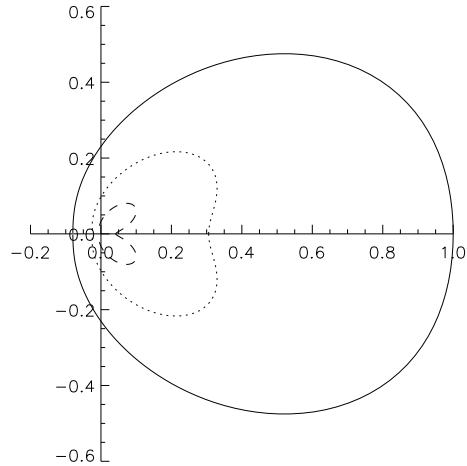}
\end{center}
\caption{Polar diagram of the bremsstrahlung cross-section \citep{1953PhRv...90.1026G}
for an electron of energy $E=100$~keV and photon energies
$\epsilon=30$~keV (solid line), $\epsilon=50$~keV (dotted line),
and $\epsilon=80$~keV (dashed line); the radial coordinate is
proportional to the size of the cross-section and the angle from
the $x$-axis corresponds to the angle between the incoming
electron direction and the line to the observer. Note that at
energies $E \gg \epsilon$, the cross-section peaks at $\theta = 0$,
while for $E \simeq \epsilon$, the cross-section peaks at $\theta
\simeq (30-40)^\circ$ \citep[after][]{2004ApJ...613.1233M}.}
\label{fig:Massone_fig1}
\end{figure}

Using Equation~(\ref{eqn:intro2}), it is straightforward to show
that the emitted bremsstrahlung in the direction ${\bf \Omega}$
toward the observer can be written
\begin{equation}
I(\epsilon,{\bf \Omega}) = {\bar{n}V \over 4\pi R^2}
\int_\epsilon^\infty \int_{\Omega'} {\overline F}(E,{\bf \Omega'})
Q(\epsilon, E,\theta') \, \mbox{d}E \, \mbox{d}{\bf\Omega'},
\label{eqn:bremss_anisotropic}
\end{equation}
where $Q(\epsilon, E,\theta')$ is the cross-section differential
in photon energy $\epsilon$ and the angle
$\theta'$ between the precollision electron direction ${\bf
\Omega}'$ and the emitted photon direction ${\bf \Omega}$, summed
over the polarization state of the emitted photon
\citep{1953PhRv...90.1026G}. The bremsstrahlung cross-section is
more angle-dependent for higher photon energies
(Figure~\ref{fig:Massone_fig1}), which is qualitatively
consistent with the findings of the statistical analysis.

In Equation~(\ref{eqn:bremss_anisotropic}) the mean source
electron flux spectrum has been generalized to the form
${\overline F}(E,{\bf \Omega'})$, which takes into account the
angular distribution of electrons. If we
denote by $(\theta, \phi)$ the polar coordinates of the electron
velocity vector ${\bf \Omega}'$ relative to the mean direction of
the electron velocity distribution (usually the same direction as
the guiding magnetic field line), and assume azimuthal symmetry,
i.e., no dependence on $\phi$, so that ${\overline F}(E,{\bf \Omega}') \equiv
f(E,\theta)$, and the angle of the mean direction
of the electron velocity distribution is $\theta _0$,
Equation (\ref{eqn:bremss_anisotropic}) can be written:
\begin{equation}\label{eqn:Massone_my_eq3}
I(\epsilon) = \frac{\bar{n}V}{4\pi R^2} \int_{E=\epsilon}^\infty
\int_{\theta=0}^\pi \int_{\phi=0}^{2\pi}
Q\left[\epsilon,E,\theta'(\theta,\phi;\theta_0)\right] f(E,\theta)
\sin\theta \, d\phi \, d\theta \, dE \quad,
\end{equation}
where the directivity angle $\theta'$ between electron and photon
directions is given by $\cos\theta'= \cos \theta \cos\theta_0 +
\sin\theta \sin\theta_0 \cos\phi$. If we now define:
\begin{equation}\label{eqn:Massone_my_eq4}
K_0\left(\epsilon,E,\theta\right)=\int_{\phi=0}^{2\pi}
Q\left[\epsilon,E,\theta'(\theta,\phi;\theta_0)\right] d\phi
\quad,
\end{equation}
we can write
\begin{equation}\label{eqn:Massone_my_eq5}
I(\epsilon) =\frac{\bar{n}V}{4\pi R^2}\int_{E=\epsilon}^\infty
\int_{\theta=0}^\pi  K_0\left(\epsilon,E,\theta\right) f(E,\theta)
\sin\theta  \, d\theta \, dE \quad.
\end{equation}
Inversion of Equation (\ref{eqn:Massone_my_eq5}) requires
construction of the bivariate function $f(E,\theta)$ from
knowledge of the (noisy) univariate function $I(\epsilon)$.
The problem is significantly more tractable if we make the further
assumption that $f(E,\theta)$ is separable in $E$ and $\theta$. We
can then choose a particular form for the angular dependence of
$f(E,\theta)$ and reconstruct the part of $f(E,\theta)$ that
depends only on the electron energy (univariate problem in $E$)
or, analogously, we can assume the $E$-dependence for
$f(E,\theta)$ and recover the $\theta$-dependence (univariate
problem in $\theta$).


In \cite{2004ApJ...613.1233M}, the $\theta$-dependence of
$f(E,\theta)$ is a prescribed (normalized) form
$g(\theta)/\int_{\bf \Omega'}g(\theta) \, d{\bf \Omega}^\prime$.
Specifically, it was assumed that at all energies the
pre-collision electron velocities are uniformly distributed over a
solid angle within a cone of half-angle $\alpha$ centered on a
direction corresponding to a photon emission direction of
$\theta_0$. With such an assumption, Equation
(\ref{eqn:Massone_my_eq5}) becomes
\begin{equation}\label{eqn:Massone_my_eq6}
I(\epsilon) =\frac{\bar{n}V}{4\pi R^2}\int_{E=\epsilon}^\infty
{\overline{K_0}}\left(\epsilon,E\right) \, {\overline{F}}(E) \,
dE,
\end{equation}
where
\begin{equation}\label{eqn:Massone_my_eq7}
{\overline{K_0}}\left(\epsilon,E\right)
=\frac{1}{2\pi(1-\cos\alpha)}\int_{\theta=0}^\alpha
K_0\left(\epsilon,E,\theta\right) \, \sin\theta \, d\theta.
\end{equation}
Equation (\ref{eqn:Massone_my_eq6}) is formally identical to
Equation (\ref{eqn:emslie_fbar}). Application of the usual
Tikhonov regularization\index{inverse problem!Tikhonov regularization} technique to equation
(\ref{eqn:Massone_my_eq6}) then yields the mean source energy
spectrum ${\overline{F}}(E)$.

\begin{figure}    
\begin{center}
\psfig{file=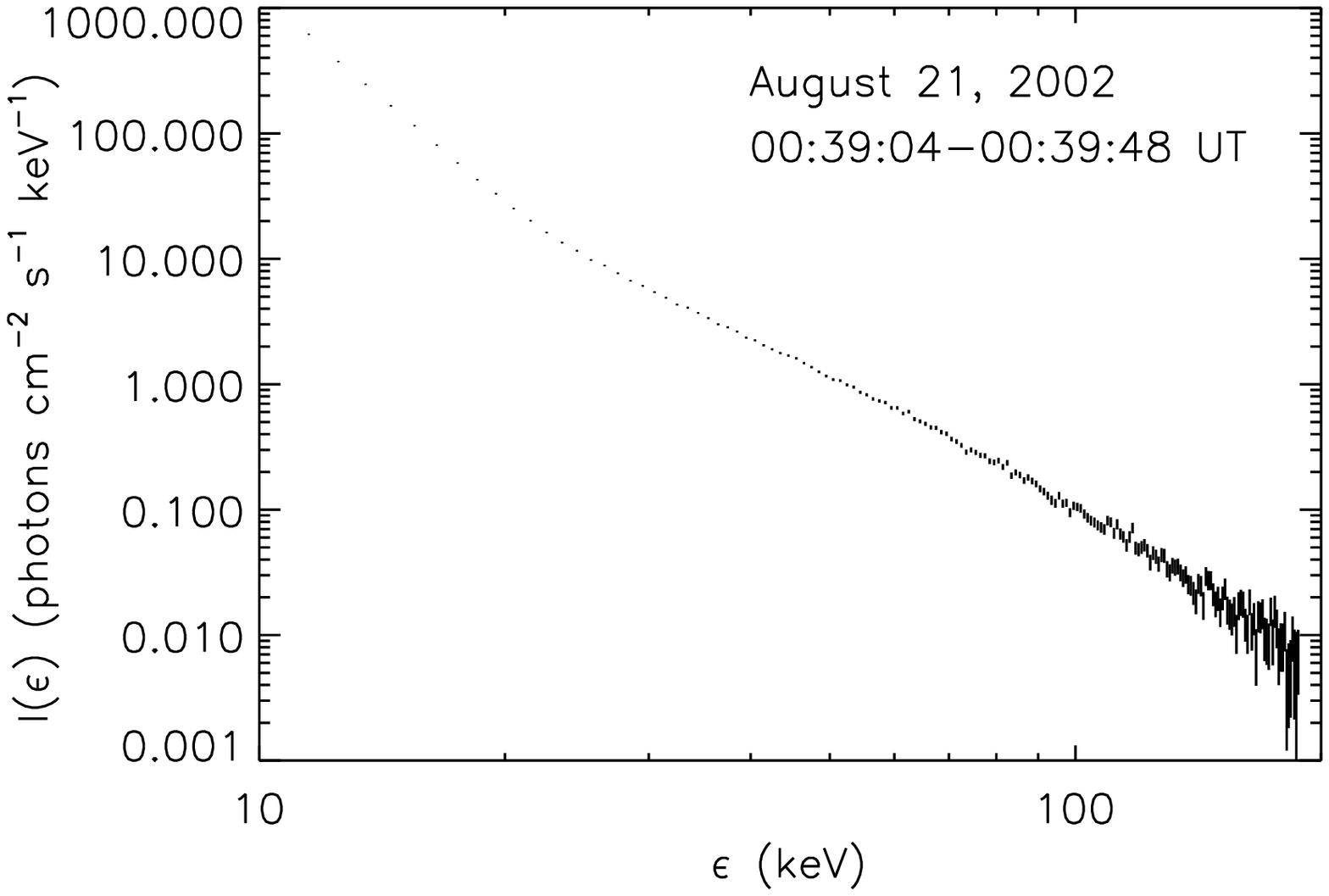,width=5.4cm,clip=}
\psfig{file=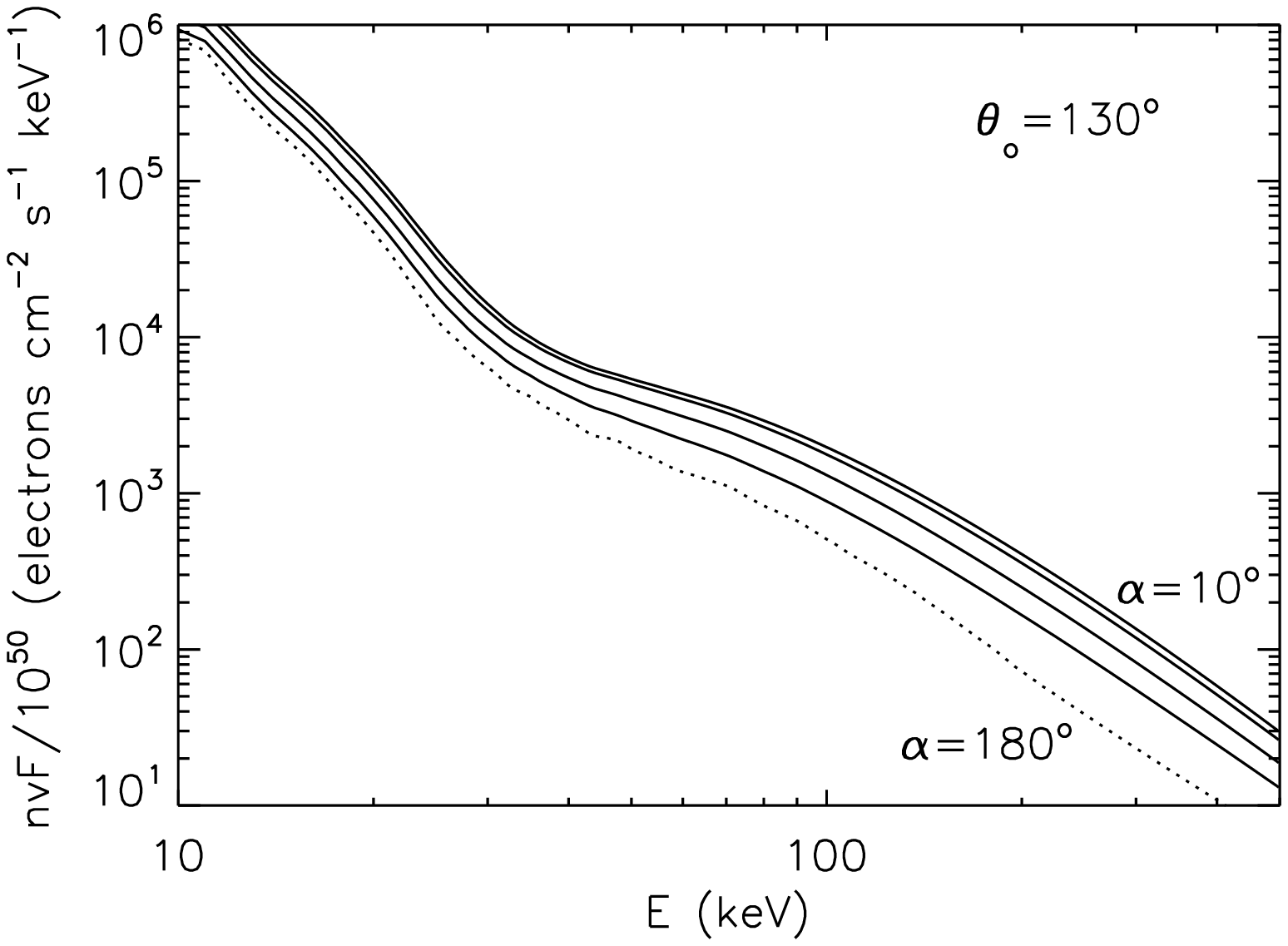,width=5.4cm,clip=}
\caption{\textit{Left:} photon flux spectrum. \textit{Right:}
regularized electron flux spectra corresponding to
$\theta_0=130^o$ (angle between an observer and the electron beam direction)
and (from top to bottom) $\alpha=10^o$, $30^o$,
$60^o$, $90^o$, $180^o$. The choice of $\theta_0=130^o$ is
justified by the fact that this is the value corresponding to a
vertically downward electron beam at the location of the selected
flare \citep[after][]{2004ApJ...613.1233M}.} \label{fig:Massone_fig2}
\index{spectrum!photons!illustration}
\end{center}
\end{figure}

\citet{2004ApJ...613.1233M} applied this analysis to a photon spectrum recorded by the
\textit{RHESSI} instrument from SOL2002-08-21T01:41 (M1.4)\index{flare (individual)!SOL2002-08-21T01:41 (M1.4)!anisotropic distribution} in the time
interval 00:39:04-00:39:48~UT (left panel of
Figure~\ref{fig:Massone_fig1}). Values for the observation angle
$\theta_0$ ranging from $0^o$ to $180^o$, and for the cone
semi-angle $\alpha$ ranging from $10^o$ to $180^o$, were
considered. The results (right panel of
Figure~\ref{fig:Massone_fig2}) demonstrated clearly that the
angular dependence cannot be neglected: use of the anisotropic
cross-section yields electron spectra that are significantly
different from the ones reconstructed by using the
solid-angle-averaged cross-section (corresponding to
$\alpha=180^o$). As the electron distribution becomes more
anisotropic (decrease in $\alpha$), the cross-section for emission
in the direction of the observer decreases for some parts of the
cone and increases for others. The overall effect is a reduction
in the cross-section, so that more electrons are required to
produce the given photon spectrum. This effect is more pronounced
at high energies (see Figure~\ref{fig:Massone_fig2}), so the
spectrum for small values of $\alpha$ is flatter than for
$\alpha=180^o$ (isotropic distribution).

As the viewing angle $\theta_0$ increases (source moves closer to
the disk center), the cross-section for emission in direction of
the observer in general decreases, especially at high values of
$E$, and accordingly the required electron spectrum increases and
also flattens.

    \subsection{Statistical results on X-ray anisotropy}\label{sec:7stat} 


\begin{figure}
\begin{center}
\includegraphics[width=60mm]{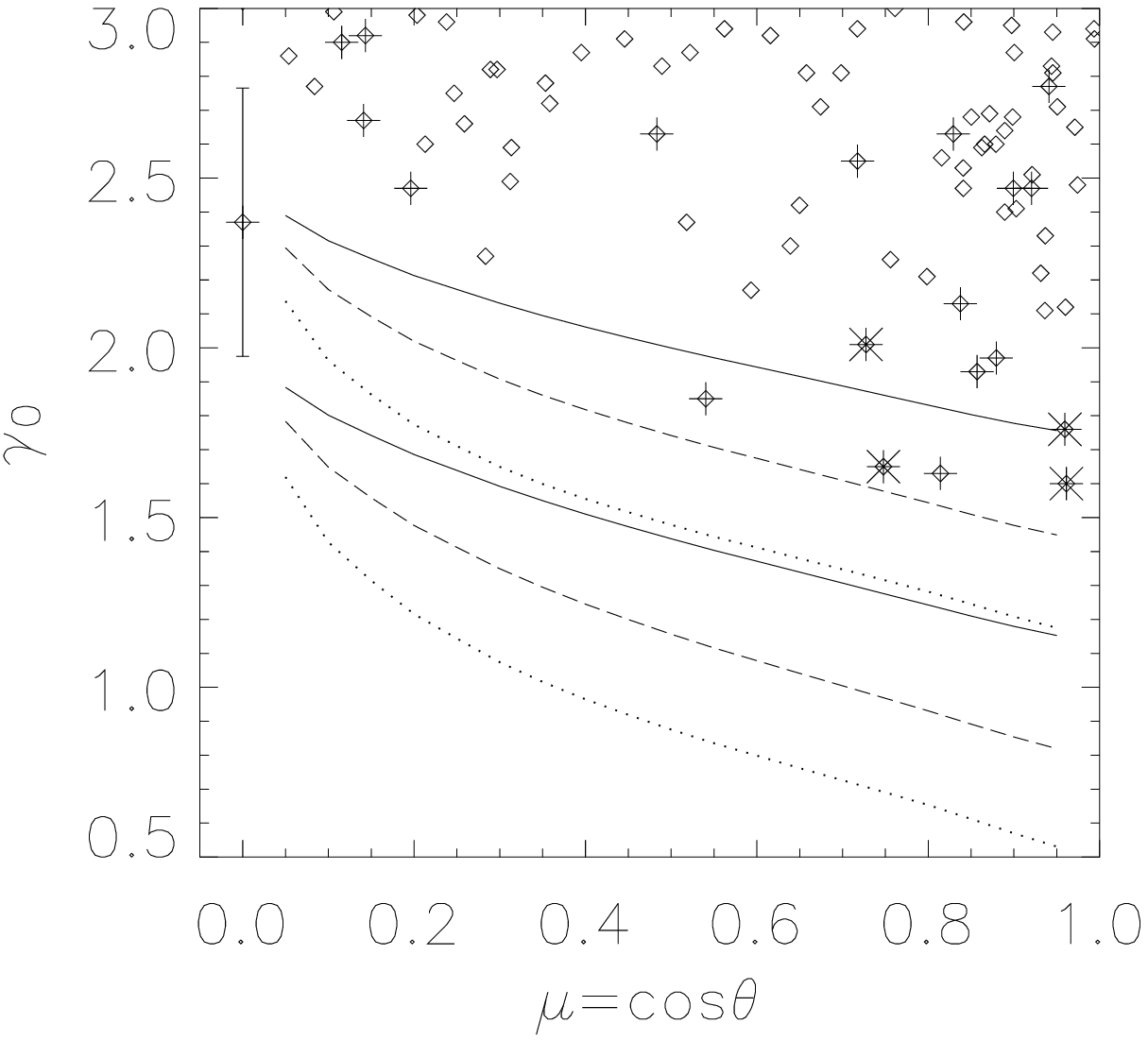}
\includegraphics[width=60mm]{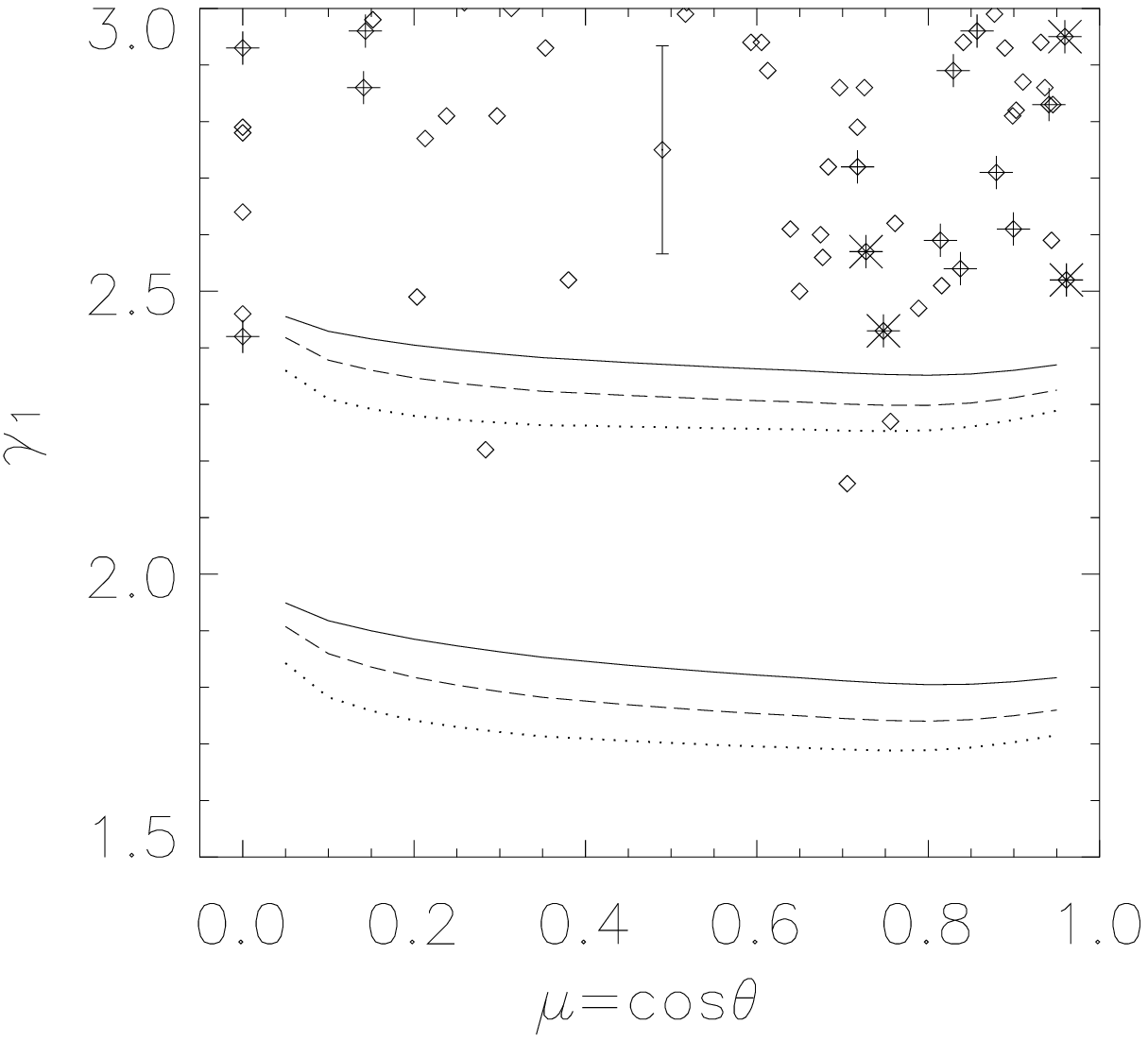}
\caption{Spectral index
\index{spectrum!photons!spectral index}
\index{center-to-limb variation!illustration}
$\gamma_0$ (15-20 keV) and $\gamma_1$ (20-35 keV) versus cosine
$\mu$ of heliocentric angle $\theta$. Vertical error bars indicate
average uncertainties on the values as determined from single
power-law fits. Lines show the predicted dependency for single
power-law primary spectra with spectral indices 2.0, 2.5 and
for directivity\index{hard X-rays!directivity} $\alpha(\mu)=1,2,4$
(solid, dashed, and dotted
lines respectively). Flares with a dip in the mean electron
distribution are denoted by stars; see also Section~\ref{sec:7cutoff}
\citep[after][]{2007A&A...466..705K}.}
\label{fig:kasparova_directivity_gamma0_mu}
\end{center}
\end{figure}

As a result of the heliocentric angle dependence of albedo
\index{albedo}\index{photospheric albedo}
(Section \ref{sec:7albedo}), the shape of photon spectra should vary as a
function of their position on the solar disk
\citep{1978ApJ...219..705B, 2006A&A...446.1157K}. A statistical
analysis of \textit{RHESSI} flares \citep[][]{2007A&A...466..705K} demonstrates a clear center-to-limb
variation\index{center-to-limb variation} of photon spectral indices in the 15~-~20 keV energy
range and a weaker dependency in the 20~-~50 keV range. The
observed spectral variations were found to be consistent with the
predictions of albedo-induced\index{albedo}\index{photospheric albedo}  spectral
index\index{spectrum!photons!spectral index} changes
(Figure~\ref{fig:kasparova_directivity_gamma0_mu}).

\begin{figure}
\begin{center}
\includegraphics[width=80mm]{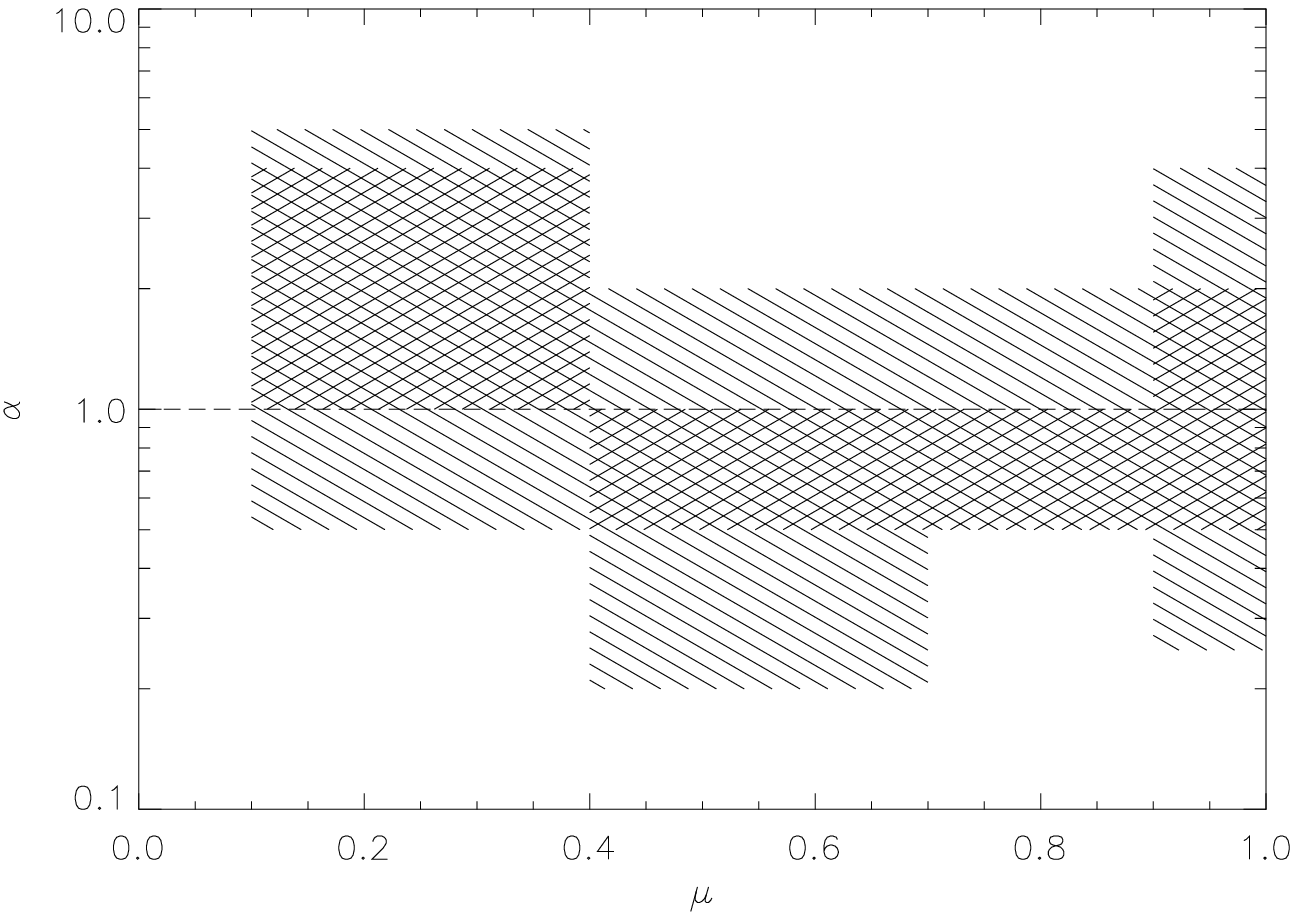}
\caption{Directivity
\index{directivity!X-rays} at different heliocentric angle $\mu$
determined from the distributions of spectral index\index{spectrum!photons!spectral index}\index{center-to-limb variation!illustration}
$\gamma_0$ in
the 15~-~20~keV energy range at the limb and a given range of
$\mu$. Cross-hatched areas show the range of directivity
\index{hard X-rays!directivity} values
that are consistent (at the 0.05 significance level) with the
hypothesis that the observed distribution at a given range of
$\mu$ and the modeled distribution are drawn from the same parent
distribution; crossed areas correspond to the 0.01 significance
level. The dashed line shows the isotropic case, i.e.,
$\alpha(\mu)=1$ \citep[after][]{2007A&A...466..705K}. }
\label{fig:kasparova_directivity_aniso_15_20}
\end{center}
\end{figure}

Because the number of albedo photons  depends on the amount of
downward directed primary emission, the characteristics of the albedo
component can be used to get an estimate of the anisotropy
of the primary hard X-ray emission.
\index{hard X-rays!anisotropy}\index{anisotropy!X-ray}
\index{albedo}\index{photospheric albedo}
In this manner,
\citet{2007A&A...466..705K} obtained values for $\alpha$ (the
ratio of downward-directed primary emission to observer-directed
primary emission) in the 15-20~keV energy range;
see Figure~\ref{fig:kasparova_directivity_aniso_15_20}.
The values of
directivity $\alpha$ range from 0.3~to~3; while they imply downward-collimated
emission in some cases, overall they are also consistent with
isotropic X-ray emission \citep[see also][]{2006ApJ...653L.149K}.
The model of \citet[][Figure 4]{1983ApJ...269..715L} predicts
$\alpha\lapprox 3$ at 22~keV for disk-center events.

    \subsection{Albedo as a probe of electron angular distribution}\label{sec:7stereo} 



The albedo\index{hard X-rays!albedo}\index{albedo}\index{photospheric albedo}
 spectral ``contaminant'' in fact offers very valuable
insight into the anisotropy\index{electrons!anisotropy}\index{anisotropy!electron}
of the flare fast electron
distribution\index{hard X-rays!directivity}. It does so by providing a view of the hard X-ray
flare from behind, like a dentist's mirror.
\index{dentist's mirror}
Moreover, the solar albedo ``mirror'' is spectrally distorting, so that its
contribution to the overall spectrum can be distinguished. The
observed spectrum in the observer's direction should contain an
albedo ``bump'' feature, the strength of which is an indicator of
the degree of downward beaming of the electron distribution. By
use of this solar ``mirror'' we can achieve a degree of knowledge
about the directionality of the primary photon distribution, and
so the accelerated electron distribution, from single-spacecraft
photon spectrometry \citep{2006ApJ...653L.149K}.

The required integration over solid angle in Equation
(\ref{eqn:bremss_anisotropic}) can be approximated using a
bi-directional representation. Invoking axial symmetry about a beam
assumed to be perpendicular to the solar surface, we
introduce the cross-sections
\begin{eqnarray}\label{kontar:angle_av}
Q(\epsilon,E,\theta _0)=\frac{1}{\cos(\theta _0-\Delta
\theta)-\cos(\theta _0+\Delta \theta)}\int _{\theta _0-\Delta
\theta}^{\theta _0+\Delta \theta}Q(\epsilon, E, \theta
^{\prime})\sin(\theta ^{\prime})\, d \theta ^{\prime},
\end{eqnarray}
averaged over $[\theta _0-\Delta \theta, \theta _0+\Delta \theta]$
and centered at angle $\theta _0$. We define
$Q^F(\epsilon,E)\equiv Q(\epsilon,E,\theta _0 = \theta)$ and
$Q^B(\epsilon,E)\equiv Q(\epsilon,E,\theta _0 = 180^o-\theta )$,
where $\theta$ is the heliocentric angle of the source. The
electron spectrum ${\overline F}(E, \theta)$ is defined in a
similar bi-directional approximation, through introduction of the
quantities
\begin{eqnarray}\label{kontar:F_ud}
\overline F_{u,d} = \frac{1}{\bar n V}\int F_{u,d}(E,{\bf r}) \,
n({\bf r})dV
\end{eqnarray}
for electrons propagating either upward ($u$) or downward ($d$)
toward the scattering photosphere: ${\overline F}_u(E)$ and
${\overline F}_d(E)$ are the density-weighted volumetric mean flux
spectra of electrons directed towards the observer (upward and
downward, respectively), also averaged over $\Delta \theta$.

With these assumptions and definitions, the discretized hard X-ray
spectrum ${\bf I}(\epsilon _i); i=1...N$, accounting both for the
primary spectrum ${\bf I_o}$ and the reflected spectrum ${\bf
I_r}$, can be written \citep{2006ApJ...653L.149K}
\begin{equation}\label{IP}
{\bf I}={\bf I_o}+{\bf I_r}=\left(
    \begin{array}{cc}
{\bf Q^F}+{\bf G}(\mu){\bf Q^B}\;\;\;\; {\bf Q^B}+{\bf G}(\mu){\bf
Q^F}
\end{array}
\right) \left(
\begin{array}{c}
{\bf {\bar F}_u }  \\
{\bf {\bar F}_d }
\end{array}
\right),
\end{equation}
where ${\bf {\bar F}_{u,d}}(E_j); j=1...M$ are the electron data
vectors for downward- and upward-directed electrons, respectively
and ${\bf Q}^{B,F}$ are matrix representations of the kernels of
integral equations (\ref{kontar:angle_av}). The Green's matrices
${\bf G}(\mu)$ depend on the heliocentric angle of the source $\mu =
\cos \theta$ and have been calculated in
\citet{2006A&A...446.1157K}. \citet{2006ApJ...653L.149K} have
solved the inverse problem (Equation~\ref{IP}) using first-order Tikhonov regularization
method.
\index{inverse problem}
\index{inversion algorithms!Tikhonov regularization}

\begin{figure}
\begin{center}
\includegraphics[width=0.7\textwidth]{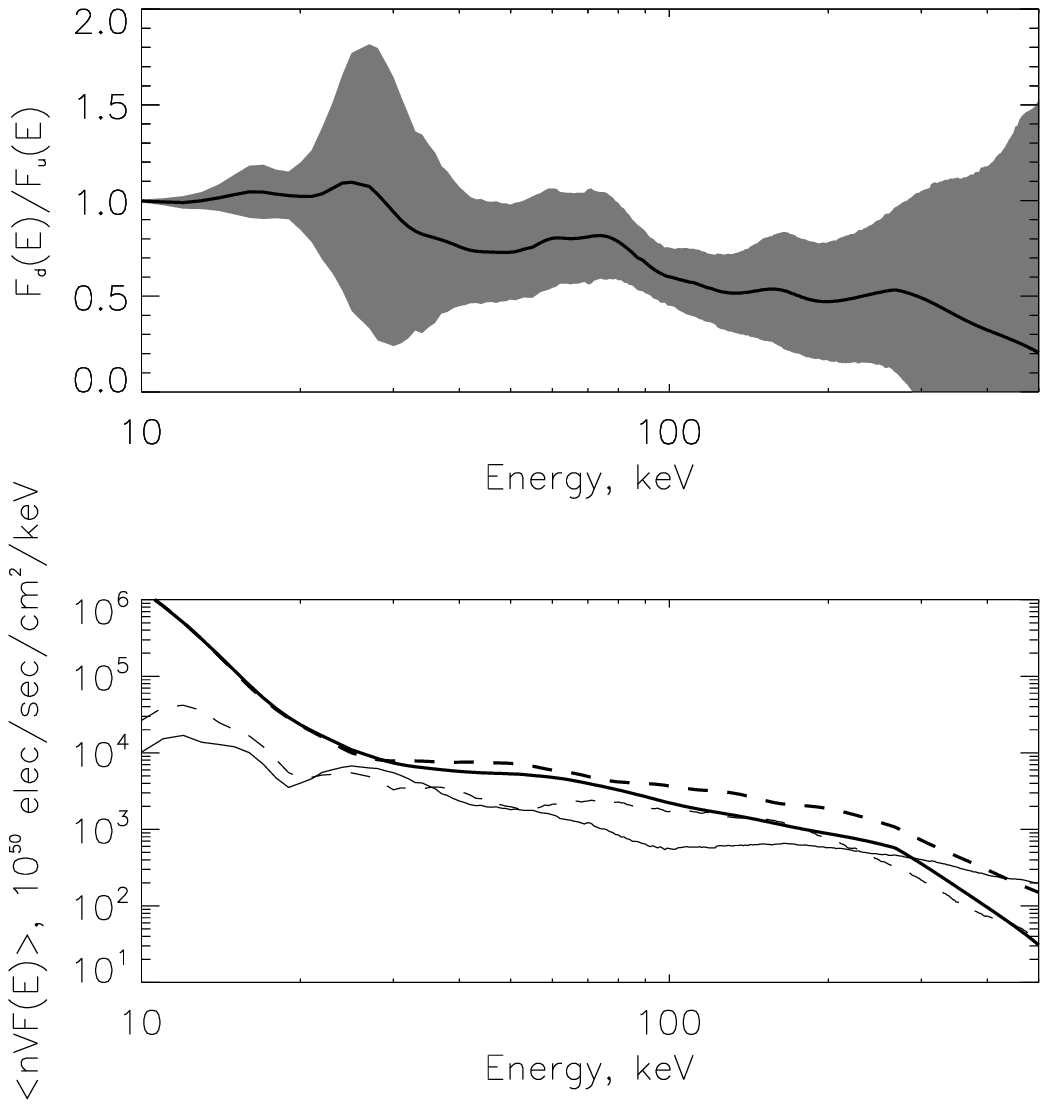}
\end{center}
\caption{{\it Lower panel:} The recovered mean electron flux
spectra defined by Equation \ref{kontar:F_ud} (thick lines) for SOL2002-08-20T08:25 (M3.4)
\index{flare (individual)!SOL2002-08-20T08:25 (M3.4)!albedo}
\index{flare (individual)!SOL2002-08-20T08:25 (M3.4)!anisotropy}
\index{flare (individual)!SOL2002-08-20T08:25 (M3.4)!illustration}
(accumulation time
interval 08:25:20-08:25:40 UT). Downward-directed ${\overline
F}_d(E)$ (solid line) and observer-directed ${\overline F}_u(E)$
(dashed line) are shown, with corresponding $2\sigma$ errors (thin lines).
{\it Upper panel:} electron anisotropy defined as ${\bar
F}_d(E)/{\overline F}_u(E)$ with confidence values within the
shaded area \citep[after][]{2006ApJ...653L.149K}.} \label{Ko1_Aug20}
\index{electrons!anisotropy}\index{anisotropy!electron}
\index{electrons!directivity}
\end{figure}

Figure \ref{Ko1_Aug20} shows electron spectral solutions (${\bar
F}_u(E), {\bar F}_d(E)$) for the M3.4 flare SOL2002-08-20T08:25.
\index{flare (individual)!SOL2002-08-20T08:25 (M3.4)!electron spectra}
\index{electrons!anisotropy}\index{anisotropy!electron}
The results are consistent with isotropy up to $E \approx 100$~keV, with some
indication of upward anisotropy above 100~keV. These findings show a near-isotropic distribution
(low electron directivity) of the mean electron
spectrum of accelerated electrons in solar flares \citep{2006ApJ...653L.149K,chapter3}.
\index{electrons!directivity}
This strongly contradicts the models with purely collisional transport in solar flares \citep[e.g.,][]{1972SoPh...26..441B,2006ApJ...651..553Z,2009A&A...506.1437K,chapter3}.
\index{flare models!collisional transport!and directivity}

    \subsection{X-ray polarization and electron angular distribution}\label{sec:7polar} 


\subsubsection{Model predictions}
\index{magnetic field!and polarization orientation}

The difficulties of statistical and stereoscopic observations for
measuring hard X-ray directivity evoke the need for a technique
that can measure anisotropies for individual flares.\index{hard X-rays!directivity}\index{polarization}\index{hard X-rays!polarization}
Because the bremsstrahlung cross-section $Q(\epsilon, E, \theta')$ is in
general polarization-dependent, polarization is a diagnostic that can, in principle, meet these requirements.
\index{flare models!thick-target!and polarization}
Models of nonthermal
(e.g., thick-target) hard X-ray production predict a clear and
significant polarization signal
with polarization levels $>$10~\% for beam-like distributions of electrons
\citep{1978ApJ...219..705B,1972SoPh...26..441B,1996A&A...308..924C,
1980ApJ...242..359E,1977ApJ...215..666L,1983ApJ...269..715L,1995A&A...304..284Z}
and an orientation parallel to the plane containing the guiding
magnetic field direction and the direction to the observer.
\index{electrons!nonthermal}
\index{thick-target model}
\index{electron beams!and polarization}
\index{polarization}\index{hard X-rays!polarization}
For vertical guiding fields, this orientation direction projects onto
a radial line on the solar disk \citep{1978ApJ...219..705B,
1983ApJ...269..715L, 1995A&A...304..284Z}; for other orientations
of the guiding field, other polarization\index{polarization}\index{hard X-rays!polarization} vector orientations are
possible \citep{2008ApJ...674..570E}. For strong
electron beaming, the polarization degree can reach values up to
60~\% at energies above 50~keV, or even higher for lower energies
\citep{1972SoPh...25..425H}.

All models predict a strong dependence between the observed value
of polarization\index{polarization}\index{hard X-rays!polarization} and the viewing angle. The highest polarization
degrees are expected for large angles of view, when the line of
sight is perpendicular to the magnetic field line.
Thus, most theories predict higher polarization
for flares located near the solar limb.
\index{polarization}
\index{hard X-rays!polarization}
\index{polarization}\index{hard X-rays!polarization}
The direction of the polarization
depends on the energy: low energy hard X-rays are negatively polarized
whereas, above 350~keV, the sign turns to positive
\citep{1978ApJ...219..705B}.
\index{collisions!and polarization}
Because collisions of beam electrons
with ambient particles tend to isotropize the distribution, the
highest polarizations (up to 85\%) are expected at energies around
100~keV generated in the coronal portions
of the top of the flare loop \citep{1983ApJ...269..715L}.
\index{coronal sources}
\index{hard X-rays!coronal sources}
Photons observed from the footpoints (in the region of the
dense chromosphere) would be polarized only to the level of around 20\%.
It should also be noted that even thermal models of the hard X-ray
source predict a finite polarization of order a few percent, due to thermal
gradients in the source \citep{1980ApJ...237.1015E}.
\index{polarization!thermal source}
\index{hard X-rays!thermal models!polarization degree}
\index{flare models!thermal!and polarization}
The thermal component, with its rather low polarization, tends to dominate the
emission from all flares at energies below about 25~keV.

All these theoretical predictions, while clearly testable, could
be criticized on the grounds that the modeling assumptions they
contain may be oversimplistic.
\index{magnetic structures!overly simplistic models}
For example, each model to date
assumes a single, simple magnetic field structure about which the
emitting electrons spiral.  It could be argued that any real
flare, particularly one sufficiently large to produce a signal of
sufficient strength to enable a polarization measurement, will in
all probability contain a mix of structures that would average out
any polarization signal present.  However, hard X-ray imaging
observations in the impulsive phase generally show a fairly simple
geometry, consisting of two footpoint sources and perhaps a
loop-top source \citep[e.g.,][]{1995PASJ...47..677M,1992PASJ...44L..83S}.
These observations suggest that simple magnetic structures are
responsible for the energetic emissions and give support to the
possibility that a statistically significant polarization signal
could be produced in a large event.

As noted in Section~\ref{sec:7albedo}, and in other places
throughout this article, a substantial fraction of the observed
hard X-ray flux is backscattered from the solar photosphere, as the
so-called {\it photosphericalbedo}.
\index{hard X-rays!albedo}\index{albedo}\index{photospheric albedo}
The precise magnitude of this backscattered fraction depends, in part,
on the polarization of the primary flux.
\index{polarization!primary flux}\index{hard X-rays!polarization}
\index{polarization}\index{hard X-rays!polarization}
Further, the reflected
component will influence the degree of polarization of the total
observed flux, since backscattering will tend to introduce
polarization fractions of a few percent at energies below 100~keV
\citep[e.g.,][]{1978ApJ...219..705B,1977ApJ...215..666L}.  Direct
imaging of this albedo patch would place a constraint on the
contribution of such backscattered photons to the primary signal.
\cite{2003SoPh..214..171H} have also suggested that Compton scattering in
the corona may lead to measurable polarization effects.
\index{polarization}\index{hard X-rays!polarization}
Clearly, a simultaneous hard X-ray imaging capability (such as that
provided by {\em RHESSI}) represents a major advantage for
interpretation of a hard X-ray polarization measurement.
\index{polarization}\index{hard X-rays!polarization}

\subsubsection{History of observations}

The history of observations of hard X-ray polarization from solar
flares is a fascinating subject in its own. The first measurements
of X-ray polarization from solar flares (at energies of $\sim$15
keV) were made by Soviet experimenters using polarimeters aboard
the {\it Intercosmos} satellites.
\index{satellites!Intercosmos@\textit{Intercosmos}}
Their polarimeters were made of a
hexagonal Be-scatterer surrounded by six counters located in front
of the Be-prism faces\index{Compton scattering!Be-scatterer}.
\index{RHESSI@\textit{RHESSI}!Be-scatterer}
Later versions of the instrument were mounted on a turnable drum
to reduce systematical errors.
In their initial study,
\cite{1970SoPh...14..204T} reported an average polarization of P = 40\% ($\pm20$\%) for
three 1969 X-ray flares:
SOL1969-10-20T10:50 (C9.0),
SOL1969-10-23T05:15,
and SOL1969-10-30T09:30 (M7.3).
\index{flare (individual)!SOL1969-10-20T10:50 (C9.0)!polarization}
\index{flare (individual)!SOL1969-10-23T05:15 (pre-\textit{GOES})!polarization}
\index{flare (individual)!SOL1969-10-30T09:30 (M7.3)!polarization}
This study was
followed by an analysis of two flares in 1970: SOL1970-10-24T05:41 (M6.4)
\index{flare (individual)!SOL1970-10-24T05:41 (M6.4)!polarization}
and SOL1970-11-05T03:21 (X2.3)\index{flare (individual)!SOL1970-11-05T03:21 (X2.3)!polarization} \citep{1972SoPh...24..429T,1972SoPh...27..426T} that showed
polarizations of approximately 20\% during the impulsive
phase.
These reports were met with considerable skepticism, on
the grounds that they did not adequately allow for proper detector
cross-calibration and had limited photon statistics
\citep{1974Natur.247..448B}.
Subsequent observations with an
instrument on the {\it OSO-7} satellite
\index{OSO-7@\textit{OSO-7}}
\index{satellites!OSO-7@\textit{OSO-7}}
seemed to confirm the existence
and magnitudes of the polarizations ($\sim$10\%), but these data
were compromised by in-flight gain shifts
\citep{1974SoPh...37..429N}.  In a later study using a polarimeter
on {\it Intercosmos} 11,\index{satellites!Intercosmos@\textit{Intercosmos}}
\index{satellites!Intercosmos@\textit{Intercosmos}}
\cite{1976SoPh...46..219T} measured
polarizations
\index{polarization}\index{hard X-rays!polarization}
of only a few percent at $\sim$15 keV for two flares in July 1974.
\index{flare (individual)!SOL1974-07-05T21:52 (pre-\textit{GOES})!X-ray polarization}
\index{flare (individual)!SOL1974-07-06T18:56 (pre-\textit{GOES})!X-ray polarization}
This small but finite polarization is consistent
with the predictions for purely thermal emission
\index{soft X-rays!and polarization} that contains an
admixture of polarized backscattered radiation
\citep{1978ApJ...219..705B}.

A decade later, a new polarimeter, designed to measure in the
energy range from 5 keV to 20 keV (with about 1.5 keV energy
resolution), was flown on the Space Shuttle \textit{Columbia}
\index{Space Shuttle \textit{Columbia}!polarimeter}
\index{satellites!Space Shuttle \textit{Columbia}}
\citep{1984ApJ...280..440T}. This design was based on metallic lithium
scattering elements surrounded by gas-proportional counters.
Contamination of the Li scattering elements invalidated the
pre-flight instrument calibration. An in-flight calibration was
performed using two flares near the center of the Sun, under the
assumption of null polarization.  The upper limits to the
polarization\index{polarization}\index{hard X-rays!polarization} derived from this calibration for 6 flares (C and
M-classes) were in the range from 2.5\% and 12.7\% (99\%
confidence).


Recent measurements at energies below 100~keV have been performed
with the SPR-N instrument on the {\it CORONAS-F} satellite
\citep{2006SoSyR..40..272Z}.
\index{satellites!CORONAS-F@\textit{CORONAS-F}}
\index{CORONAS-F@\textit{CORONAS-F}}
The SPR-N instrument included a solar
X-ray radiation monitor and a polarimeter capable of detecting
signals in the energy ranges of 20--40, 40--60 and 60--100 keV.
The polarization detector consisted of a hexahedral Be-scatterer
and three pairs of CsI(Na) scintillation detectors located on the
faces of the Be prism. With a total effective area ranging between
$\sim$0.3 cm$^2$ at 20 keV to $\sim$1.5 cm$^2$ at 100 keV, it
detected hard X-rays from more than 90 flares between 2001 and
2005. From a sample of 25 solar flares, one could determine the
upper limits of the polarization degree from 8 to 40\%
($3\sigma$).
\index{flare (individual)!SOL2003-10-29T20:49 (X10.0)!X-ray
polarization}
Only for the single case of SOL2003-10-29T20:49 (X10.0), located near Sun center, was a significant
polarization level measured: its value increased from 50\% (20--40
keV) to greater than 70\% ($60-100$~keV). Although the same flare
was observed by {\em RHESSI}, an independent polarization analysis
was not possible because {\em RHESSI} was located at
high magnetic latitude and was experiencing a high level of
charged particle events \citep{2006SoPh..239..149S}.

\subsubsection{{\it RHESSI} polarization measurements}

{\em RHESSI} enables polarization measurements in a wide range
of energies from 20 keV to 1 MeV. Its detection system of 9 cylindrical coaxial
germanium detectors\index{RHESSI@\textit{RHESSI}!germanium detectors} \citep{2002SoPh..210...33S},
coupled with a satellite rotation every 4 seconds, strongly reduces systematic effects
in the polarization measurements. However, the \textit{RHESSI} design is not optimized
for studies of polarization and its small effective area for this kind of study,
together with the high background contribution, have led to measurements
with limited statistical significance.

For the energy range between 20~keV and 100~keV, polarization\index{polarization}
\index{hard X-rays!polarization} can
be measured using the photons that are scattered into the rear
segments of the Ge detectors by a (passive) Be scattering block that
is placed between four of the germanium detectors within {\it RHESSI}'s
spectrometer array \citep{2002SoPh..210..125M}.
\index{RHESSI@\textit{RHESSI}!Be-scatterer}
At energies above
$\sim$100~keV, polarization measurements are performed using the
so-called ``coincidence mode'':
photons that Compton-scatter from
one Ge detector to another are identified by a suitable
coincidence timing window.
In this way, one detector plays the role
of an {\it active} scattering element, significantly reducing the
background level.
\index{RHESSI@\textit{RHESSI}!germanium detectors}

\citet{2003SPD....34.2205M} reported on a measurement of
polarization covering the energy range from 20~keV up to~100 keV
performed with the Be~scattering block.
\index{RHESSI@\textit{RHESSI}!Be-scatterer}
\index{flare (individual)!SOL2002-07-23T00:35 (X4.8)!X-ray polarization}
The initial results from an analysis of SOL2002-07-23T00:35 (X4.8) showed evidence
for polarization at a level of 15$\pm4$\%.   The measured
polarization angle (as measured counterclockwise from solar west)
was $\Psi$ = 79$^{o}$ ($\pm5^{o}$), implying that the polarization
vector is inclined $\sim$64$^{o}$ with respect to the radial
direction at the flare site.
\index{flare models!and polarization}
\index{flare models!electron beam}
\index{precipitation}
\index{beams}
While the magnitude of the observed
polarization is broadly consistent with the prediction of solar
flare models which invoke the precipitation of a nonthermal
electron beam
\index{electrons!nonthermal} into a dense chromospheric target,
the orientation of the polarization vector is somewhat surprising.
\citet{2008ApJ...674..570E} suggest that the orientation of the
polarization vector in this case could be explained by a tilt of
the flaring loop with respect to the local vertical.  Such a tilt
is also consistent with gamma-ray line\index{gamma-rays!nuclear line radiation} observations for this flare
\citep{2003ApJ...595L..81S}. Unfortunately, recent analysis of
these data \citep{2007AAS...210.9301M} suggests that at least a
part of the reported signal may be a result of systematic effects
in the data analysis, underscoring the difficulty of making
reliable polarization measurements\index{polarization}\index{hard X-rays!polarization}.

\begin{figure}      
\centering
\includegraphics[width=0.7\textwidth]{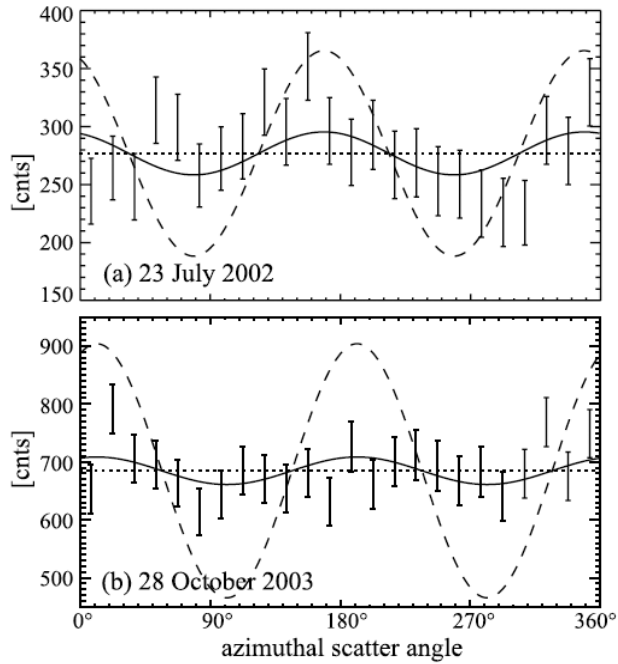}
\caption{Solar flare polarization data from
\cite{2006ApJ...638.1129B} for photon energies between 200~keV and
1~MeV.  The best fit curve is shown as a solid line.  The case for
100\% polarization is shown as a dotted line.  The measured
polarization for SOL2002-07-23T00:35 (X4.8) is 21\% $\pm$ 9\%. The
measured polarization for SOL2003-10-28T11:10 (X17.2) is -11\% $\pm$
5\%.
\index{flare (individual)!SOL2002-07-23T00:35 (X4.8)!X-ray polarization}
\index{flare (individual)!SOL2003-10-28T11:10 (X17.2)!X-ray polarization}}
\label{fig:mcconnell_fig1}
\end{figure}

Using the \textit{RHESSI} coincidence mode, two results have been recently
published for energies above 100~keV.
The first one (see Figure~\ref{fig:mcconnell_fig1}) describes the polarization measurements of two X-class
solar flares, one located close to the solar limb and the other close to the center
of the disk \citep{2006ApJ...638.1129B}.
The energy range selected was between 200~keV and 1~MeV.
The polarization degrees found were 21\% $\pm$
10\% and $-11$\% $\pm$ 5\% in 1$\sigma$ respectively.
Assuming that the measured polarization was significant in each case,
they measured a radial polarization direction for the flare close to the solar center,
and an azimuthal direction for the flare near the limb.
\index{polarization}\index{hard X-rays!polarization}
The levels of polarization,
as well as their directions, are consistent with beamed electron distribution
models \citep{1978ApJ...219..705B} and contrary to the results
using the albedo method \citep{2006ApJ...653L.149K,2007A&A...466..705K}.
However, the level of inconsistency with more isotropic distributions is rather weak.
Given that a report of GRB (gamma-ray burst) polarization by this same team has been met with
considerable skepticism in the literature, one might also question these flare results.
It should be noted, however, that the flare analysis differed significantly from the GRB analysis
and has not been contested in the literature.
\index{gamma-rays!GRB!polarization}

\begin{figure}      
\centering
\includegraphics[width=0.9\textwidth]{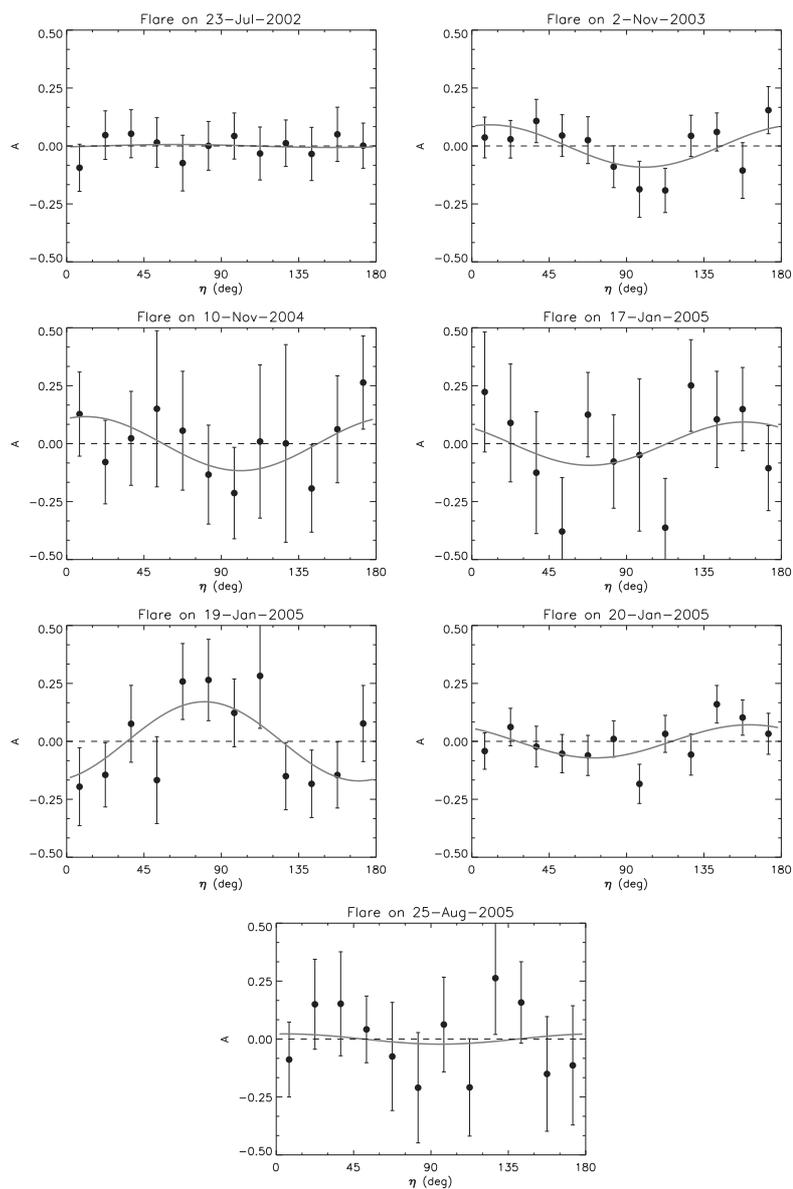}
\caption{Solar flare polarization data from
\cite{2006SoPh..239..149S} for photon energies between 100 and 350
keV.  The best fit curve is shown as a solid line.  The best case
for polarization is from SOL2005-01-19T08:22 (X1.3), where a
polarization value of 54\%$\pm$21\% was measured.
\index{flare (individual)!SOL2005-01-19T08:22 (X1.3)!X-ray polarization}}
\index{flare (individual)!SOL2005-01-19T08:22 (X1.3)!illustration}
\label{fig:mcconnell_fig2}
\end{figure}

\begin{figure}          
\begin{center}
\includegraphics[width=0.6\textwidth]{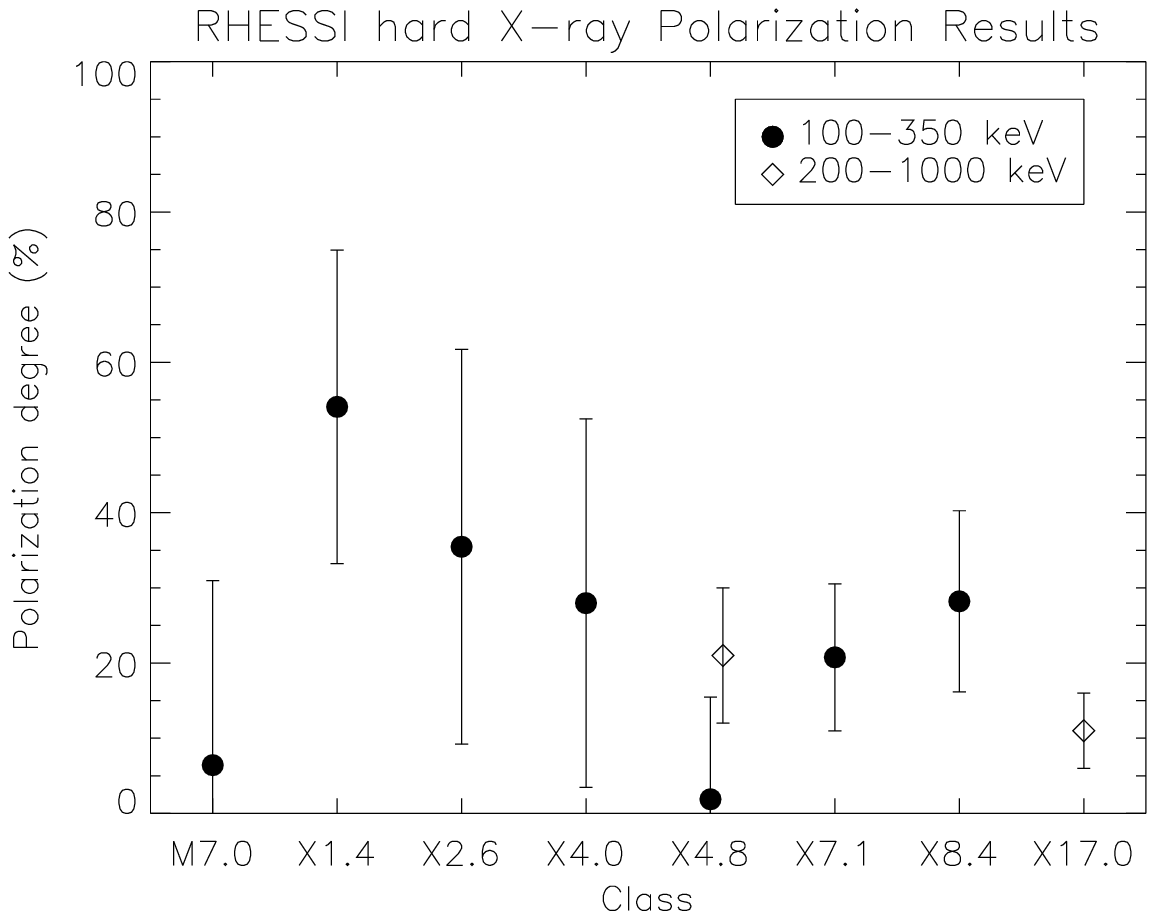}
\end{center}
\caption{Results on the absolute value of the degree of hard X-ray polarization
above 100 keV achieved with \textit{RHESSI} ($1\sigma$ errors). The diamonds and the filled
circles correspond to two independent measurements (see text for references).
From low to high flare class (left to right) the flares represented are: SOL2005-08-25T04:40 (M6.4), SOL2005-01-19T08:22 (X1.3), SOL2004-11-10T02:13 (X2.5), SOL2005-01-17T09:52 (X3.8),
SOL2002-07-23T00:35 (X4.8) (two measurements), SOL2005-01-20T07:01 (X7.1),
SOL2003-11-02T17:25 (X8.3), and SOL2003-10-28T11:10 (X17.2).
None of these results are significant at a level of greater than $3\sigma$,
suggesting that higher quality data are still needed.
\index{flare (individual)!SOL2002-07-23T00:35 (X4.8)!X-ray polarization}
\index{flare (individual)!SOL2002-07-23T00:35 (X4.8)!illustration}
\index{flare (individual)!SOL2003-10-28T11:10 (X17.2)!X-ray polarization}
\index{flare (individual)!SOL2003-10-28T11:10 (X17.2)!illustration}
\index{flare (individual)!SOL2003-11-02T17:25 (X8.3)!X-ray polarization}
\index{flare (individual)!SOL2003-11-02T17:25 (X8.3)!illustration}
\index{flare (individual)!SOL2005-01-20T07:01 (X7.1)!X-ray polarization}
\index{flare (individual)!SOL2005-01-20T07:01 (X7.1)!illustration}
\index{flare (individual)!SOL2005-01-17T09:52 (X3.8)!X-ray polarization}
\index{flare (individual)!SOL2005-01-17T09:52 (X3.8)!illustration}
\index{flare (individual)!SOL2005-01-19T08:22 (X1.3)!X-ray polarization}
\index{flare (individual)!SOL2005-01-19T08:22 (X1.3)!illustration}
\index{flare (individual)!SOL2005-08-25T04:40 (M6.4)!X-ray polarization}
\index{flare (individual)!SOL2005-08-25T04:40 (M6.4)!illustration}
\index{flare (individual)!SOL2004-11-10T02:13 (X2.5)!X-ray polarization}}
\index{flare (individual)!SOL2004-11-10T02:13 (X2.5)!illustration}
\label{fig:suarez_fig1}
\end{figure}

The second study (see Figure~\ref{fig:mcconnell_fig2}) applies the coincidence
method to the impulsive phase of seven solar flares in the 100~keV to 350~keV
energy range \citep{2006SoPh..239..149S}. The flare sample consisted
of six X-class (from X1.4 to X8.4) flares and one M7.0-class flare located
either on the limb or in the outer part of the solar disc.
Values for the polarization degree were 2\% and 54\%.
The lowest degree of polarization obtained was found to be fully compatible
with a 0\% polarization measurement. The highest degree of polarization,
which was the most statistically significant of the seven results obtained,
was found to be 2.6 sigma away from 0. The angles of polarization $\Psi$ were
distributed between $35^\circ$ and $85^\circ$ independent of the flare location,
contrary to both the results mentioned in \citet{2006ApJ...638.1129B},
and to the expectation based on simple geometrical modeling. Additional attempts
to correlate various parameters (e.g., polarization level, polarization angle,
heliocentric angle, footpoint orientation, flare intensity) were also inconclusive.
The results were compared with the theoretical predictions from \cite{1978ApJ...219..705B}
and \cite{1983ApJ...269..715L} and with the 0\% polarization hypothesis.
The $\chi^2$ analysis allowed only for rejection (90\% of confidence) of one of the models
from \cite{1983ApJ...269..715L}. In this model, predicting very high polarization
values up to 85\%, the magnetic field strength is constant along the loop
and the electrons spiral at pitch angles close to $90^\circ$.
Due to the statistical uncertainties, for the rest of the models the  $\chi^2$
values were very close to unity, making it impossible to distinguish between them.
The polarization amplitudes measured by both
\citet{2006ApJ...638.1129B} and \citet{2006SoPh..239..149S} are
combined in Figure~\ref{fig:suarez_fig1} with diamonds and filled circles,
respectively.

The results to date do not yet provide unambiguous evidence for solar flare
polarization at hard X-ray energies. It is clear, however, that such data would
constitute a unique probe into the electron acceleration process in solar flares,
providing new constraints on theoretical models and allowing for more detailed
studies of the acceleration processes and geometries.

\section{The electron spatial distribution}\label{sec:7spatial}
\index{imaging spectroscopy}

Hard X-ray imaging spectroscopy
is a powerful tool with which to explore the underlying physics of
particle acceleration and transport in solar flares, and has been
a central component of the {\em RHESSI} concept since its
beginnings.
\index{RHESSI@\textit{RHESSI}!imaging spectroscopy}\index{imaging spectroscopy}
\index{hard X-rays!imaging spectroscopy}

\subsection{Early results}

In its most basic form, imaging spectroscopy involves simply
constructing and comparing two-dimensional (count) maps of the source for
different energy bands. The earliest imaging of solar hard
X-ray sources was carried out using the Hard X-Ray Imaging
Spectrometer (HXIS) \citep{1980SoPh...65...39V} on (\textit{SMM}).
\index{satellites!SMM@\textit{SMM}!HXIS}
This instrument provided imaging information between $3.5$ and $30$~ keV though
the use of an array of subcollimator-defining grids;
it had an angular resolution of
$8$~arcsec over a $160$~arcsec field of view, and $32$~arcsec over
a wider, $444$~arcsec field. \citet{1981ApJ...244L.153H} showed
that the hard X-ray emission in SOL1980-04-07T01:07 (M4)\index{flare (individual)!SOL1980-04-07T01:07 (M4)!imaging spectroscopy}
was located ``in two patches;'' the patch with the harder spectrum
``coincided with the brightest H$\alpha$ emission.''
\index{flare (individual)!SOL1980-04-10T09:23 (M4)!imaging spectroscopy}
\index{flare models!thick-target}
On the other hand, in SOL1980-04-10T09:23 (M4),
the hard X-ray emission was ``concentrated in a looplike structure, with the
softer spectrum at the top of the loop and the harder spectrum in
the legs, thus indicating preference for the thick-target model of hard X-ray production.''
\citet{1981ApJ...246L.155H} studied the
hard X-ray emission from a large two-ribbon\index{flare types!two-ribbon}\index{ribbons!H$\alpha$} flare SOL1980-05-21T20:50 (X1)
\index{flare (individual)!SOL1980-05-21T20:50 (X1)!imaging spectroscopy} and
concluded that the higher-energy (16-30~keV) emission
originated in ``separate locations of $\sim$8~arcsec width,
coinciding in position with H$\alpha$ flare kernels,'' while the
softer ($3.5-8$~keV) emission originated from a ``broader region
in between.''\index{flare kernels!H$\alpha$}

Treating these observations as evidence for a hot
coronal\index{coronal sources}\index{hard X-rays!coronal sources}
region \citep[][]{2008A&ARv.tmp....8K,2009A&A...502..665T} produced by the primary energy
release process, and electron
precipitation from this primary acceleration region into the dense
chromosphere, \citet{1982SoPh...79...85M} studied the energetics
of SOL1980-04-10T09:23 (M4)
and concluded that ``only a fraction of the
total flare energy'' was present in the accelerated electrons.\index{flare (individual)!SOL1980-04-10T09:23 (M4)!imaging spectroscopy}\index{flare (individual)!SOL1980-04-10T09:23 (M4)!coronal sources}\index{flare (individual)!SOL1980-05-21T20:50 (X1)!imaging spectroscopy}\index{flare (individual)!SOL1980-11-15T15:53 (X1)!imaging spectroscopy}
\citet{1982SoPh...81..137D} continued the study of this event, in
addition to SOL1980-05-21T20:50 (X1)
and SOL1980-11-15T15:53 (X1).
By comparing the energy in the accelerated electrons with an
estimate of the thermal conductive flux out of the hot coronal
acceleration region, they concluded that ``a large fraction of the
dissipated flare power has to go into electron acceleration.''

The Solar X-ray Telescope (SXT) instrument
\citep{1983ApJ...270L..83T} on the {\em Hinotori} satellite
\index{satellites!Hinotori@\textit{Hinotori}/SXT}
\index{Hinotori@\textit{Hinotori}!Solar X-ray Telescope (SXT)} used the rotating modulation
collimator (RMC)\index{Hinotori@\textit{Hinotori}!rotating modulation collimator}
technique with two pairs of rotating grids to make hard X-ray
images \citep{1976SSI.....2..141O}; the FWHM angular resolution was
$\sim$30~arcsec.
\index{flare (individual)!SOL1981-05-13T06:10 (X1.5)!imaging spectroscopy}
\citet{1983SoPh...86..313T} and
\citet{1984ApJ...280..887T} report observations of a near-limb
flare, SOL1981-05-13T06:10 (X1.5),
which exhibited a diffuse hard X-ray
source situated some 40,000~km above the photosphere.
\citet{1986SoPh..107..109T} reported that a hard X-ray ($20-4$0~keV range)
 image for the impulsive component of SOL1982-02-22T04:44 (M2.7)
was an extended source elongated along the solar
limb with a source height of 7,000~km.
\index{flare (individual)!SOL1982-02-22T04:44 (M2.7)!imaging spectroscopy}
\index{flare (individual)!SOL1981-07-20T14:41 (M5.5)!imaging spectroscopy}
\citet{1983SoPh...86..313T} also report imaging hard X-ray
observations of SOL1981-07-20T14:41 (M5.5), for which ``each hard X-ray source
in the initial phase coincides with each
H$\alpha$ bright region,'' further evidence for a thick-target
interpretation of the hard X-ray emission at such energies.

In contrast to the RMC  technique
used by {\em Hinotori} and {\em RHESSI}, the Hard X-ray Telescope
(HXT) instrument on \textit{Yohkoh} \citep{1991SoPh..136...17K} used
information from 64 pairs of occultation grids; each pair of
grids yields a single Fourier component of the source.
\index{rotating modulation collimator}
\index{Yohkoh@\textit{Yohkoh}!HXT}
\index{satellites!Yohkoh@\textit{Yohkoh}}
The FWHM resolution achievable was $\sim$8~arcsec.
\citet{1992PASJ...44L..89M}, using observations of about a hundred
flares observed with the {\it Yohkoh}/HXT,
reported that, on average,
the hard X-ray source height ``decreased with increasing X-ray
energy,'' consistent with the deeper penetration of higher-energy
electrons
\citep[e.g.,][]{1975SoPh...41..135B,1978ApJ...224..241E,1981ApJ...245..711E}.
With nine RMC components, an angular resolution down to $\sim$2~arcsec, and high spectral resolution, the information available
from {\em RHESSI} \citep{2002SoPh..210....3L} is far superior for
imaging spectroscopy studies.\index{imaging spectroscopy}

\subsection{Imaging spectroscopy with \textit{RHESSI}}
\index{RHESSI@\textit{RHESSI}!imaging spectroscopy}\index{imaging spectroscopy}

One of the earliest results using {\em RHESSI} imaging
spectroscopy data was by \citet{2002SoPh..210..383A};
see also \cite{2002SoPh..210..229K},
who analyzed hard X-ray source height as a function
of energy, improving the previous statistical
analysis  \citep{1992PASJ...44L..89M}.
With the superior data available from {\em RHESSI}, \citet{2002SoPh..210..383A}
were able to determine accurate source locations as a function of photon energy
for the single event SOL2002-02-20T11:07 (C7.5), rather than a statistical average for an ensemble of events used by \citet{1992PASJ...44L..89M}.
\index{flare (individual)!SOL2002-02-20T11:07 (C7.5)!imaging spectroscopy}
\index{flare (individual)!SOL2002-02-20T11:07 (C7.5)!footpoint heights}
\index{footpoints!heights}
This was done by determining the
centroid location of a circular Gaussian forward-fit to each of
the two clearly resolved footpoints
in this event.
\index{hard X-rays!footpoints}
\index{footpoints}
Then, using thick-target modeling of the expected centroid location vs.
photon energy in a plane-parallel-stratified atmosphere model
\citep{2002SoPh..210..373B,2006AdSpR..38..962M,2008A&A...489L..57K},
they were able to deduce a density vs. height structure for the atmosphere that
was consistent with empirical models of flaring atmosphere density.
\index{flare models!thick-target!and HXR source height}

As mentioned at the outset of this section, imaging spectroscopy\index{imaging spectroscopy}
is, in principle, straightforwardly accomplished by constructing
two-dimensional maps of the source at different energies
\citep[e.g.,][]{2002SoPh..210..229K,2002SoPh..210..287S,2002SoPh..210..245S,2003ApJ...595L.107E,
2004ApJ...605..938J,2006ApJ...640..505A,2007A&A...471..705J,2007ApJ...666.1268A,2008ApJ...676..704L,
2009ApJ...698.2131D,2009ApJ...706..917P,2009ApJ...691..299S,2010ApJ...712L.131P}.
These maps are produced by applying image processing algorithms
\citep[e.g., back-projection, CLEAN, Maximum Entropy or Pixon; for details see][]{2002SoPh..210...61H}
to the temporally modulated
fluxes from each of {\em RHESSI}'s detectors, in which spatial
information on the source is encoded.
\index{imaging algorithms!back-projection}
\index{imaging algorithms!CLEAN}
\index{imaging algorithms!maximum entropy}
\index{imaging algorithms!pixons}
Then,``interesting'' regions
in the field of view are selected and the intensity in such
regions is determined in the map corresponding to each energy
range. There results from this a set of intensity-versus-energy
profiles, i.e., a count spectrum, for each feature.  Using
knowledge of the instrument response permits {\it photon} spectra
for each feature to be determined.
Finally, the corresponding spatially-resolved {\it electron}
spectra are constructed through forward fitting
or by applying regularized spectral inversion methods
\citep[e.g.,][]{2006ApJ...643..523B} to the spatially resolved
photon spectra.
Analysis of the variation of the electron spectrum throughout the target is
a powerful diagnostic of the physical processes affecting the
bremsstrahlung-producing electrons
\citep[see][]{2001ApJ...557..921E}.
\index{electrons!source spectrum}

\begin{figure}
\begin{center}
\includegraphics[width=0.9\textwidth]{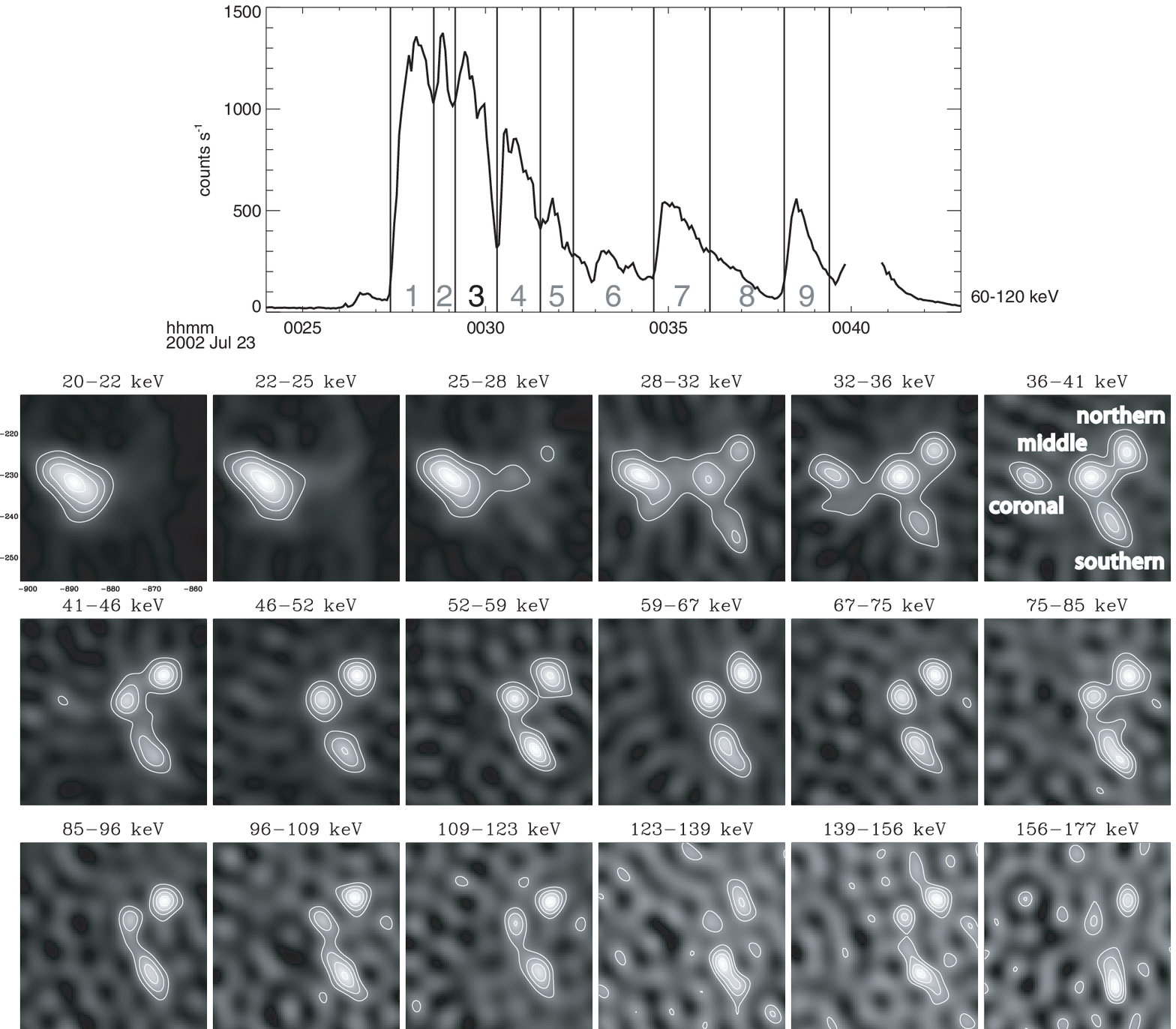}
\end{center}
\caption[]{{\it Top panel}: time profile for event. {\it Lower
panels}: Images of SOL2002-07-23T00:35 (X4.8)
in different
count energy channels, for time interval 3 of the event (cf., flux
profile in top panel).
\index{flare (individual)!SOL2002-07-23T00:35 (X4.8)!spectroscopic images}
\index{flare (individual)!SOL2002-07-23T00:35 (X4.8)!illustration}
The four main regions in the event are
labeled. Since this event was located near the East limb,
structures to the left are higher in the atmosphere
\citep[after][]{2003ApJ...595L.107E}.}
\label{fig:emslie_jul_23_images}
\end{figure}

\begin{figure}
\begin{center}
\includegraphics[width=0.9\textwidth]{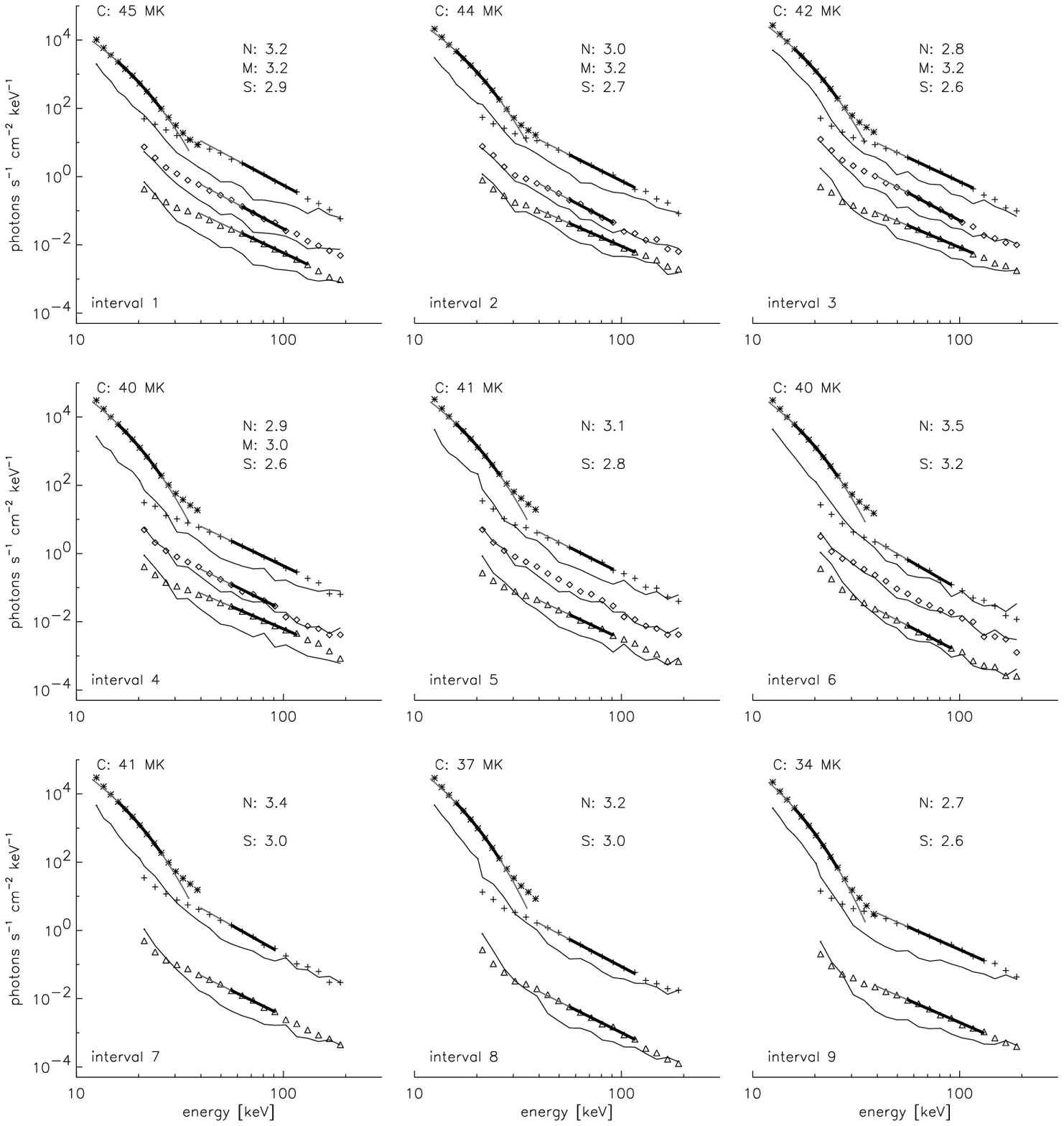}
\end{center}
\caption[]{Spectra for each of the features
identified in Figure~\ref{fig:emslie_jul_23_images}. For each time
interval, the ``coronal'' source\index{coronal sources}
\index{hard X-rays!coronal sources} is well-fit to a thermal spectrum
with the temperatures shown, and the ``footpoint''
\index{hard X-rays!footpoints}
\index{footpoints} sources are well-fit by power laws with the
spectral indices shown. The spectral indices
\index{hard X-rays!spectral index} of the various
footpoints\index{hard X-rays!footpoints}\index{footpoints} vary with
time; however, the spectral index {\it difference} between the
``North'' and ``South'' footpoints is relatively constant with
time \citep[after][]{2003ApJ...595L.107E}.}
\label{fig:emslie_jul_23_analysis}
\index{hard X-rays!footpoints}\index{footpoints}
\end{figure}

\citet{2003ApJ...595L.107E} performed such a
``stacked-image'' analysis for different energy bands via an imaging spectroscopy
analysis of SOL2002-07-23T00:35 (X4.8).
\index{flare (individual)!SOL2002-07-23T00:35 (X4.8)!imaging spectroscopy}
They identified four ``interesting'' features in the images
(Figure~\ref{fig:emslie_jul_23_images}) and constructed a count
spectrum for each (Figure~\ref{fig:emslie_jul_23_analysis}).
\index{coronal sources}\index{hard X-rays!coronal sources}
They identified one of these features as a ``coronal'' source,
and showed that its spectrum was consistent with a thermal source at a
temperature of order $45 \times 10^6$~K.  The spectra of the other
three sources were more power-law-like in form. This, plus their
location near the lower altitudes in the flaring structure, led
\citet{2003ApJ...595L.107E} to identify these features as chromospheric
footpoints.
\index{hard X-rays!footpoints}
\index{footpoints}\index{footpoints!heights}\index{footpoints!sizes}

Interestingly, the spectra of two of these
footpoints\index{hard X-rays!footpoints}\index{footpoints} (labeled ``North'' and ``South'' in
Figure~\ref{fig:emslie_jul_23_images}), while significantly
different and varying with time, maintained a relatively constant
{\it difference} throughout the event.
\index{magnetic structures!footpoint connections}
\index{hard X-rays!footpoints}\index{footpoints}
\index{hard X-rays!spectral index}
\citet{2003ApJ...595L.107E} interpreted the similar time
variation to a magnetic coupling of the two footpoints,
and the systematic difference in spectral indices as due to differential
spectral hardening associated with an asymmetric location of the
electron acceleration region.
From the magnitude ($\sim$0.3) of
the spectral-index difference, they obtained an estimate of the
differential column density between the acceleration region
\index{acceleration region} and each of the footpoints
\index{hard X-rays!footpoints}\index{footpoints}.  More
recently, however, \citet{2008SoPh..250...53S}, in
a statistical study of some 50 events, have pointed out that such
a large intervening column density should in general (but,
somewhat ironically, not necessarily in SOL2002-07-23T00:35)
produce more emission in the legs of the loop than was observed by
{\em RHESSI}; they conclude that electron precipitation in
asymmetric magnetic field geometries is a more reasonable
\index{magnetic structures!asymmetric footpoints}
explanation for the observed footpoint\index{footpoints!spectrum}
spectral differences.

While the above imaging spectroscopy technique appears relatively
straightforward to implement, there are a number of difficulties
that warrant some commentary.  First, for large events, pulse
pileup \citep{1976SSI.....2..239D,1976SSI.....2..523D,2002SoPh..210...33S} is
an issue.\index{pulse pileup}\index{RHESSI@\textit{RHESSI}!pulse pileup}\index{imaging spectroscopy!pulse pileup}
Pairs of low energy photons arriving nearly
simultaneously are detected as a single energy count. Since the
modulation of such counts corresponds to the low energy source, an
image made at the higher energy returns a ``ghost" image of the low
energy source. Second, the dynamic range \index{RHESSI@\textit{RHESSI}!dynamic
range} of {\em RHESSI} is such that features containing a few
percent of the total flux cannot be reliably imaged, so that
spectral information in the relevant energy ranges is not reliably
recovered.  For example, in the \citet{2003ApJ...595L.107E}
analysis of SOL2002-07-23T00:35, the spectrum of the
``coronal'' source\index{coronal sources!spectrum} was determined only at energies $\lapprox
40$~keV, while the spectrum of the footpoints\index{hard X-rays!footpoints}\index{footpoints}\index{flare (individual)!SOL2002-07-23T00:35 (X4.8)!footpoints}
was determined only at energies $\gapprox 30$~keV. Only in the
relatively narrow energy range from 30~to 40~keV could the
spectra of all the features in the source be obtained.

\subsection{Visibilities and imaging spectroscopy}
\index{imaging spectroscopy!visibilities}

It is important to realize that, because of the RMC technique used
by {\em RHESSI}, spatial information is encoded in the {\em
RHESSI} data in a distinctive way, namely in rapid time variations
of the detected counts in each of the {\em RHESSI} subcollimators.
\citet{2007SoPh..240..241S} have developed a technique in which
the observed temporal modulations produced are interpreted in
terms of a set of {\it visibilities}\index{visibilities!count}\index{hard X-rays!visibilities} (calibrated
measurements of specific spatial Fourier components of the source
distribution). As with image reconstruction in radio
interferometry, the set of visibilities\index{visibilities}\index{hard X-rays!visibilities}
thus determined can then be used to infer the spatial properties
of the X-ray source.  It is important to note (1) that
visibilities\index{visibilities}\index{hard X-rays!visibilities} are a ``first-order''
product of the {\em RHESSI} data, and (2) that the data can be
used to determine not only the values
of the visibilities\index{visibilities}\index{hard X-rays!visibilities}, but
also their quantitative uncertainties.

\citet{2008ApJ...673..576X} have analyzed a set of ten events,
each of which exhibits a rather simple, single-extended-source,
geometry.
They performed their imaging spectroscopy analysis by
assuming a parametric form of the source structure (a seven-parameter
curved elliptical Gaussian) and then forward fit not to the
actual images themselves, but rather to the corresponding source.
{\it visibilities}\index{visibilities}\index{visibilities!X-ray}\index{hard X-rays!visibilities}\index{imaging spectroscopy!forward fit}
Because both the visibilities and their
uncertainties were determined quantitatively, they were able to
deduce not only the best-fit values of the source parameters, but
also the uncertainties on the value of each parameter.
\citet{2008A&A...489L..57K} have performed such an analysis with circular Gaussian fits
(four parameters) of footpoints for the SOL2004-01-06T06:29 (M5.8)
flare and have measured the sizes and heights of the hard X-ray
sources in a few energy ranges between 18~and 250~keV.
\index{flare (individual)!SOL2004-01-06T06:29 (M5.8)!footpoint sizes}
\index{flare (individual)!SOL2004-01-06T06:29 (M5.8)!visibilities}
The height variations of footpoint emission with energy and the source size
with height have been found with an unprecedented vertical resolution
of $\sim$150~km at chromospheric heights of $400-1500$~km \citep[cf.,][]{2002SoPh..210..373B,2002SoPh..210..383A}.
The interpretation of these findings in terms of electron transport
models is given by \citet{chapter3}.
\index{chromospheric density model!and hard X-ray imaging}


One of the major goals of hard X-ray imaging spectroscopy is the
determination not just of the source structure as a function of
photon (or count) energy, and hence the variation of the hard
X-ray spectrum throughout the source, but rather of the variation
of the corresponding {\it electron} spectrum throughout the
source.  It is the spectral variation of this electron spectrum
that provides the key insight into the physics of electron
acceleration and propagation in solar flares.

\begin{figure}
\begin{center}
\includegraphics[width=0.9\textwidth]{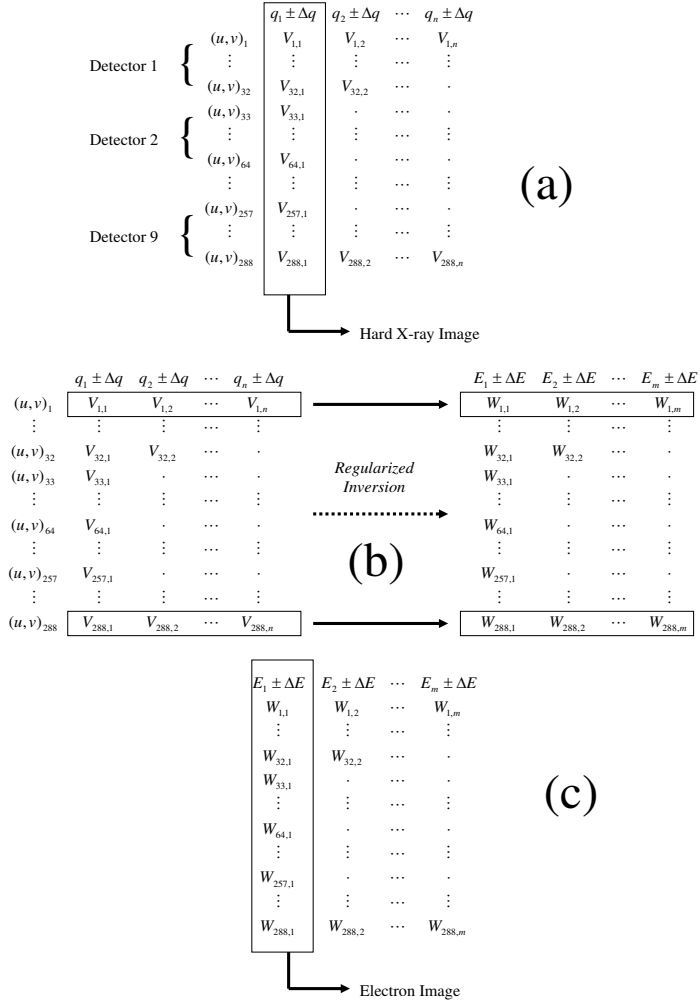}
\end{center}
\caption[]{Construction of electron flux images from measured
visibilities\index{visibilities!electron}\index{hard X-rays!visibilities}. Panel (a) shows the ``traditional'' approach to
image reconstruction, in which the visibility information in count
space from each of {\em RHESSI}'s nine detectors is used to
construct an image based on hard X-ray counts.  Note that there is
no imposition of energy smoothing in this process: images in
adjacent count energy channels $(q_i \pm \Delta q_i)$ can exhibit
significant differences due to count statistics.  These
differences would be further magnified if the count-based images
were inverted to yield electron images.  Panel (b) shows the
innovative approach of \citet{2007ApJ...665..846P}, in which the
energy spectrum for each count visibility (left panel) is
subjected to a regularized inversion procedure
\citep[see, e.g.,][]{2003ApJ...595L.127P} to yield the corresponding (smooth)
electron spectrum for each visibility.
\index{electrons!visibilities}\index{visibilities!electron}
Once all the electron visibilities
have been determined (right panel), they can be used
to yield electron images using the same image reconstruction
algorithms as used to produce count-based images (Panel (c)).  These {\it
electron flux images} are, by construction, necessarily smooth
across electron energy ranges $(E_j \pm \Delta E_j)$, and so are more
amenable to further analysis.}
\label{fig:emslie_visibility_matrices}
\end{figure}

Recognizing this, \citet{2007ApJ...665..846P} introduced a new
approach to imaging spectroscopy which is optimized to the
distinctive way in which spatial information is encoded in the
{\em RHESSI} data. Although it is possible to use the count
visibilities\index{visibilities!electron}\index{hard X-rays!visibilities}
determined rather straightforwardly from the raw data
to construct count images and then proceed to do imaging
spectroscopy analysis in the ``stacked image'' manner described
above, \citet{2007ApJ...665..846P} point out that such
``traditional'' imaging algorithms are completely ineffective in
smoothing in the energy direction, with the result that recovered
images corresponding to adjacent energy bins can exhibit
substantial differences. Further, {\em RHESSI}'s Fourier-component
approach to imaging detects ``patterns'' of emission, rather than
information in a ``pixel-by-pixel'' format, so that analysis of a
particular sub-region of a source is affected by the signal (and
noise) contained in all other features in the source.
\citet{2007ApJ...665..846P} therefore argue that imaging
spectroscopy analysis using {\em RHESSI} data is best accomplished
through conversion of the temporal modulations in terms of {\it
count visibilities}
and subsequent analysis of these count
visibilities to obtain information on the electron spectrum in the
spatial-frequency (rather than spatial) domain.\index{visibilities!X-ray}\index{hard X-rays!visibilities}
\index{visibilities!electron}\index{frequency!spatial}
This leads to a set of {\it electron visibilities},
which contain all the information on the variation of the electron
spectrum throughout the source (albeit as a function of spatial
frequency, rather than position).
\index{visibilities!electron}
\index{hard X-rays!visibilities}
If desired, such electron visibilities
can then be used to construct {\it electron flux
maps} using the same algorithms used to convert count visibilities
into count maps.\index{imaging spectroscopy!electrons}

Figure~\ref{fig:emslie_visibility_matrices} illustrates the
essence of this procedure.  Visibilities in count space are used
to determine the corresponding set of electron visibilities using the
Tikhonov regularization
technique \citep{ti63} previously used
\citep[e.g.,][]{2003ApJ...595L.127P} to ascertain
spatially-integrated electron spectra ${\overline F} (E)$ from
observations of spatially-integrated count (or photon) spectra
$I(\epsilon)$ (see Section \ref{sec:7inversion}).
\index{inversion algorithms!Tikhonov regularization}
\index{inverse problem!Tikhonov regularization}
Applied to visibilities, the Tikhonov regularization
method permits the determination of information on
electron fluxes at energies above the maximum count energy
observed \citep{2004SoPh..225..293K}. More importantly, it {\it
forces smoothness in the inferred electron visibility spectra at
each point in the spatial frequency domain} and so enhances real
features that persist over a relatively wide energy band, while
suppressing noise-related features that show up only over a narrow
range of energies.\index{frequency!spatial!electron visibilities}
The combination of visibility data and
the Tikhonov regularization methodology allows the derivation of the
most robust information on the spatial structure of the electron
flux spectrum image, the key quantity of physical interest.

\citet{2007ApJ...665..846P} have illustrated the power of the
electron visibility method by applying it to data obtained near
the peak of SOL2002-02-20T11:07 (C7.5)\index{flare (individual)!SOL2002-02-20T11:07 (C7.5)!imaging spectroscopy}.
Using visibilities\index{flare (individual)!SOL2002-02-20T11:07 (C7.5)!visibilities}
\index{flare (individual)!SOL2002-02-20T11:07 (C7.5)!footpoints}
from {\em RHESSI} RMCs
3~through~9, corresponding to spatial resolutions from $\sim$7$''$
to $\sim$183$''$, they construct the amplitude and phase of the
count visibilities $V(u,v;q)$
(counts~cm$^{-2}$~s$^{-1}$~keV$^{-1}$) as a function of Fourier
components $(u,v)$ and count energy~$q$.
They then apply the Tikhonov inversion algorithm to construct the corresponding set of
electron visibilities $W(u,v;E)$.\index{visibilities!electron}\index{hard X-rays!visibilities}\index{inversion algorithms!Tikhonov regularization}\index{u,v@(u,v)-plane}
By construction, these electron visibilities vary smoothly with electron energy $E$.

\begin{figure}
\begin{center}
\includegraphics[width=\textwidth]{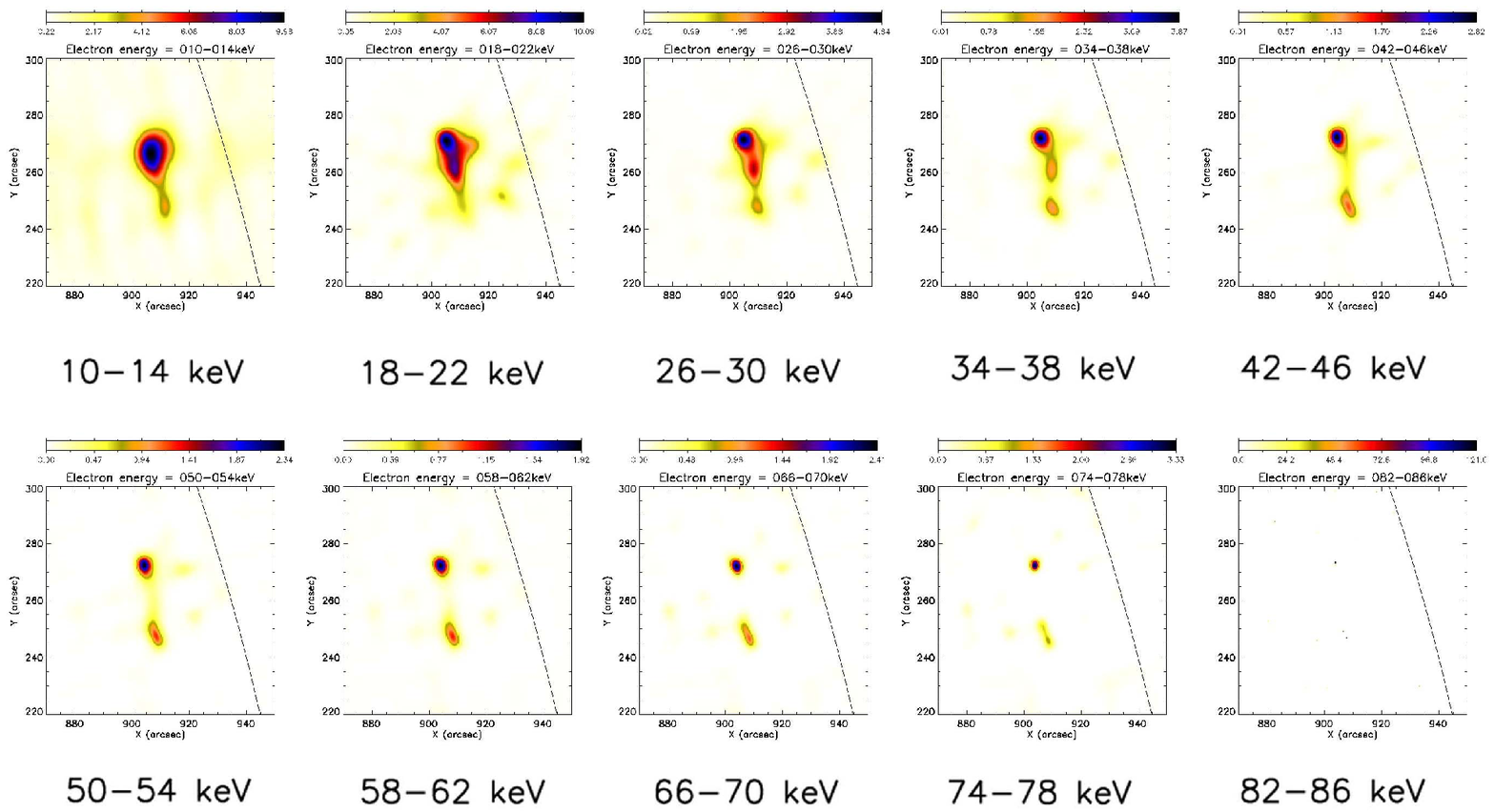}
\end{center}
\centering \caption{\footnotesize Electron flux spectral images
corresponding to the regularized electron flux spectral
visibilities,\index{visibilities!X-ray}\index{hard X-rays!visibilities}
obtained through application of
\citeauthor{2006ApJ...636.1159B}'s (2006) MEM-NJIT algorithm
\citep[after][]{2007ApJ...665..846P}.}
\label{fig:emslie_electron_images}
\end{figure}

\begin{figure}
\begin{center}
\includegraphics[width=0.9\textwidth]{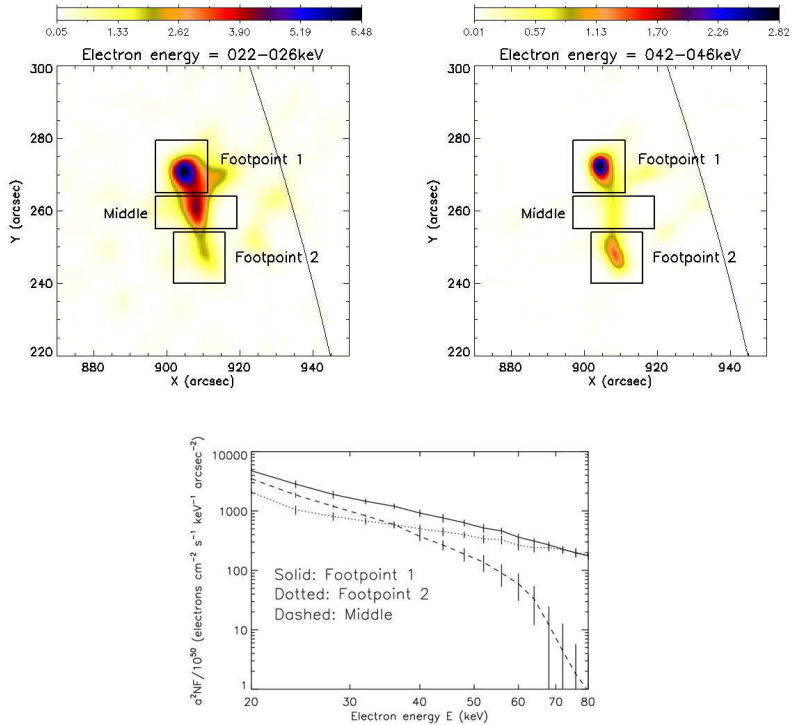}
\end{center}
\caption[]{{\it Top panels}: Electron images in the energy ranges
$22 - 26$~keV and $42-46$~keV, respectively. Three sub-regions of
interest are labeled on each image. Two of these correspond to
bright footpoint-like sources and one to a region midway between
the footpoints.  {\it Bottom panel}: Areally-averaged electron
flux spectra
(electrons~cm$^{-2}$~s$^{-1}$~keV$^{-1}$~arcsec$^{-2}$) for each
of the three sub-regions shown \citep[after][]{2007ApJ...665..846P}.}
\label{fig:emslie_image_spectra}
\index{hard X-rays!footpoints}\index{footpoints}
\end{figure}

Figure~\ref{fig:emslie_electron_images} shows electron flux images
for the illustrative SOL2002-02-20T11:07 (C7.5)\index{flare (individual)!SOL2002-02-20T11:07 (C7.5)!spectroscopic images} event. They
show evidence for two footpoints\index{hard X-rays!footpoints}\index{footpoints}, connected by a ``strand'' of
coronal flux. Because of the inherent smoothness demanded by the
regularized algorithm used to construct electron visibilities\index{visibilities!electron}\index{hard X-rays!visibilities} from
count visibilities, the electron flux images vary smoothly with
energy. As pointed out by \citet{2007ApJ...665..846P}, this contrasts
markedly with the behavior in the count images, for which the
image in each energy range is independent, so that statistical
fluctuations, including those in sidelobes from neighboring
features, result in a set of images that do {\it not} vary
smoothly with energy.
Inversion of the count spectra obtained from
such images leads to amplification of such noise and electron
spectra that exhibit large (and most probably) unphysical
features.

Three different spatial subregions in the source are highlighted
in Figure~\ref{fig:emslie_image_spectra}. Two of these regions
correspond to the footpoint sources visible at higher energies and
the other one to similarly-sized regions located approximately
midway between the two footpoints.\index{hard X-rays!footpoints}\index{footpoints}
The lower panel of
Figure~\ref{fig:emslie_image_spectra} shows the
areally-averaged\footnote{To get the total count spectrum for each
region [counts~cm$^{-2}$~s$^{-1}$~keV$^{-1}$], simply multiply the
areally-averaged spectrum by the area of that region, viz., $14.4
\times 14.4 = 207.36$~arcsec$^2$ (Footpoint~1), $22.8 \times 9 =
205.2$~arcsec$^2$ (Middle), and $14.4 \times 14.4 =
207.36$~arcsec$^2$ (Footpoint~2), respectively.} electron-flux
spectra (electrons~cm$^{-2}$~s$^{-1}$~keV$^{-1}$~arcsec$^{-2}$),
for each of these three subregions.
These spectra are sufficiently
smooth that significant conclusions regarding the variation of the
electron spectrum throughout the source can be made.

\section{Summary}\label{sec:7discussion} 



\subsection{Hard X-Ray emission processes}

With the launch of {\em RHESSI}, not only have new types of hard
X-ray data analysis become possible, but also the in-depth study of
X-ray producing processes has been triggered (Section
\ref{sec:7emission}). The roles of electron-ion and
electron-electron bremsstrahlung, free-bound electron-ion
emission, and Compton backscatter of primary photons, have been
highly scrutinized in view of the unprecedented quality of the
{\em RHESSI} data. Examples include:

$\bullet$ As pointed out in Section \ref{sec:7ee}, pure
electron-ion bremsstrahlung spectra have a spectral index $\gamma
\simeq \delta+1$\index{hard X-rays!spectral index}\index{spectral index},
for pure electron-electron bremsstrahlung
the resulting X-ray spectrum has a significantly shallower photon
spectrum, with $\gamma \simeq \delta$.  Hence, the importance of
the electron-electron bremsstrahlung contribution increases with
photon energy and the enhanced emission per electron leads to a
flattening of the photon spectrum $I(\epsilon)$ produced by a
given ${\overline F}(E)$, or, equivalently, a steepening of the
${\overline F}(E)$ form required to produce a given $I(\epsilon)$.

$\bullet$ The recent work on the importance of long-neglected
free-bound emission by energetic electrons (Section
\ref{sec:7fb}), with its emphasis on emission due to electron
capture onto various ionization states of various elements,
necessitates further work on the ionization structure of the solar
atmosphere and its time dependence during flares.

$\bullet$
The solar atmosphere above an X-ray emitting region can be safely treated
as an optically-thin medium, whereas the lower levels
of the atmosphere are optically thick for X-rays.
X-rays at energies below $\sim$11~keV are mostly photoelectrically
absorbed in the photosphere, while Compton scattering dominates
at the energies above.
As a result, X-rays emitted downwards can be Compton-backscattered
toward the observer; as discussed in Sections
\ref{sec:7albedo_spectr} and~\ref{sec:7albedo_image}, they play a
major role in the spectral and imaging characteristics of the
emission.

\subsection{Electron source spectrum}

The spatially-integrated photon spectrum is a key source of
information about the mean electron flux spectrum in solar flares.
High-energy-resolution data from {\em RHESSI} have permitted, for the
first time, not only the extraction of basic parameters through
forward-fitting the observed spectra (Section \ref{sec:7ff}), but
have also permitted the first reliable model-independent inversion
of observed spectra (Section \ref{sec:7inversion}). Although the
general form of a large number of nonthermal flare spectra can be
adequately approximated by an isothermal Maxwellian for the
low-energy component (Section \ref{sec:7thermal}), plus a broken
power law for the high-energy component, {\em RHESSI} data have
clearly demonstrated a wealth of features beyond such simple
parametric models: the presence of high-energy and low-energy
cutoffs (Sections~\ref{sec:7hecutoff} and~\ref{sec:7cutoff}),
spectral breaks (Section~\ref{sec:7knee}), and the presence
of an albedo component leading to ``dips'' in the mean electron spectrum
(Section~\ref{sec:7inversion}). The inclusion of an isotropic
albedo correction removes the need for low-energy cutoffs
and, if low-energy cutoffs exist in the mean electron spectrum,
they should be below $\sim$12~keV (Sections~\ref{sec:7albedo_spectr}
and~\ref{sec:7cutoff}).

\subsection{Anisotropy}

The electron-ion bremsstrahlung emission cross-section is generally anisotropic,
with a dipole-like diagram at low energies
(Section~\ref{sec:7angular}). The anisotropy or angular
distribution of X-ray emitting electrons can be measured in a
number of different ways, and various possibilities have been
employed with {\em RHESSI} data.
Parameter-free regularized electron flux spectra, reconstructed by assuming a parameterized
form of the electron angular distribution, highlight the need to
consider anisotropy in determining the true shape of the electron
flux spectrum (Section~\ref{sec:7stereo}).
A statistical study of
center-to-limb variations in the 15-50~keV energy range has
shown that an anisotropy factor $\alpha$, or the ratio of downward
and upward directed fluxes, for hard X-ray
emission that lies outside the range $[0.2,5]$ can be rejected
with 99\% confidence (Section~\ref{sec:7stat}).

The albedo portion of the observed spectrum can be rather
effectively used to infer the mean electron flux in
two directions simultaneously (Section~\ref{sec:7stereo}).
\index{hard X-rays!albedo}
\index{albedo}\index{photospheric albedo}
The reconstructions of the mean electron spectra in the downward and
upward directions suggest that X-ray emitting electrons need not
be significantly anisotropic\index{mean electron flux!lack of electron beaming} in a broad range of energies from tens of~keV to about 200~keV.
This imposes significant challenges
to solar flare models based on collisional transport; see
\cite{chapter2} and \cite{chapter3} for details.
\index{collisions!and transport}
\index{flare models!collisional transport}

{\em RHESSI} has been used to attempt polarization measurements in
two different energy bands using two different scattering processes.
The results are, however, of limited statistical significance.
If real, they would suggest that X-ray polarization
above 50~keV is significant (with values up to 50\% or more)
and that high quality polarization measurements would provide
significant constraints on particle acceleration models
(Section ~\ref{sec:7polar}).
\index{polarization!RHESSI@\textit{RHESSI} results}
The {\em RHESSI} design is not optimized for studies of polarization,
and attempts to measure it have been hampered by the small effective
area and the high background contribution. Future progress will require a
polarimetric instrument able to measure polarization with errors on the level
of 1 to 2\%.
Such measurements would allow detailed studies of the electron
beaming, quantifying its magnitude, and placing significant constraints on
the acceleration geometry.
\index{acceleration region!and polarization}

\subsection{Spatial variation of electron flux}


Imaging spectroscopy results (Section~\ref{sec:7spatial}) from
{\em RHESSI} have (not surprisingly) revealed that the hard X-ray
emission from solar flares is far from homogeneous: the centroid of
high energy emission moves downward with higher energy, presumably
due to the increased penetration of higher energy electrons;
coronal sources have radically different spectra than chromospheric
footpoints; and even systematic differences exist between different
footpoint sources in the same event.
\index{coronal sources}
\index{hard X-rays!coronal sources}
\index{hard X-rays!footpoints}
\index{footpoints}

Since {\em RHESSI} provides spatial information on detected hard
X-ray emission through a (Fourier-transform-based) rotating
modulation collimator technique, it follows that the
highest-fidelity spatial information is contained in the finite
number of Fourier components (``visibilities'') sampled.  Both
forward-fit and inversion techniques have been applied to {\em
RHESSI} imaging spectroscopy data, with very positive results,
including the first empirical estimates of the density and volume
of the electron acceleration region, and the recovery of {\it
electron flux maps} that, by construction, vary smoothly with
energy and so provide valuable information on the variation of the
emitting electron spectrum throughout the flare source.

\begin{acknowledgements} 

EPK, JCB and PCVM acknowledge the support of a PPARC/STFC Rolling
Grant, UC Berkeley NASA \textit{RHESSI} Visitor funds (JCB), a PPARC/STFCr
UK Advanced Fellowship and Royal Society Conference Grant (EPK)
and a Dorothy Hodgkin Scholarship (PCVM). JK acknowledges
support from Grant 205/06/P135 of the Grant Agency of the Czech
Republic and the research plan AV0Z10030501 of the Astronomical
Institute AS CR, v.~v.~i. AGE acknowledges support through NASA's
Office of Space Science and through a grant from the University of
California-Berkeley.  EPK, JCB, AGE, GDH, GJH, JK, AMM, MP and MP
have been supported in part by a grant from the International
Space Science Institute (ISSI) in Bern, Switzerland. EJS
acknowledges a grant from NASA Goddard Space Flight Center to the
University of Maryland and a Heliophysics GI grant 06-HGI06-15
from NASA HQ to NWRA for research in \textit{RHESSI} albedo determination
and applications. Financial support by the European Commission
through the SOLAIRE Network (MTRN-CT-2006-035484) is gratefully
acknowledged by EPK and JCB. AMM, MP and MP acknowledge a grant
by the Italian MIVR.

\end{acknowledgements}

\addcontentsline{toc}{section}{References}
\bibliography{book_chapter_references,ch7}
\bibliographystyle{ssrv}

\printindex
\end{document}